\documentclass[12pt,a4paper]{article}

\usepackage{ifthen} 
\newboolean{pdflatex}
\setboolean{pdflatex}{true} 

\newboolean{articletitles}
\setboolean{articletitles}{true} 

\newboolean{uprightparticles}
\setboolean{uprightparticles}{false} 

\def\paperauthors{LHCb collaboration} 
\def\paperasciititle{Angular analysis of the decay Bs -> phi e+ e-} 
\def\papertitle{Angular analysis of the decay $B_{s}^{0}\to\phi e^+e^-$} 
\def\paperkeywords{{High Energy Physics}, {LHCb}} 
\def\papercopyright{\the\year\ CERN for the benefit of the LHCb collaboration} 
\def\paperlicence{CC BY 4.0 licence}
\def\paperlicenceurl{https://creativecommons.org/licenses/by/4.0/}

\newif\ifEnableSectionTOCLinks
\EnableSectionTOCLinksfalse 

\usepackage[top=1in, bottom=1.25in, left=1in, right=1in]{geometry}

\usepackage{microtype}
\usepackage{lineno}  
\usepackage{xspace} 
\usepackage{caption} 

\usepackage{graphicx}  
\usepackage{color}
\usepackage{colortbl}
\graphicspath{{./figs/}} 

\usepackage{amsmath} 
\usepackage{amssymb}
\usepackage{amsfonts}
\usepackage{upgreek} 

\newcommand*\patchAmsMathEnvironmentForLineno[1]{%
\expandafter\let\csname old#1\expandafter\endcsname\csname #1\endcsname
\expandafter\let\csname oldend#1\expandafter\endcsname\csname
end#1\endcsname
 \renewenvironment{#1}%
   {\linenomath\csname old#1\endcsname}%
   {\csname oldend#1\endcsname\endlinenomath}%
}
\newcommand*\patchBothAmsMathEnvironmentsForLineno[1]{%
  \patchAmsMathEnvironmentForLineno{#1}%
  \patchAmsMathEnvironmentForLineno{#1*}%
}
\AtBeginDocument{%
\patchBothAmsMathEnvironmentsForLineno{equation}%
\patchBothAmsMathEnvironmentsForLineno{align}%
\patchBothAmsMathEnvironmentsForLineno{flalign}%
\patchBothAmsMathEnvironmentsForLineno{alignat}%
\patchBothAmsMathEnvironmentsForLineno{gather}%
\patchBothAmsMathEnvironmentsForLineno{multline}%
\patchBothAmsMathEnvironmentsForLineno{eqnarray}%
}

\usepackage[pdftex,
            pdfauthor={\paperauthors},
            pdftitle={\paperasciititle},
            pdfkeywords={\paperkeywords}]{hyperref}

\usepackage{hyperxmp}
\hypersetup{pdfcopyright={Copyright (C) \papercopyright}}
\hypersetup{pdflicenseurl={\paperlicenceurl}}

\usepackage[bottom,flushmargin,hang,multiple]{footmisc}

\usepackage[all]{hypcap} 

\usepackage{xspace} 
\usepackage{upgreek}

\def\lhcb   {\mbox{LHCb}\xspace}

\def\MagUp {\mbox{\em Mag\kern -0.05em Up}\xspace}

\ifthenelse{\boolean{uprightparticles}}%
{

 \def\Pmu         {\ensuremath{\upmu}\xspace}                 
 \def\Pnu         {\ensuremath{\upnu}\xspace}                 
                  
 \def\Ppi         {\ensuremath{\uppi}\xspace}

 \def\Ppsi        {\ensuremath{\uppsi}\xspace}

 \def\PDelta      {\ensuremath{\Delta}\xspace}                 
 \def\PXi         {\ensuremath{\Xi}\xspace}                 
 \def\PLambda     {\ensuremath{\Lambda}\xspace}                 
 \def\PSigma      {\ensuremath{\Sigma}\xspace}                 
 \def\POmega      {\ensuremath{\Omega}\xspace}                 
 \def\PUpsilon    {\ensuremath{\Upsilon}\xspace}
 \let\oldPi\Pi
 \def\PPi         {\ensuremath{\oldPi}\xspace}

 \def\PB      {\ensuremath{\mathrm{B}}\xspace}                 
 \def\PD      {\ensuremath{\mathrm{D}}\xspace}                 
 \def\PJ      {\ensuremath{\mathrm{J}}\xspace}                 
 \def\PK      {\ensuremath{\mathrm{K}}\xspace}                 
 \def\Pb      {\ensuremath{\mathrm{b}}\xspace}                 
 \def\Pc      {\ensuremath{\mathrm{c}}\xspace}                 
                  
 \def\Pe      {\ensuremath{\mathrm{e}}\xspace}                 
 \def\Ps      {\ensuremath{\mathrm{s}}\xspace}

 \def\thebaroffset{0.0em}
}
{

 \def\Pmu         {\ensuremath{\mu}\xspace}                 
 \def\Pnu         {\ensuremath{\nu}\xspace}                 
                  
 \def\Ppi         {\ensuremath{\pi}\xspace}

 \def\Ppsi        {\ensuremath{\psi}\xspace}                 
                  
 \mathchardef\PDelta="7101
 \mathchardef\PXi="7104
 \mathchardef\PLambda="7103
 \mathchardef\PSigma="7106
 \mathchardef\POmega="710A
 \mathchardef\PUpsilon="7107
 \mathchardef\PPi="7105
 \def\PB      {\ensuremath{B}\xspace}                 
 \def\PD      {\ensuremath{D}\xspace}                 
 \def\PJ      {\ensuremath{J}\xspace}                 
 \def\PK      {\ensuremath{K}\xspace}                 
 \def\Pb      {\ensuremath{b}\xspace}                 
 \def\Pc      {\ensuremath{c}\xspace}                 
                  
 \def\Pe      {\ensuremath{e}\xspace}                 
 \def\Ps      {\ensuremath{s}\xspace}

 \def\thebaroffset{0.18em}
}
\newcommand{\offsetoverline}[2][\thebaroffset]{\kern #1\overline{\kern -#1 #2}}%

\makeatletter
\ifcase \@ptsize \relax
  \newcommand{\miniscule}{\@setfontsize\miniscule{4}{5}}
\or
  \newcommand{\miniscule}{\@setfontsize\miniscule{5}{6}}
\or
  \newcommand{\miniscule}{\@setfontsize\miniscule{5}{6}}
\fi
\makeatother

\DeclareRobustCommand{\optbar}[1]{\shortstack{{\miniscule (\rule[.5ex]{1.25em}{.18mm})}
  \\ [-.7ex] $#1$}}

\def\en         {{\ensuremath{\Pe^-}}\xspace}   
\def\ep         {{\ensuremath{\Pe^+}}\xspace}
\def\epm        {{\ensuremath{\Pe^\pm}}\xspace} 
 
\def\epem       {{\ensuremath{\Pe^+\Pe^-}}\xspace}

\def\mumu       {{\ensuremath{\Pmu^+\Pmu^-}}\xspace}

\def\ellm       {{\ensuremath{\ell^-}}\xspace}
\def\ellp       {{\ensuremath{\ell^+}}\xspace}

\def\ellell     {\ensuremath{\ell^+ \ell^-}\xspace}

\def\neu        {{\ensuremath{\Pnu}}\xspace}
\def\neub       {{\ensuremath{\overline{\Pnu}}}\xspace}

\def\squark    {{\ensuremath{\Ps}}\xspace}

\def\cquark    {{\ensuremath{\Pc}}\xspace}
\def\cquarkbar {{\ensuremath{\overline \cquark}}\xspace}
\def\ccbar     {{\ensuremath{\cquark\cquarkbar}}\xspace}
\def\bquark    {{\ensuremath{\Pb}}\xspace}

\def\pion   {{\ensuremath{\Ppi}}\xspace}

\def\pim    {{\ensuremath{\pion^-}}\xspace}

\def\kaon    {{\ensuremath{\PK}}\xspace}

\def\KorKbar {\kern \thebaroffset\optbar{\kern -\thebaroffset \PK}{}\xspace}

\def\Kp      {{\ensuremath{\kaon^+}}\xspace}
\def\Km      {{\ensuremath{\kaon^-}}\xspace}
\def\Kpm     {{\ensuremath{\kaon^\pm}}\xspace}

\def\Kstarz  {{\ensuremath{\kaon^{*0}}}\xspace}

\def\Kstar   {{\ensuremath{\kaon^*}}\xspace}

\def\D       {{\ensuremath{\PD}}\xspace}

\def\DorDbar {\kern \thebaroffset\optbar{\kern -\thebaroffset \PD}\xspace}

\def\Dp      {{\ensuremath{\D^+}}\xspace}
\def\Dm      {{\ensuremath{\D^-}}\xspace}

\def\DpDm    {\ensuremath{\Dp {\kern -0.16em \Dm}}\xspace}

\def\Dsm     {{\ensuremath{\D^-_\squark}}\xspace}

\def\B       {{\ensuremath{\PB}}\xspace}
\def\Bbar    {{\ensuremath{\offsetoverline{\PB}}}\xspace}

\def\BorBbar {\kern \thebaroffset\optbar{\kern -\thebaroffset \PB}\xspace}
\def\Bz      {{\ensuremath{\B^0}}\xspace}

\def\Bd      {{\ensuremath{\B^0}}\xspace}

\def\BdorBdbar {\kern \thebaroffset\optbar{\kern -\thebaroffset \Bd}\xspace}

\def\Bs      {{\ensuremath{\B^0_\squark}}\xspace}
\def\Bsb     {{\ensuremath{\Bbar{}^0_\squark}}\xspace}
\def\BsorBsbar {\kern \thebaroffset\optbar{\kern -\thebaroffset \Bs}\xspace}

\def\jpsi     {{\ensuremath{{\PJ\mskip -3mu/\mskip -2mu\Ppsi}}}\xspace}
\def\psitwos  {{\ensuremath{\Ppsi{(2S)}}}\xspace}

\def\Y#1S{\ensuremath{\PUpsilon{(#1S)}}\xspace}

\def\LorLbar     {\kern \thebaroffset\optbar{\kern -\thebaroffset \PLambda}\xspace}

\newcommand{\decay}[2]{\mbox{\ensuremath{#1\!\to #2}}\xspace} 

\def\to                 {\ensuremath{\rightarrow}\xspace}

\def\qsq       {{\ensuremath{q^2}}\xspace}

\def\CP                {{\ensuremath{C\!P}}\xspace}

\def\FL       {\ensuremath{F_{\mathrm{L}}}\xspace}
\def\AT#1     {\ensuremath{A_{\mathrm{T}}^{#1}}\xspace}           

\def\C#1      {\ensuremath{\mathcal{C}_{#1}}\xspace}                       
\def\Cp#1     {\ensuremath{\mathcal{C}_{#1}^{'}}\xspace}                    
\def\Ceff#1   {\ensuremath{\mathcal{C}_{#1}^{\mathrm{(eff)}}}\xspace}        
\def\Cpeff#1  {\ensuremath{\mathcal{C}_{#1}^{'\mathrm{(eff)}}}\xspace}       
\def\Ope#1    {\ensuremath{\mathcal{O}_{#1}}\xspace}                       
\def\Opep#1   {\ensuremath{\mathcal{O}_{#1}^{'}}\xspace}                    

\newcommand{\aunit}[1]{\ensuremath{\text{\,#1}}}       

\newcommand{\tev}{\aunit{Te\kern -0.1em V}\xspace}
\newcommand{\gev}{\aunit{Ge\kern -0.1em V}\xspace}
\newcommand{\mev}{\aunit{Me\kern -0.1em V}\xspace}
\newcommand{\kev}{\aunit{ke\kern -0.1em V}\xspace}
\newcommand{\ev}{\aunit{e\kern -0.1em V}\xspace}
 
\newcommand{\mevc}{\ensuremath{\aunit{Me\kern -0.1em V\!/}c}\xspace}
\newcommand{\gevc}{\ensuremath{\aunit{Ge\kern -0.1em V\!/}c}\xspace}
\newcommand{\mevcc}{\ensuremath{\aunit{Me\kern -0.1em V\!/}c^2}\xspace}
\newcommand{\gevcc}{\ensuremath{\aunit{Ge\kern -0.1em V\!/}c^2}\xspace}
\newcommand{\gevgevcccc}{\ensuremath{\gev^2\!/c^4}\xspace} 

\def\fb   {\ensuremath{\aunit{fb}}\xspace}
\def\invfb   {\ensuremath{\fb^{-1}}\xspace}

\def\deriv {\ensuremath{\mathrm{d}}}

\def\gsim{{~\raise.15em\hbox{$>$}\kern-.85em
          \lower.35em\hbox{$\sim$}~}\xspace}
\def\lsim{{~\raise.15em\hbox{$<$}\kern-.85em
          \lower.35em\hbox{$\sim$}~}\xspace}

\def\evtgen     {\mbox{\textsc{EvtGen}}\xspace}

\def\geant      {\mbox{\textsc{Geant4}}\xspace}

\def\photos     {\mbox{\textsc{Photos}}\xspace}

\def\pythia     {\mbox{\textsc{Pythia}}\xspace}

\def\tell1  {TELL1\xspace}
\def\ukl1   {UKL1\xspace}

\newcommand{\eg}{\mbox{\itshape e.g.}\xspace}
\newcommand{\ie}{\mbox{\itshape i.e.}\xspace}

\newcommand{\lhcborcid}[1]{\href{https://orcid.org/#1}{\hspace*{0.1em}\raisebox{-0.45ex}{\includegraphics[width=1em]{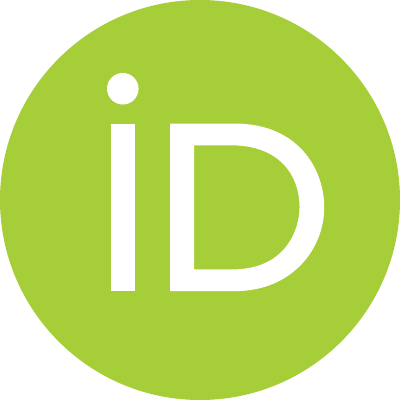}}}}

\hypersetup{
  colorlinks   = true, 
  urlcolor     = blue, 
  linkcolor    = blue, 
  citecolor    = red   
}

\ifEnableSectionTOCLinks
    \usepackage[explicit]{titlesec} 
    
    \let\oldcontentsline\contentsline
    \renewcommand

    \titleformat{\section}{\normalfont\Large\bf}{\hyperlink{tocsection.\thesection}{{\thesection} \parbox[t]{\dimexpr\textwidth-1pc}{#1}}}{1pc}{}

    \titleformat{\subsection}{\normalfont\bf}{\hyperlink{tocsubsection.\thesubsection}{{\thesubsection} \parbox[t]{\dimexpr\textwidth-1pc}{#1}}}{1pc}{}

    \titleformat{name=\section,numberless}[display]{}{}{0pt}{\normalfont\Huge\bfseries #1}
\fi

\usepackage{cite} 
\usepackage{mciteplus}

\def\BsPhiLL     {\decay{\Bs}{\phi\ellp\ellm}} 
\def\BsPhiEE     {\decay{\Bs}{\phi\ep\en}} 
\def\BsPhiMM     {\decay{\Bs}{\phi\mumu}} 
\def\BsPsiTwoSPhi{\decay{\Bs}{\psitwos\phi}}
\def\BsJpsiPhi   {\decay{\Bs}{\jpsi\phi}}
\def\KKEE        {\ensuremath{\Kp\Km\ep\en}\xspace}
\def\mKKEE       {\ensuremath{m(\Kp\Km\ep\en)}\xspace}

\def\Gammas      {\ensuremath{\Gamma_s}\xspace}
\def\DeltaGammas {\ensuremath{\Delta\Gamma_s}\xspace}
\def\thetak      {\ensuremath{\theta_K}\xspace}
\def\thetae      {\ensuremath{\theta_e}\xspace}
\def\costhetak   {\ensuremath{\cos\theta_K}\xspace}
\def\costhetae   {\ensuremath{\cos\theta_e}\xspace}
\def\IntFL       {\ensuremath{\langle F_{\rm L} \rangle}\xspace}
\def\FL          {\ensuremath{F_{\rm L}}\xspace}
\def\IntSthree   {\ensuremath{\langle S_{3} \rangle}\xspace}
\def\Sthree      {\ensuremath{S_{3}}\xspace}
\def\IntAnine    {\ensuremath{\langle A_{9} \rangle}\xspace}
\def\Anine       {\ensuremath{A_{9}}\xspace}
\def\IntAsix     {\ensuremath{\langle A_{6} \rangle}\xspace}
\def\Asix        {\ensuremath{A_{6}}\xspace}
\def\IntAsixp    {\ensuremath{\langle A^\prime_{6} \rangle}\xspace}

\def\qsqtrue     {\ensuremath{q^2_{\mathrm{true}}}\xspace}

\usepackage{booktabs}

\begin{document}

\renewcommand{\thefootnote}{\fnsymbol{footnote}}
\setcounter{footnote}{1}


\begin{titlepage}
\pagenumbering{roman}

\vspace*{-1.5cm}
\centerline{\large EUROPEAN ORGANIZATION FOR NUCLEAR RESEARCH (CERN)}
\vspace*{1.5cm}
\noindent
\begin{tabular*}{\linewidth}{lc@{\extracolsep{\fill}}r@{\extracolsep{0pt}}}
\ifthenelse{\boolean{pdflatex}}
{\vspace*{-1.5cm}\mbox{\!\!\!\includegraphics[width=.14\textwidth]{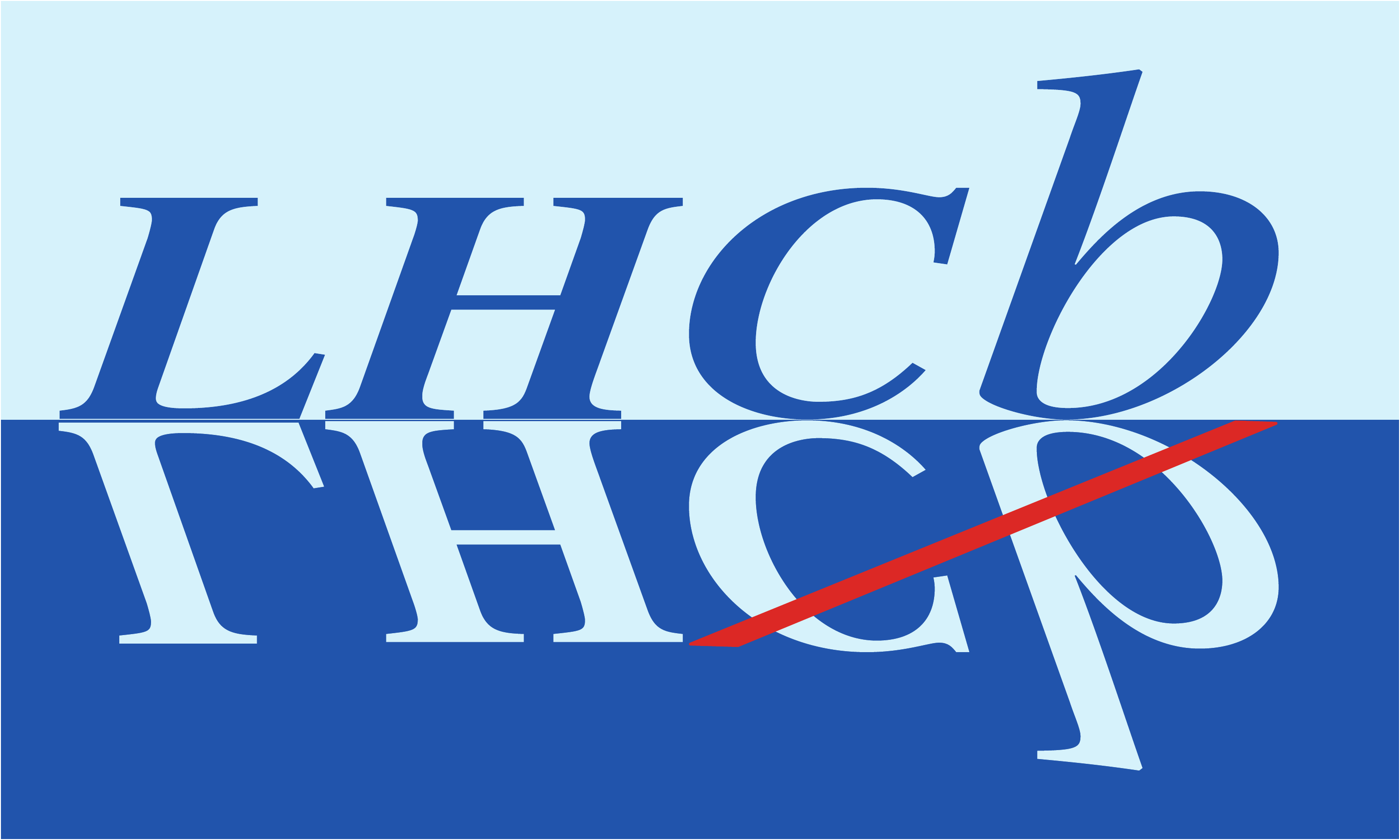}} & &}%
{\vspace*{-1.2cm}\mbox{\!\!\!\includegraphics[width=.12\textwidth]{figs/lhcb-logo.eps}} & &}%
\\
 & & CERN-EP-2025-064 \\  
 & & LHCb-PAPER-2025-006 \\  
 & & July 17, 2025 \\ 
 & & \\
\end{tabular*}

\vspace*{4.0cm}

{\normalfont\bfseries\boldmath\huge
\begin{center}
  \papertitle 
\end{center}
}

\vspace*{2.0cm}

\begin{center}
\paperauthors\footnote{Authors are listed at the end of this paper.}
\end{center}

\vspace{\fill}

\begin{abstract}
  \noindent
   An angular analysis of the decay \decay{\Bs}{\phi\ep\en} is presented, using proton-proton collision data collected with the LHCb detector between 2011 and 2018 at centre-of-mass energies of 7, 8 and 13\tev. The combined dataset corresponds to an integrated luminosity of 9\invfb. Observables are determined by fitting time-integrated projections of the angular distribution in three bins of dielectron mass squared, $q^2$, corresponding to $[0.1,1.1]$, $[1.1,6.0]$ and $[15.0,19.0]\gevgevcccc$. 
   The results are compatible with predictions based on the Standard Model of particle physics.  
\end{abstract}

\vspace*{2.0cm}

\begin{center}
  Published in JHEP {\bf 07} (2025) 069
\end{center}

\vspace{\fill}

{\footnotesize 
\centerline{\copyright~\papercopyright. \href{\paperlicenceurl}{\paperlicence}.}}
\vspace*{2mm}

\end{titlepage}


\newpage
\setcounter{page}{2}
\mbox{~}



\renewcommand{\thefootnote}{\arabic{footnote}}
\setcounter{footnote}{0}


\cleardoublepage


\pagestyle{plain} 
\setcounter{page}{1}
\pagenumbering{arabic}


\section{Introduction}
\label{sec:introduction}

The \BsPhiLL decays, where $\ell$ represents an electron, a muon or a tau lepton, proceed through a \bquark- to \squark-quark flavour-changing neutral-current transition.
Such processes are suppressed in the Standard Model (SM) of particle physics. 
The rate and angular distribution of these decay can be modified in many SM extensions, with distinct effects possible in final states involving different lepton flavours. 
Interestingly, measurements of a variety of decays involving \decay{\bquark}{\squark\ellell} transitions are found to be in tension with SM predictions. 
The largest discrepancies between measurements and predictions are seen in the rates of \decay{\B}{\kaon\mumu}~\cite{LHCb-PAPER-2014-006,CMS:2024syx,Belle:2019xld} and \BsPhiMM~\cite{LHCb-PAPER-2021-014} decays, as well as in the rate and angular distribution of \decay{\Bz}{\Kstarz\mumu} decays~\cite{LHCb-PAPER-2020-002,LHCb-PAPER-2024-011, LHCb-PAPER-2023-032, LHCb-PAPER-2023-033,CMS:2024atz,ATLAS:2018gqc,Belle:2019oag,Belle:2016fev}. 
Unfortunately, the interpretation of these discrepancies is complicated by long-distance contributions to the decay rate, which are difficult to estimate reliably in the SM (see for example Refs.~\cite{Ciuchini:2015qxb,Gubernari:2022hxn}). 
Such contributions do not impact comparisons between decays to different lepton flavours and significant discrepancies between measurements of observables in dielectron and dimuon final states would indicate a breakdown of the SM. 

The consistency between the rates of the \decay{B^{(+,0)}}{K^{(+,*0)}\mumu} and \decay{B^{(+,0)}}{K^{(+,*0)}\epem} decays has been studied extensively in recent years.\footnote{
    The symbols $\Kstar$ and $\phi$ are used to refer to the $K^*(892)$ and $\phi(1020)$ mesons, respectively, throughout the paper.
} 
Measurements of the rates of decays to dielectron and dimuon final states are found to be compatible~\cite{LHCb-PAPER-2022-045,LHCb-PAPER-2022-046}, showing a similar pattern of tensions with SM predictions, consistent with universal couplings to the different lepton flavours. 
The results of a study of the angular distributions of \decay{\Bz}{\Kstarz\epem} decays are also in agreement with similar measurements of the \decay{\Bz}{\Kstarz\mumu} final state~\cite{LHCb-PAPER-2024-022,LHCb-PAPER-2020-002}.
A recent study of the \BsPhiEE decay rate in bins of dilepton mass squared, \qsq, likewise found consistent values with those measured in \BsPhiMM decays~\cite{LHCb-PAPER-2024-032}.
A separate study of the angular distribution of the \BsPhiEE decay with \qsq\ in the range $[0.0009,0.2615]\gevgevcccc$, obtained results compatible with SM predictions~\cite{LHCb-PAPER-2024-030}. 
It is therefore of interest to study angular observables in \BsPhiEE decays at higher \qsq values, to test whether they are consistent with SM predictions and with the measurements made in \BsPhiMM decays~\cite{LHCb-PAPER-2021-022}.

This paper describes the first angular analysis of the \BsPhiEE decay in the \qsq ranges $[0.1,1.1]$, $[1.1, 6.0]$ and $[15.0, 19.0]\gevgevcccc$. 
The \qsq range $[6.0,15.0]\gevgevcccc$ is dominated by contributions from \BsJpsiPhi and \BsPsiTwoSPhi decays, involving \decay{\bquark}{\ccbar\squark} quark-level transitions. 
This \qsq range is used as a control region for the analysis. 
The results presented in this paper are based on a proton-proton collision dataset, corresponding to 9\invfb of integrated luminosity, collected with the LHCb detector between 2011 and 2018. 
Since the final state of this decay is not flavour specific, the notation \BsPhiEE is used to refer to the sum of both \Bs and \Bsb meson decays to the same final state. 

\begin{figure}[!tb]
    \centering
    \includegraphics[width=0.6\linewidth]{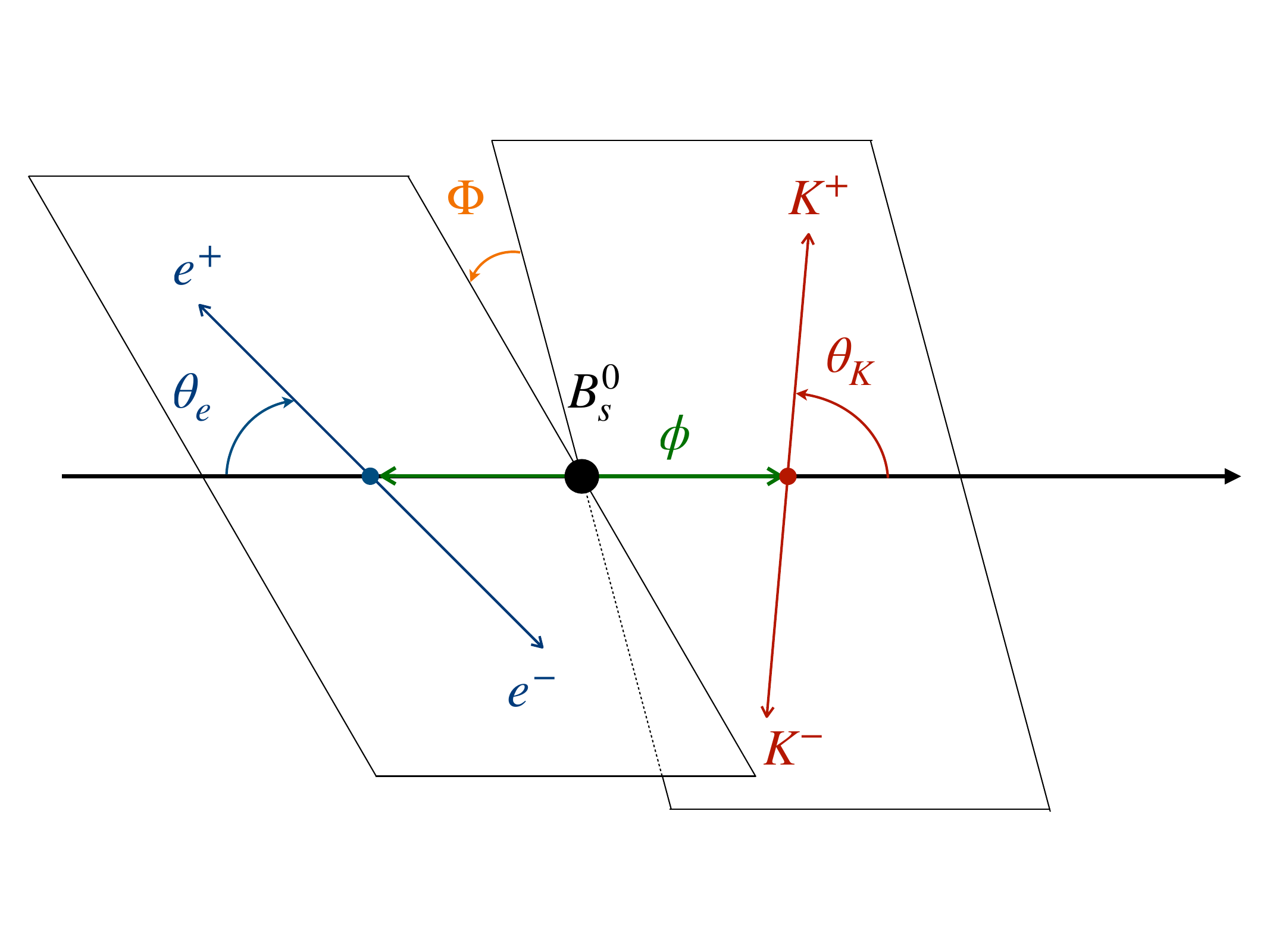}
    \caption{
    Illustration of the three decay angles that define the angular basis used for the \BsPhiEE decay. 
    }
    \label{fig:angles}
\end{figure}

The angular distribution of the \BsPhiEE decay, where the $\phi$ meson decays to $\Kp\Km$, can be parametrised by three decay angles: 
the angle \thetak between the \Kp direction and the direction opposite that of the \Bs meson in the $\phi$ meson rest frame; 
the angle \thetae between the \ep direction and the direction opposite that of the \Bs meson in the dielectron rest frame; 
and the angle $\Phi$ between the decay planes of the $\phi$ meson and dielectron pair in the rest frame of the \Bs meson. 
The angles are illustrated in Fig.~\ref{fig:angles}.
The decay-time-dependent angular distribution of the decay is described in Sec.~\ref{sec:distribution}, following the formalism of Ref.~\cite{Descotes-Genon:2015hea}. 
Due to the limited size of the data sample it is not possible to determine the full three-dimensional angular distribution, neither is it feasible to apply initial-state flavour-tagging to distinguish \Bs\ from \Bsb\ decays.
A subset of the angular observables, \FL, \Asix, \Sthree and \Anine are determined by analysing reduced angular distributions of the decay, separately in the \thetae and \thetak angles and in $\Phi$. 
The observable $F_{\rm L}$ is the longitudinal polarisation fraction of the $\phi$ meson. 
The asymmetry $A_{6}$ is even under time-reversal and generates a forward-backward asymmetry in the dilepton system.
Conversely, the asymmetry \Anine is odd under time-reversal~\cite{Bobeth:2008ij}.
The observable \Sthree is proportional to the difference between the magnitude squared of the \CP-odd and \CP-even amplitudes that contribute to the transverse polarisation of the $\phi$ meson. 
The observables \Sthree and \Anine are associated with $\cos2\Phi$ and $\sin2\Phi$ variations of the $\Phi$ distribution, respectively. 

The remainder of the paper is organised as follows: 
the parametrisation used to describe the angular distribution of the \BsPhiEE decay is introduced in Sec.~\ref{sec:distribution}; 
Sec.~\ref{sec:detector} contains a description of the LHCb detector and the production of simulated samples used in the analysis; 
in Sec.~\ref{sec:selection} the selection of \Bs candidates is discussed;
corrections for the nonuniform efficiency of the event reconstruction and candidate selection are described in Sec.~\ref{sec:efficiency:angular}; 
the determination of the observables from the candidates \KKEE mass, \mKKEE, and angular distributions is described in Sec.~\ref{sec:fitting};
Sec.~\ref{sec:validation} is focussed on the validation of the analysis using pseudoexperiments and \BsJpsiPhi decays in data; 
sources of systematic uncertainty are discussed in Sec.~\ref{sec:systematics};
and results are presented in Sec.~\ref{sec:results}.

\section{Angular distribution}
\label{sec:distribution} 

The full angular distribution of \Bs and \Bsb mesons decaying to $\phi\ep\en$ can be expressed as~\cite{Descotes-Genon:2015hea}
\begin{align}
\frac{\deriv^{4}[\Gamma(t) + \bar{\Gamma}(t)]}{\deriv\qsq\,\deriv\!\costhetak\,\deriv\!\costhetae\,\deriv\Phi} = \sum\limits_{i}\left[ J_{i}(\qsq,t) + \tilde{J}_{i}(\qsq,t) \right] f_{i}(\costhetak,\costhetae,\Phi)\,,
\label{eq:fullEquation}
\end{align} 
where the $f_i$ are functions of the decay angles. 
The decay-time- and \qsq-dependent angular coefficients have a decay-time dependence given by 
\begin{align}
\begin{split}
J_{i}(\qsq,t) + \tilde{J}_{i}(\qsq,t) = e^{-\Gammas t} & \Bigg\{ \left[ J_{i}(\qsq,0) + \tilde{J}_{i}(\qsq,0)\right] \cosh \left( \frac{\DeltaGammas}{2} t \right)  \\  & \quad - h_i(\qsq) \sinh\left( \frac{\DeltaGammas}{2} t \right) \Bigg\}\,.
\end{split}
\end{align} 
Here, \Gammas is the average of, and \DeltaGammas the difference between, the light and heavy \Bs mass eigenstates' widths.
The coefficients $J_i$, $\tilde{J}_{i}$ and $h_{i}$ at $t=0$ can be expressed in terms of bilinear combinations of decay amplitudes. 
Due to the limited size of the available dataset, the measurement is performed integrated in decay time. 
Neglecting for now decay-time acceptance effects, the term involving $h_i$ is diluted relative to that involving $J_{i} + \tilde{J}_{i}$ by a factor $y_s =  \tfrac{\DeltaGammas}{2\Gammas} \approx 0.06$~\cite{HFLAV21}, \ie
\begin{align}
\int\limits_{0}^{\infty} [J_{i}(\qsq, t) + \tilde{J}_{i}(\qsq, t)] \deriv t = \frac{1}{\left(1-y_s^2\right)\Gamma_s} \left[ \left( J_{i}(\qsq,0) + \tilde{J}_{i}(\qsq,0)\right) - y_{s} h_i(\qsq) \right]\,. 
\label{eq:hi-dilution-noAcceptance}
\end{align} 

Reduced distributions are obtained integrating either over $\Phi$ or over both $\cos\theta_K$ and $\cos\theta_e$, with the result that many of the terms in the sum of Eq.~\eqref{eq:fullEquation} vanish.
After integration over $\Phi$ and decay time, and neglecting contributions that are suppressed by the small electron mass, the distribution contains only terms with $i \in [1c, 1s, 2c, 2s, 6s]$.
In the notation of Refs.~\cite{Descotes-Genon:2015hea,Descotes-Genon:2022qce}, it can be written as 
\begin{align}
\begin{split}
\frac{1}{\deriv[\Gamma + \tilde{\Gamma}]/\deriv\qsq}\frac{\deriv^{3}[\Gamma + \tilde{\Gamma}]}{\deriv\qsq\,\deriv\!\costhetak\,\deriv\!\costhetae} = \frac{9}{16} &  \left[ \frac{3}{4}(1 - \langle F_{\rm L} \rangle) \sin^{2}\thetak (1 + \tfrac{1}{3} \cos 2\thetae) \right. \\ 
&  \left. \phantom{\frac{1}{1}} + \langle F_{\rm L} \rangle \cos^{2}\thetak (1 - \cos 2\thetae ) \right. \\ 
&  \left. \phantom{\frac{1}{1}} + \langle A_{6} \rangle \sin^{2} \thetak \cos\thetae \right]\,.
\end{split} 
\label{eq:twodimensional}
\end{align} 
The last term can also be expressed in terms of $\IntAsixp=\IntAsix/(1-\IntFL)$. 
The reduced distribution after integration over $\cos\theta_K$, $\cos\theta_e$ and decay time contains only terms with $i \in [3, 9]$, and can be expressed as 
\begin{align}
\frac{1}{\deriv[\Gamma + \tilde{\Gamma}]/\deriv\qsq}\frac{\deriv^{2}[\Gamma + \tilde{\Gamma}]}{\deriv\qsq\,\deriv\Phi} = \frac{1}{2\pi} \left[ 1 + \langle S_{3} \rangle \cos 2\Phi + \langle A_{9}\rangle \sin 2\Phi \right]\,.
\label{eq:phi} 
\end{align}
Here, the triangular braces indicate that the observables include contributions from the $h_i$ terms and are different from the observables at $t=0$, which will be referred to without the braces. 
These reduced distributions neglect contributions that are suppressed by the small size of the electron mass squared compared to \qsq. 

\begin{figure}[!tb]
    \centering
    \includegraphics[width=0.8\linewidth]{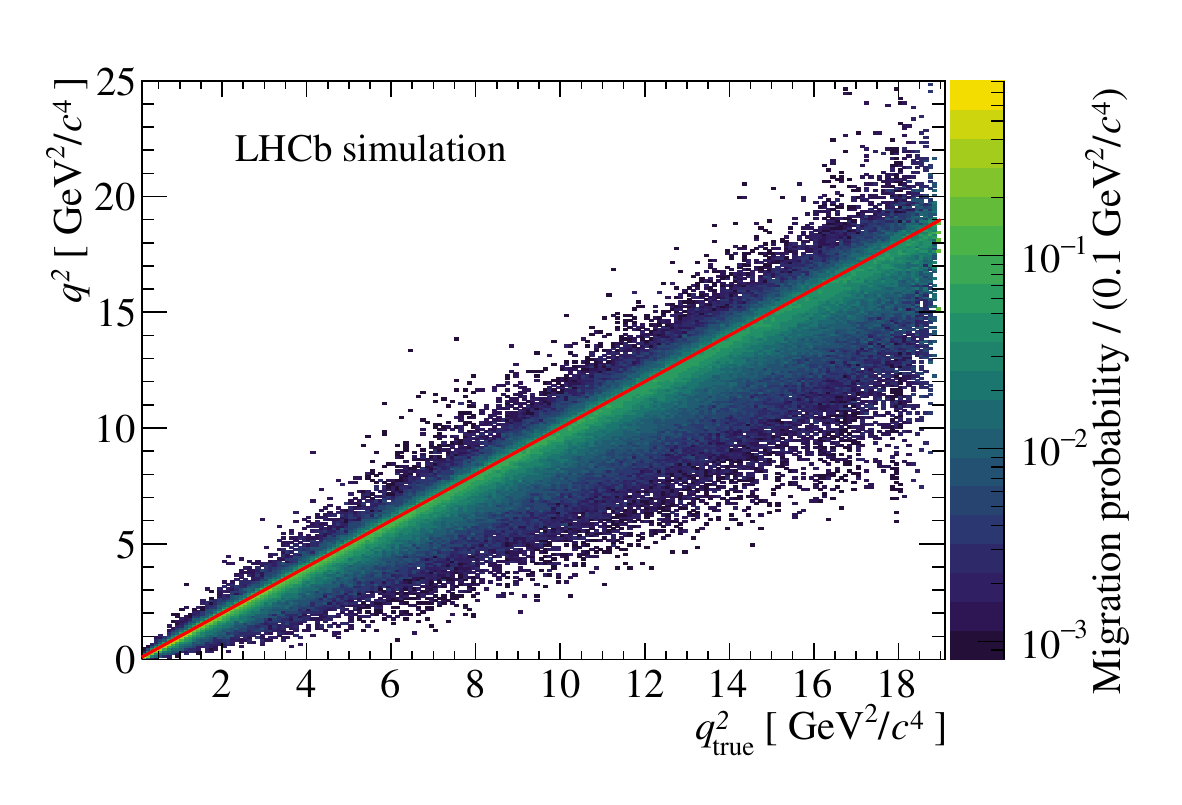}
    \caption{
    Correlation between reconstructed (\qsq) and true (\qsqtrue) values of the dilepton mass squared in simulated \BsPhiEE decays. 
    The distribution is normalised such that, at a fixed value of \qsqtrue, the migration probability in bins of reconstructed \qsq sums to unity. 
    The true value is determined from the \Bs and $\phi$ meson four-momenta and does not include effects from final-state radiation. 
    The red line corresponds to reconstructed $\qsq = \qsqtrue$.
    }
    \label{fig:migration}
\end{figure}

The analysis is performed in bins of reconstructed \qsq and the measurements correspond to rate-averages of the observables over the \qsq bin. 
The reconstructed values of \qsq differ from the true values due to experimental resolution and radiative losses from the electrons that are not recovered in the event reconstruction. 
The correlation between the reconstructed \qsq value and the true value in simulation is illustrated in Fig.~\ref{fig:migration}. 

\section{Detector and simulation} 
\label{sec:detector}

The LHCb detector~\cite{LHCb-DP-2022-002,LHCb-DP-2008-001,LHCb-DP-2014-002} is a single-arm forward spectrometer covering the pseudorapidity range $2 < \eta < 5$, designed for the study of particles containing \bquark\ or \cquark\ quarks. 
The detector elements that are particularly relevant for this analysis are: 
a silicon-strip vertex detector surrounding the proton-proton  interaction region that allows \cquark\ and \bquark\ hadrons to be identified from their characteristically long flight distance; 
a tracking system that provides a measurement of the momentum, $p$, of charged particles; 
two ring-imaging Cherenkov detectors that are able to discriminate between different species of charged hadrons; 
an electromagnetic calorimeter to reconstruct electrons and photons energy deposits; 
and a muon system to identify muons. 
Energy loss from the electrons by bremsstrahlung is recovered using a dedicated algorithm, which associates energy in the calorimeter to the electrons based on their trajectories.   
Events are selected online using a trigger~\cite{LHCb-DP-2012-004,LHCb-DP-2019-001} that consists of a hardware and a software stage. 

Simulated samples are used to determine the efficiency of the candidate selection as well as the distributions of specific sources of background. 
In the simulation, proton-proton collisions are generated using \pythia~\cite{Sjostrand:2007gs,*Sjostrand:2006za} with a specific \lhcb configuration~\cite{LHCb-PROC-2010-056}.
Decays of unstable particles are described by \evtgen~\cite{Lange:2001uf}, using \photos~\cite{davidson2015photos} to generate final-state radiation from the electrons and kaons.  
Simulated \BsPhiEE decays are generated according to two models: a SM-like model based on Refs.~\cite{Ali:1999mm,Descotes-Genon:2015hea,Ball:2004rg}; and a model that is uniform in \qsq and the decay angles. 
The interactions of the generated particles with the detector, and its response, are simulated using the \geant toolkit~\cite{Allison:2006ve, *Agostinelli:2002hh} as described in Ref.~\cite{LHCb-PROC-2011-006}.
The simulated samples are corrected to account for known differences between simulation and data. 
These discrepancies involve the production kinematics of the \Bs meson, the occupancy of the detector, and the efficiencies related to tracking and particle identification. 

\section{Candidate selection} 
\label{sec:selection} 

The candidate selection used in this analysis is identical to that in Ref.~\cite{LHCb-PAPER-2024-032}. 
Online, candidates are selected from events that are either triggered at hardware level by an electron with large transverse energy, with a threshold in the range 2.4--3.0\gev, or by particles from the event that do not form part of the signal candidate. 
In the software trigger, events are selected if they contain at least one reconstructed track with large transverse momentum and a significant impact parameter (IP) with respect to every proton-proton collision vertex (PV) in the bunch-crossing event. 
Two or more tracks are required to form a vertex that is displaced from every PV and fulfills a set of topological criteria~\cite{BBDT,LHCb-PROC-2015-018}.

Offline, \BsPhiEE candidates are formed by combining \ep and \en with \Kp and \Km candidates.  
The electrons and kaons are required to have a significant IP with respect to every PV and fulfil a set of particle identification criteria.
The kaon pair is required to have a mass within $\pm 12\mevcc$ of the known $\phi$ meson mass~\cite{PDG2024}; this narrow mass window suppresses many potential sources of misidentified background. 
The resulting \Bs candidate vertex is required to have a good vertex-fit quality and to be significantly displaced from every PV. 
The \Bs candidate is required to be consistent with originating from a PV by requiring a small \Bs IP and good alignment between the \Bs momentum and displacement vectors. 

Combinatorial background, formed by tracks from two or more different \bquark-hadron decays, is suppressed using a boosted decision tree classifier~\cite{CatBoost}. 
The classifier is trained using simulated signal decays, generated with a SM-like distribution in \qsq and the decay angles, and a background sample comprised of candidates from an upper \mKKEE sideband. 
The classifier uses kinematic and topological information that is characteristic of the signal decay. 
The classifier working point is chosen separately for each \qsq region. 
A second multivariate classifier is trained for the high-\qsq region to reject background involving poorly reconstructed electron candidates. 
This classifier is trained on simulated samples and uses kinematic information as well as information related to the bremsstrahlung recovery procedure. 

There are relatively few specific background sources that the selection criteria does not suppress to negligible levels. 
Potential background from decays with similar topologies, \eg from \decay{\Bz}{\Kstarz\epem} decays where the \Kstarz meson decays to $\Kp\pim$, is suppressed by the narrow $\phi$ mass window and particle identification requirements applied to the kaon candidates.\footnote{
    The inclusion of charge conjugate processes is implied when discussing backgrounds.
} 
Background from semileptonic \decay{\Bs}{\Dsm(\to\phi\pim)\ep\neu_{e}} decays, where the negative pion is mistakenly identified as an electron, is removed by rejecting candidates where the $\phi\en$ mass is consistent with that of a \Dsm meson, 
where a relatively broad range $[1920,2000]\mevcc$ is vetoed to account for the misidentification.
Background from doubly semileptonic \decay{\Bs}{\Dsm(\to\phi\en\neub_{e})\ep\neu_{e}} decays cannot be fully suppressed and is accounted for in the fits. 
Sources of potential background from other partially reconstructed decays typically involve either a kaon or multiple particles that are not reconstructed, and typically populate regions in $\Kp\Km\ep\en$ mass outside the range considered in the analysis. 
The decays \BsJpsiPhi and \BsPsiTwoSPhi with \decay{\jpsi\ \text{and}\ \psitwos}{\ep\en} form background in the central- and high-\qsq regions if too little or too much bremsstrahlung energy is added back to the candidate, respectively.
Background from these decays is also accounted for in the fits. 

\section{Angular and time-dependent efficiency} 
\label{sec:efficiency:angular} 

The event reconstruction and candidate selection distort the measured angular distributions of the candidates in particular due to kinematic and IP requirements. 
The decay-time-integrated efficiency is parametrised using sets of orthogonal functions as  
\begin{align}
\varepsilon(\qsq,\costhetak,\costhetae,\Phi) = \sum\limits_{klmn} c_{klmn} P_k(\tilde{q}^{2}) P_{l}(\costhetak) P_m(\costhetae) \cos( n\Phi )\,,
\label{eq:acceptance}
\end{align}
where the $P_i$ are Legendre polynomials of degree $i$ and $\tilde{q}^{2}$ is a linear transformation of \qsq to the range $[-1,+1]$.
The orthogonality of the functions used to construct the efficiency models allows for the coefficients to be extracted using a moment analysis.
This is done using simulated signal decays generated uniformly in \qsq and the decay angles, corrected for data-simulation differences following the procedures described in Refs.~\cite{LHCb-PAPER-2022-045,LHCb-PAPER-2024-032}.
The efficiency is parametrised by functions with $k \leq 6$, $l \leq 4$, $m \leq 8$, and $n \leq 2$, where the maximum orders are found empirically to be those necessary to describe the efficiency across the entire phase space. 
Only even-ordered terms are considered for \costhetak and \costhetae as differences in the interaction probabilities in the detector between \Kp and \Km mesons and between \ep and \en leptons are sufficiently small that they do not introduce significant asymmetries in the angular distributions. 
The \Dsm veto suppresses a narrow region of phase space in \qsq and \costhetae. 
This cannot be modelled without very high-degree polynomials and is therefore treated separately. 
The ratio of the \costhetae distribution of the simulated candidates with and without the veto is determined after integrating over \qsq, separately in each \qsq\ bin, and used to obtain a parametric description of the veto effect. 
The efficiency in \costhetae is multiplied by this function to account for the effect of the veto.

\begin{figure}[!bt]
    \centering
    \includegraphics[width=0.48\linewidth]{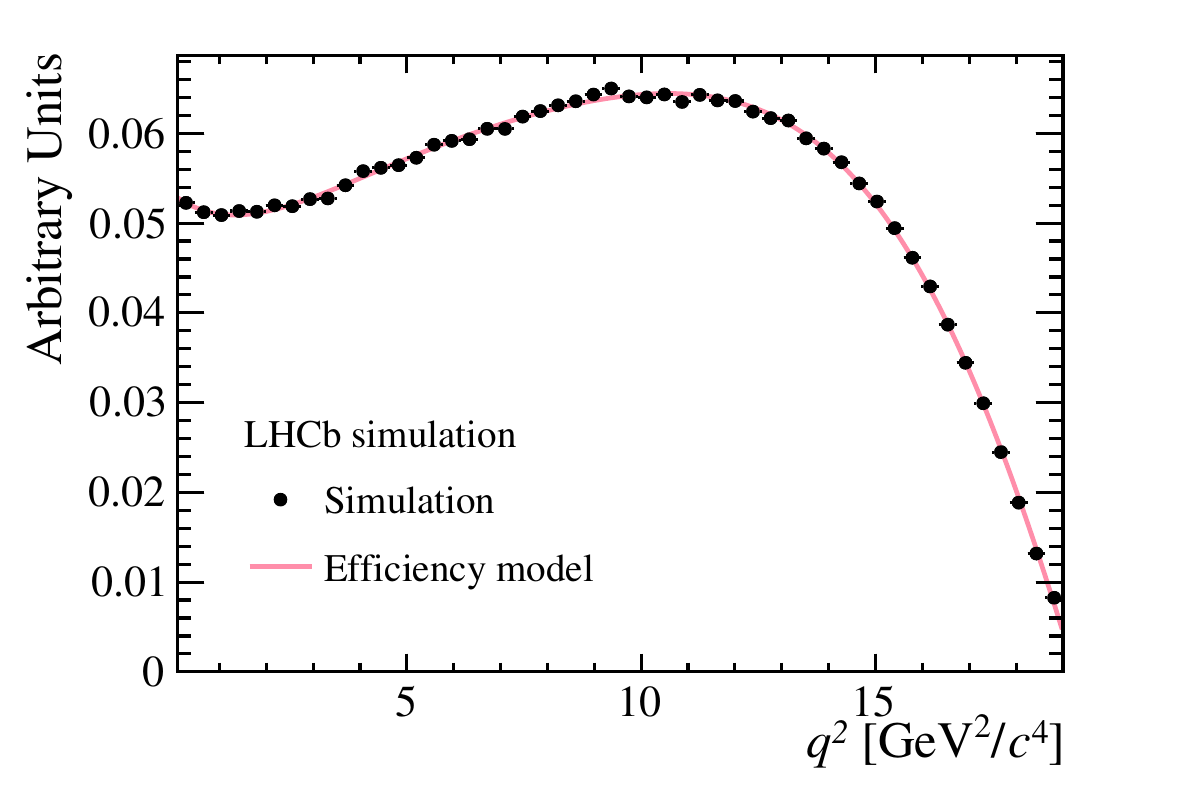} 
    \includegraphics[width=0.48\linewidth]{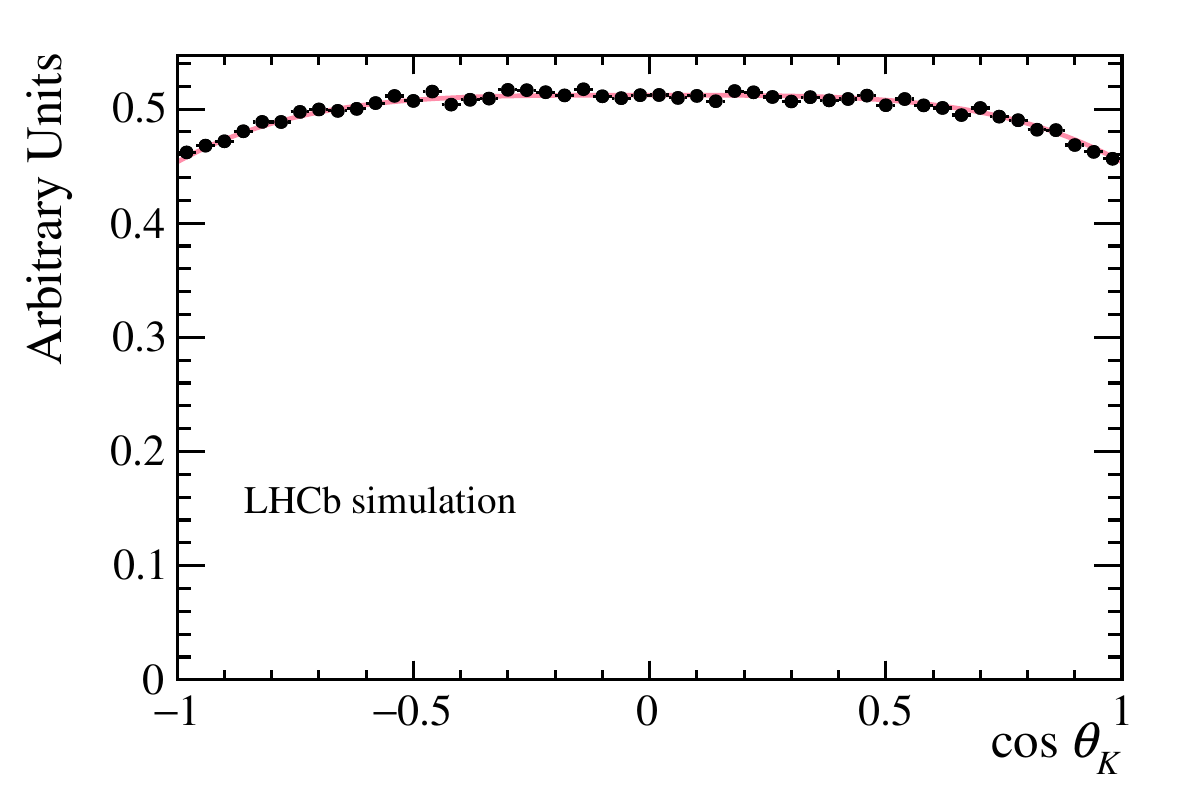} \\
    \includegraphics[width=0.48\linewidth]{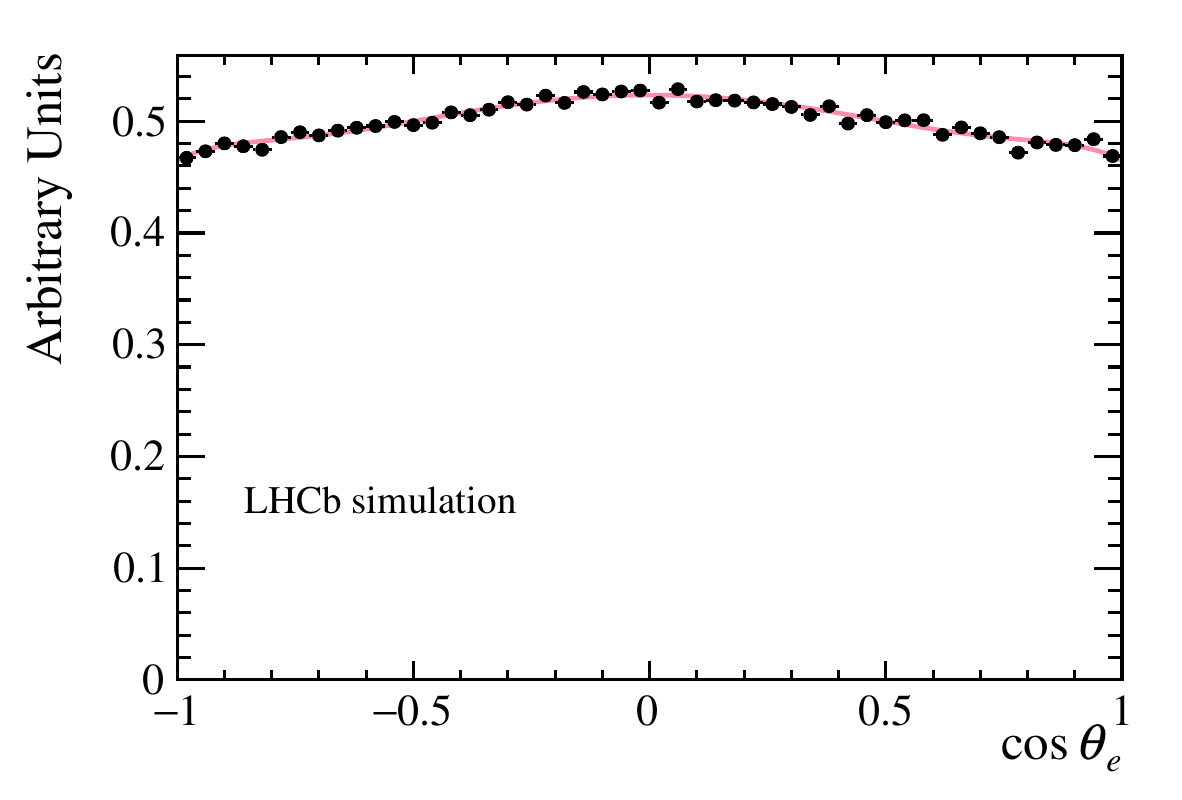} 
    \includegraphics[width=0.48\linewidth]{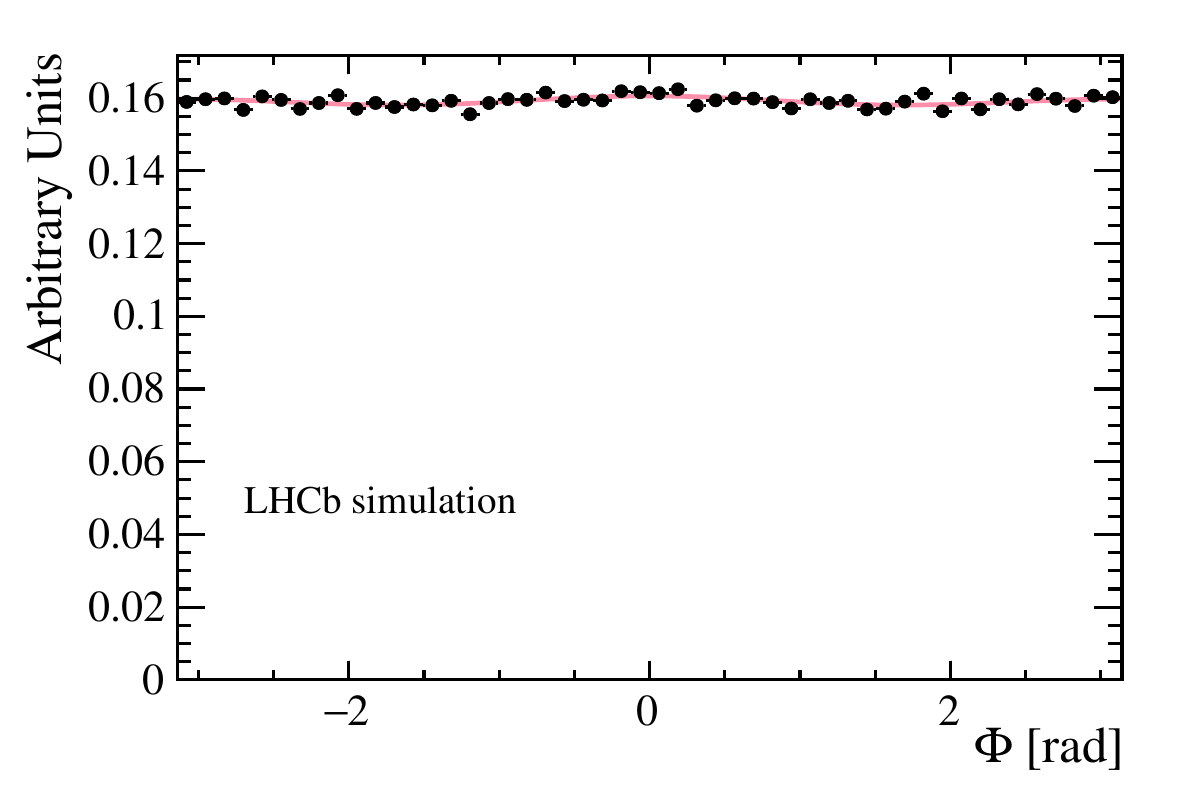}
    \caption{
    Distributions of simulated \BsPhiEE decays, generated uniformly in \qsq and the decay angles, in (top left)~\qsq, (top right)~\costhetak, (bottom left)~\costhetae, and (bottom right)~$\Phi$.
    All selection criteria except the \Dsm veto are applied to the simulation.
    The efficiency model integrated over the other variables is also shown. 
    }
    \label{fig:acceptance}
\end{figure}

One-dimensional projections of the efficiency function $\varepsilon(\qsq,\costhetak,\costhetae,\Phi)$ in each of the variables are compared in Fig.~\ref{fig:acceptance} to the distributions in the simulation.
The drop-off in efficiency at high-\qsq is due to the IP requirements on the \Kpm tracks in the candidate selection.

The reduced angular distributions are obtained by fixing \qsq to the median value of the \qsq bin and integrating the product of the three-dimensional angular distribution and $\varepsilon$ over the relevant angles. 
Due to the efficiency not being perfectly uniform, the resulting distributions feature additional terms that are not present in Eqs.~\eqref{eq:twodimensional} and~\eqref{eq:phi}.
These are suppressed by small numerical factors and are neglected, but considered as a source of systematic uncertainty. 

The selection requirements are correlated with the \Bs candidate decay time and modify the dilution factor that appears in front of the $h_{i}$ terms in Eq.~\eqref{eq:hi-dilution-noAcceptance} from the value ${\cal D} = y_s$ that is obtained with uniform decay-time acceptance. 
If $\epsilon(t)$ is the decay-time dependence of the selection efficiency, the dilution is instead 
\begin{align}
    {\cal D} = { {\displaystyle\int\limits_{0}^{\infty}}\epsilon(t) e^{-\Gammas t} \sinh\left( \tfrac{\DeltaGammas}{2} t \right) \deriv t} \bigg/ {{\displaystyle \int\limits_{0}^{\infty}}\epsilon(t) e^{-\Gammas t} \cosh\left( \tfrac{\DeltaGammas}{2} t\right) \deriv t}\,.
\end{align}
The decay-time-dependent efficiency is well described by the function~\cite{LHCb-PAPER-2014-019}  
\begin{align}
\epsilon(t) = \frac{a (t - t_0)^{c}}{1 + a (t - t_0)^{c}} ( 1 + b t )\,.
\end{align}
Using values of $t_0$, $a$, $b$ and $c$ determined from simulation, the dilution is ${\cal D} = 0.122 \pm 0.015$.

\section{Mass and angular distributions} 
\label{sec:fitting} 

The angular observables are determined by performing unbinned maximum-likelihood fits, simultaneously to the \KKEE mass and angular distributions of the selected candidates. 
The inclusion of \mKKEE in the fit aids discrimination between signal and various sources of background. 
Two separate fits are performed in each \qsq bin, one to the three dimensional distribution of candidates in \mKKEE, \costhetak and \costhetae, and the second to the two dimensional distribution of candidates in \mKKEE and $\Phi$. 
The description of the \mKKEE distribution of each fit component is identical to that in Ref.~\cite{LHCb-PAPER-2024-032}.

The signal mass distribution is described by a Gaussian function with power-law tails on the left- and right-hand side of the spectrum. 
Bremsstrahlung, and the imperfect nature of the recovery procedure, leads to a tail that is more pronounced on the left-hand side of the distribution. 
The tails are further modified by sigmoid activation functions, which are used to account for phase-space related effects.
The tail parameters are fixed from simulation. 
For each \qsq bin, the values of the peak position and width from simulation are corrected based on results of fits to the \BsJpsiPhi data sample.

In the low- and central-\qsq regions, the \mKKEE distribution of the combinatorial background is described by an exponential function. 
In the high-\qsq region, the exponential background shape is modified at small \mKKEE values to account for the limited phase space due to the large dielectron mass. 
The phase-space turn-on is modelled assuming that the background is exponentially distributed in \qsq with a shape derived at large \mKKEE values. 
The angular distribution of the combinatorial background is modelled by sums of Chebyshev polynomial functions, with each sum describing one of the \costhetak, \costhetae and $\Phi$ distributions. 
The parameters of the polynomials are constrained from same-sign $\Kp\Km\epm\epm$ or $\Kpm\Kpm\ep\en$ candidates. 
The orders of the polynomials are chosen empirically based on the distributions of candidates in the upper mass sideband and same-sign data. 

The shape of the background from doubly semileptonic decays is taken from simulation. 
The simulation samples include cases where the \Bs directly produces a \Dsm meson and cases where the \Bs produces an excited \Dsm meson that subsequently decays to the ground state. 
This component is described by a nonparametric kernel-density estimate. 

The shapes of the backgrounds from \BsJpsiPhi and \BsPsiTwoSPhi decays with too little or too much bremsstrahlung energy recovered, which leak into other \qsq regions, are taken from simulation. 
The mass distribution of the \BsJpsiPhi background in the central-\qsq region is parametrised by the tail of a Gaussian function. 
In the high-\qsq region, the backgrounds are described using kernel-density estimates.  
The yields of these backgrounds are each constrained from their respective simulated samples, normalised to the yield of \BsJpsiPhi decays in data in the range $6 < \qsq < 11\gevgevcccc$.  

The background from misidentified decays is taken from control samples in  data, where one or both electrons fail the particle-identification requirements used in the analysis. 
The control samples are weighted following the procedure in Refs.~\cite{LHCb-PAPER-2022-045,LHCb-PAPER-2022-046} to determine the shape and level of the misidentified background in the dataset. 
This component is described by a nonparametric kernel-density estimate.

\section{Validation} 
\label{sec:validation} 

The analysis procedure is validated using pseudoexperiments and the SM-like simulated samples. 
The pseudoexperiments are generated with values of the observables from Ref.~\cite{LHCb-PAPER-2021-022} for the signal. 
Fits to datasets much larger than that used in this analysis yield unbiased results with appropriate uncertainty estimates. 
Fits to pseudoexperiments generated with representative yields (from Ref.~\cite{LHCb-PAPER-2024-032}) are observed to provide unreliable uncertainty estimates in some of the \qsq bins and have a non-negligible failure rate. 
This behaviour is caused by fits converging to values close to the boundaries in the allowed parameter space: $\FL \leq 1$, $\Asix \leq (1-\FL)$ and $(\Sthree^2 + \Anine^2) \leq 1$.
The stability of the fit is improved by fitting for \IntAsixp instead of \IntAsix. 
To obtain correct coverage, the Feldman--Cousins procedure is used to determine confidence intervals on the observables~\cite{Feldman-Cousins}. 

\begin{figure}[!tb]
    \centering
    \includegraphics[width=0.48\linewidth]{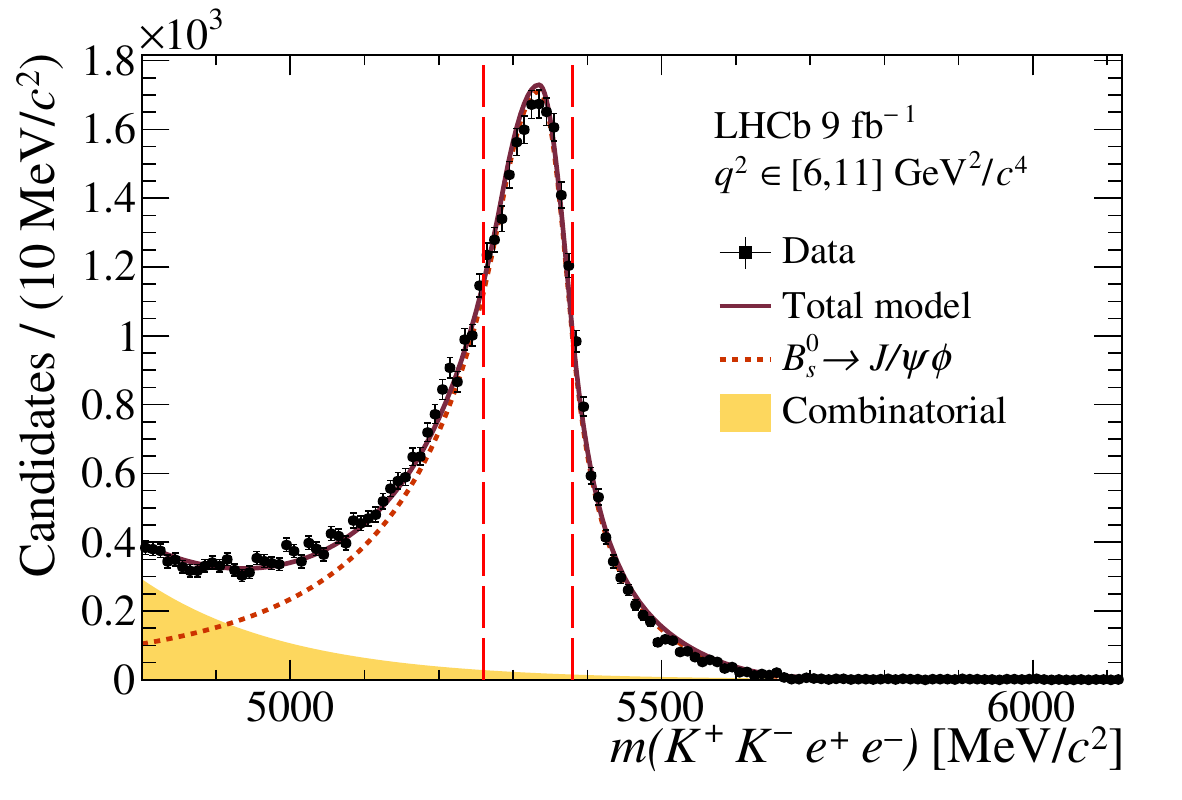} 
    \includegraphics[width=0.48\linewidth]{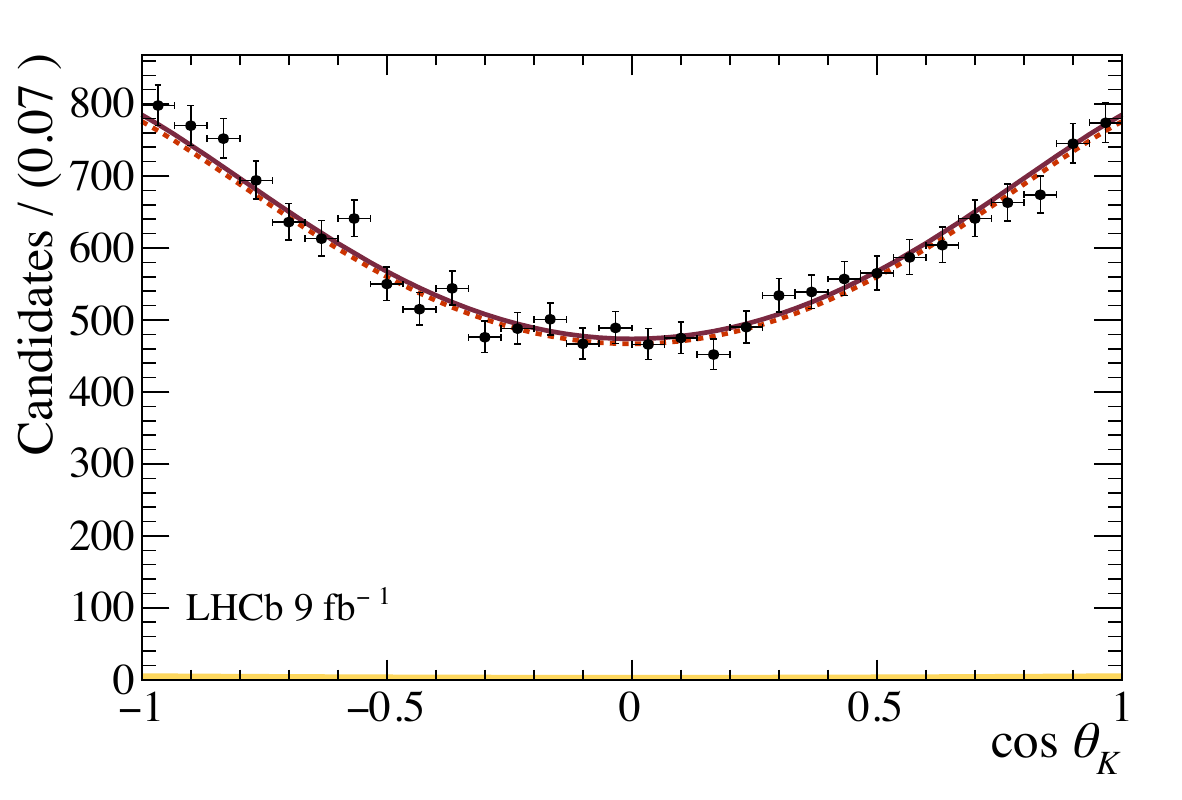} \\
    \includegraphics[width=0.48\linewidth]{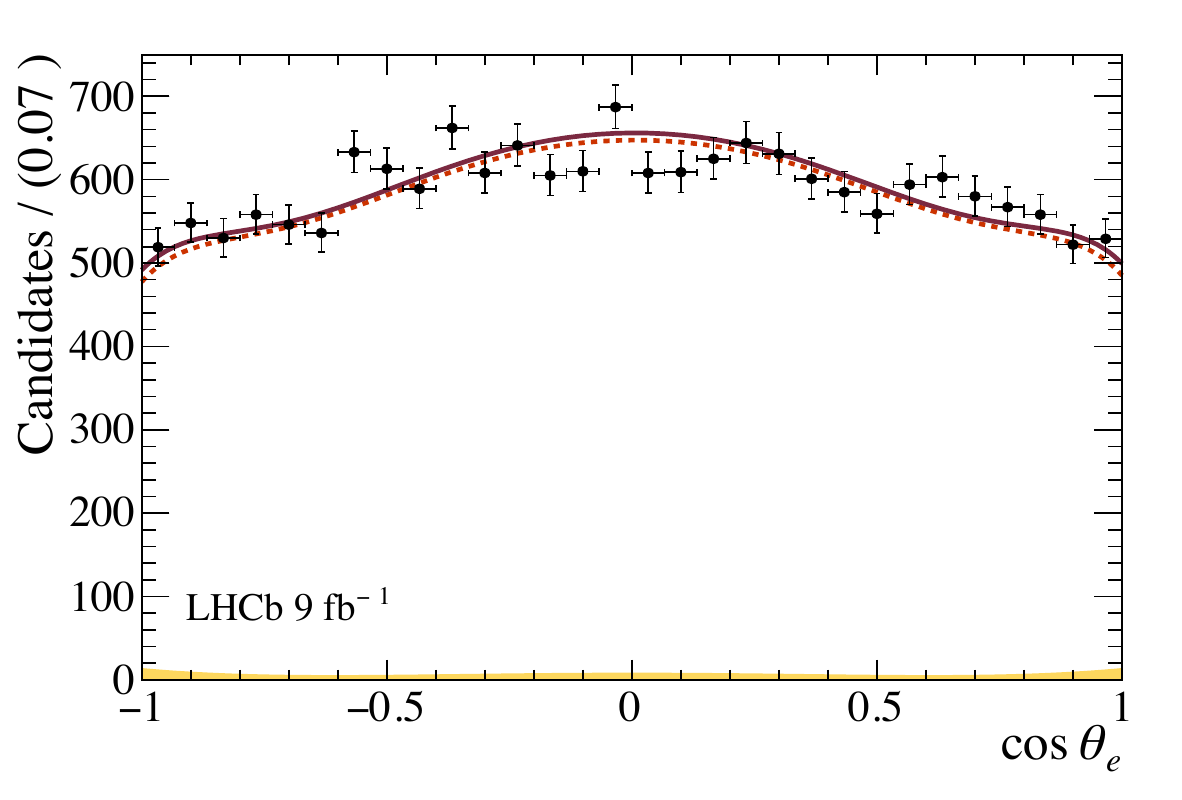}  
    \includegraphics[width=0.48\linewidth]{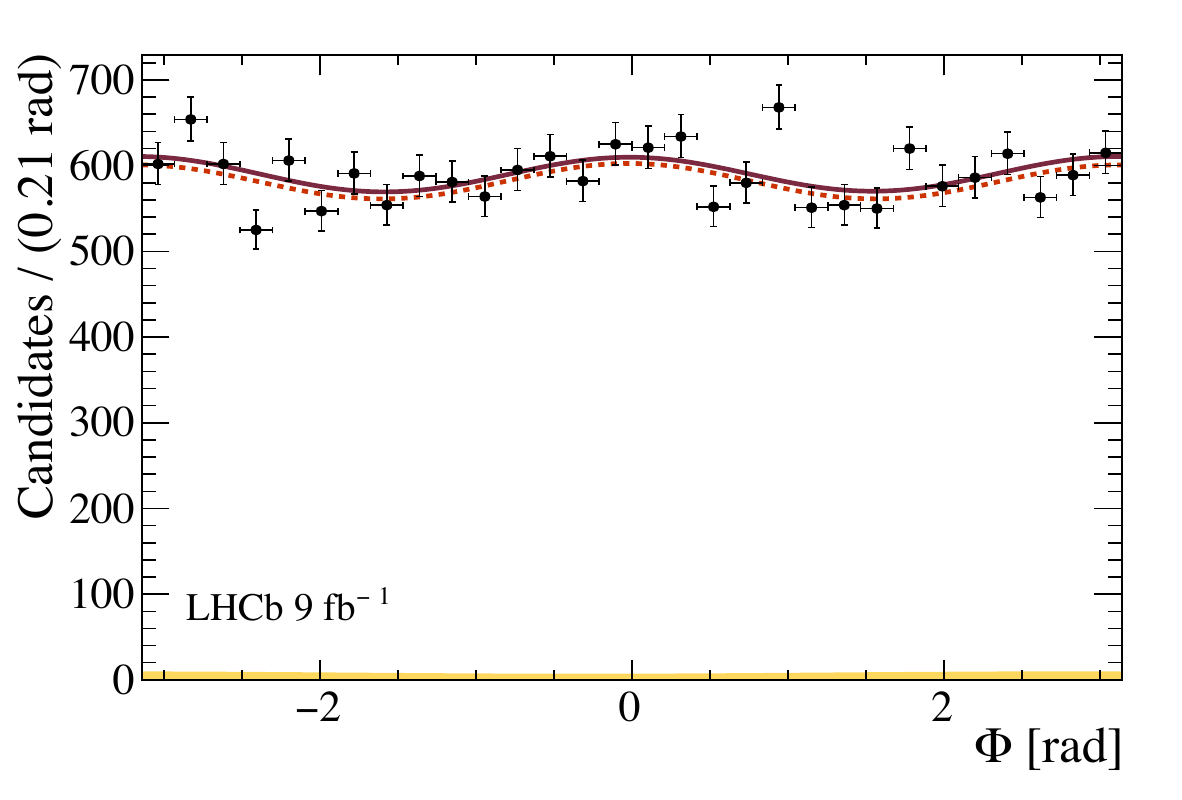} 
    \caption{
    Distributions of selected \BsJpsiPhi decays in (top left)~\mKKEE, (top right)~\costhetak, (bottom left)~\costhetae and (bottom right)~$\Phi$. 
    The \costhetak, \costhetae and $\Phi$ distributions are given in the reduced mass window indicated in the top-left figure by the vertical dashed red lines. 
    The results of the fit with the model described in the text are also shown.
    }
    \label{fig:validation:jpsi}
\end{figure}

\begin{figure}[!tb]
    \centering
    \includegraphics[width=0.48\linewidth]{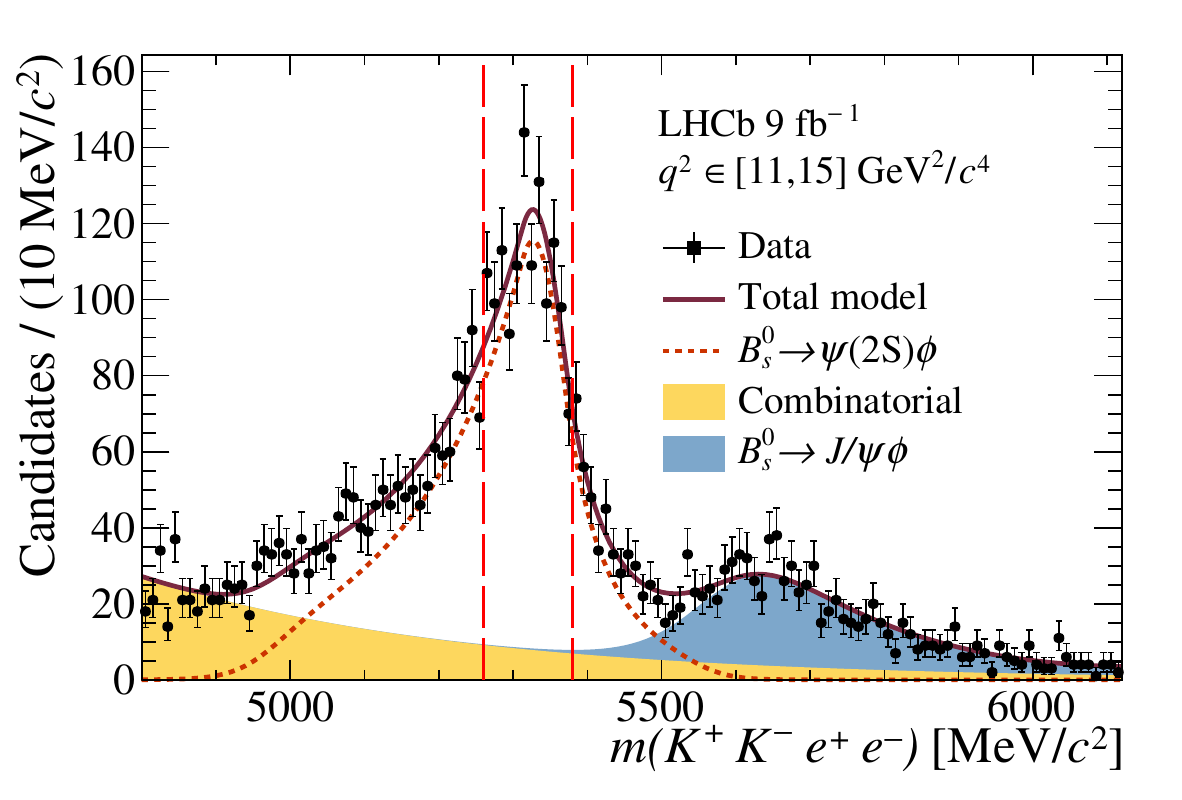} 
    \includegraphics[width=0.48\linewidth]{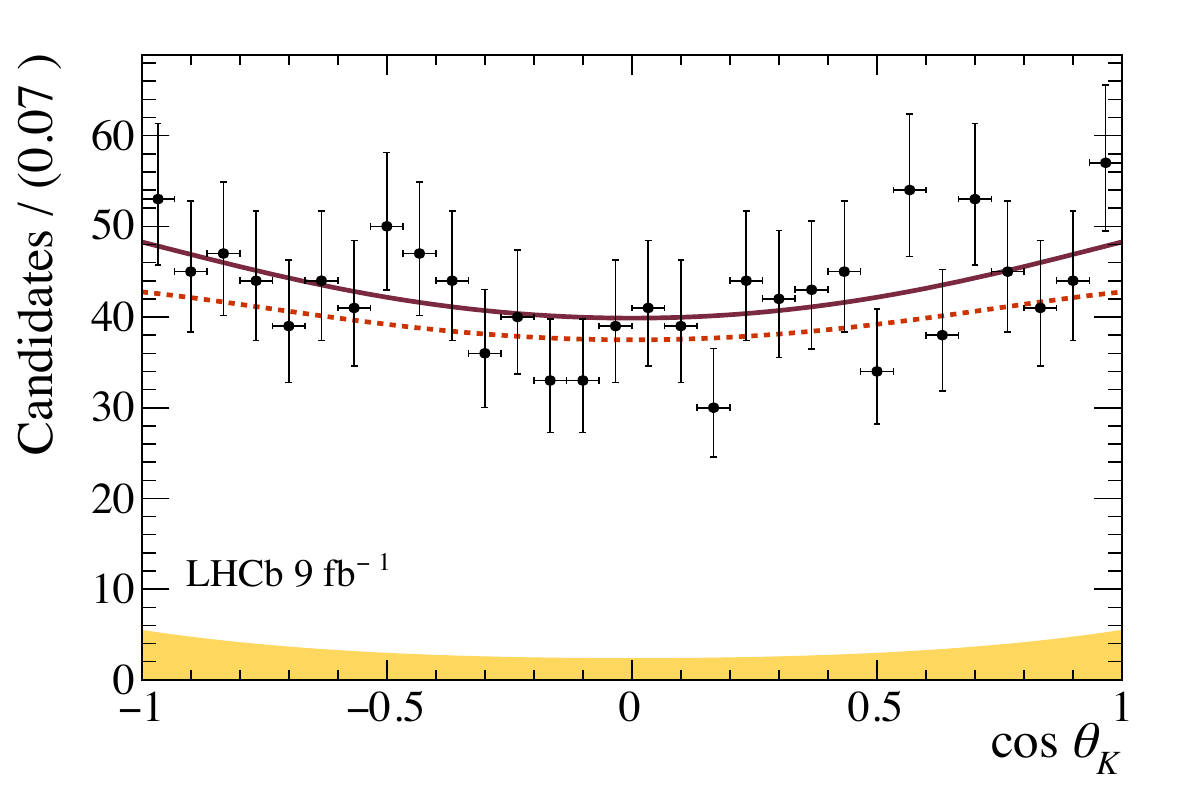} \\
    \includegraphics[width=0.48\linewidth]{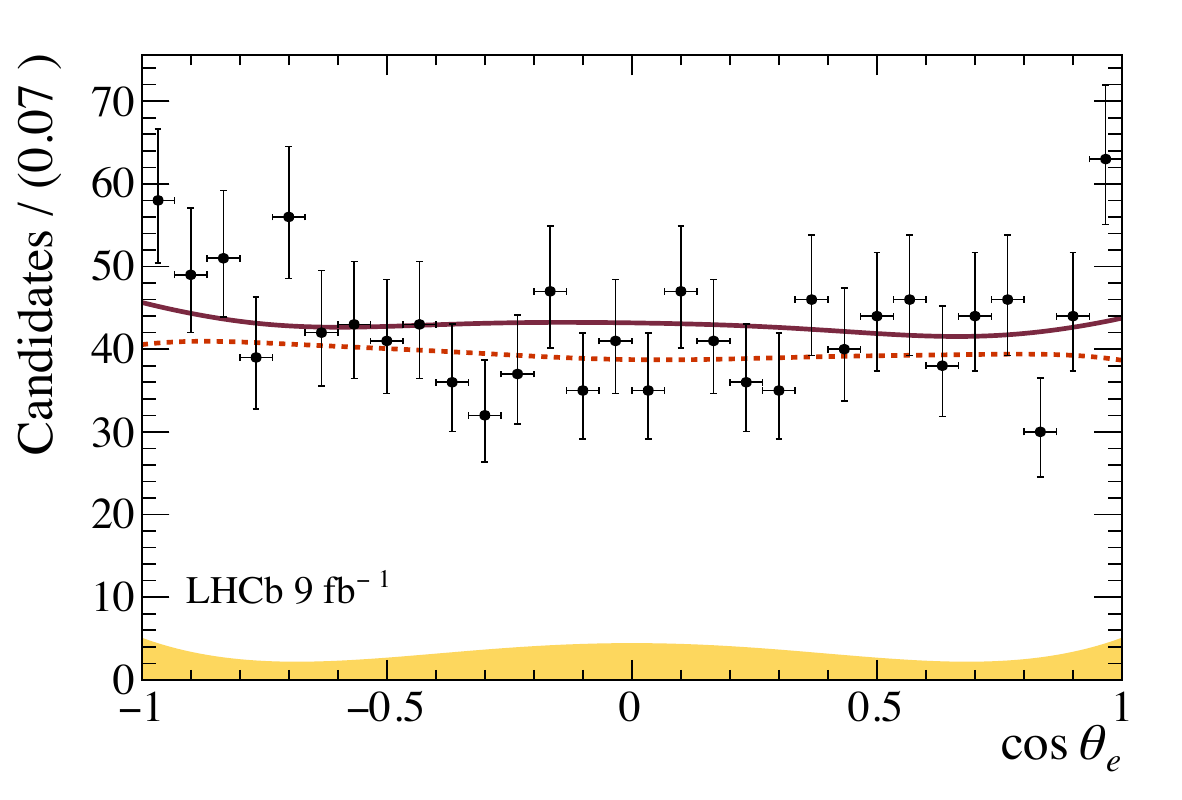}  
    \includegraphics[width=0.48\linewidth]{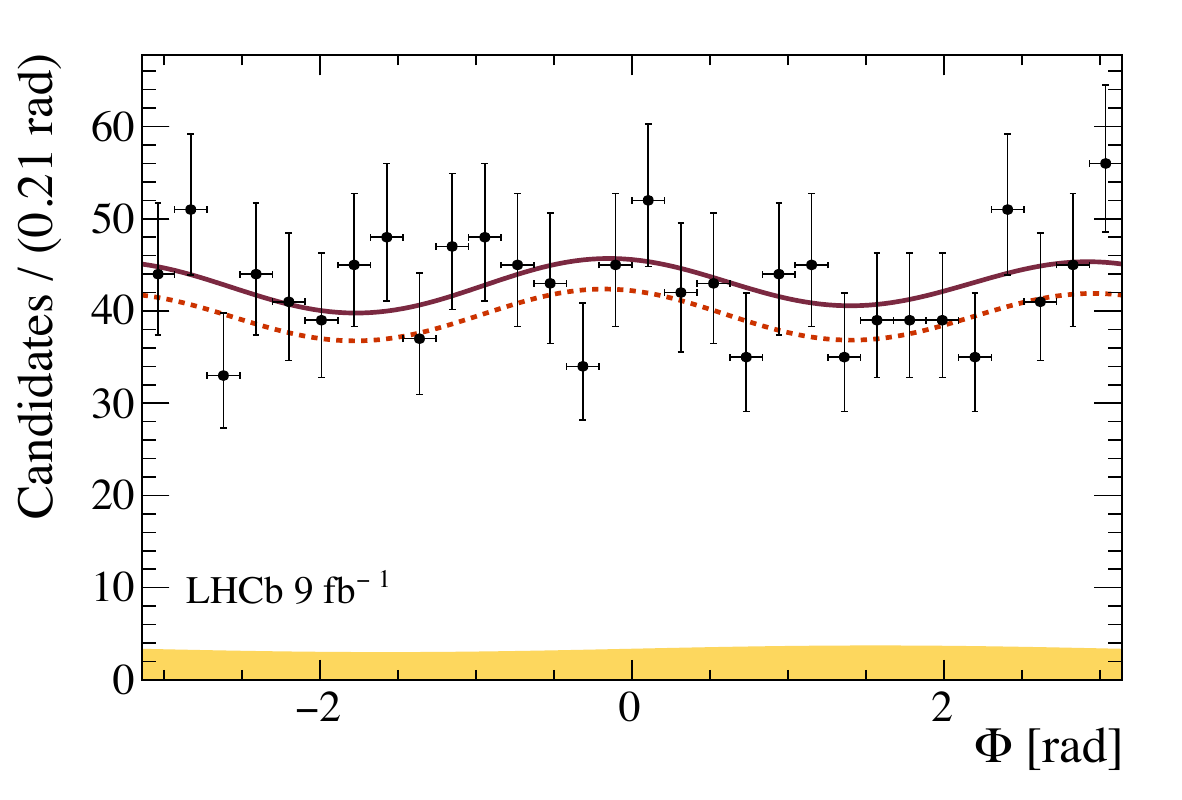} 
    \caption{
    Distributions of selected \BsPsiTwoSPhi decays in (top left)~\mKKEE, (top right)~\costhetak, (bottom left)~\costhetae and (bottom right)~$\Phi$. 
    The \costhetak, \costhetae and $\Phi$ distributions are given in the reduced mass window indicated in the top-left figure by the vertical dashed red lines. 
    The results of the fit with the model described in the text are also shown.
    }
    \label{fig:validation:psitwos}
\end{figure}

The analysis procedure is also validated using \BsJpsiPhi and \BsPsiTwoSPhi decays in data with the same final states as the signal. 
The \BsJpsiPhi and \BsPsiTwoSPhi decays are selected following the same candidate selection procedure as \BsPhiEE candidates but with $6 < \qsq < 11\gevgevcccc$ and $11 < \qsq < 15\gevgevcccc$, respectively. 
To simplify the modelling of background in the fits to the \BsJpsiPhi and \BsPsiTwoSPhi datasets, the analysis is performed in two steps. 
First, the \mKKEE distribution is fitted to determine the relative contribution of the signal and different background components. 
The angular variables are then determined over the reduced \mKKEE range, $5260 < \mKKEE < 5380\mevcc$, taking into account only contributions from signal, combinatorial background and, for \BsPsiTwoSPhi decays, leakage from \BsJpsiPhi decays. 
The angular distribution of the combinatorial background is constrained from a fit to candidates with $5600 < \mKKEE < 6800 \gevcc$. 

\begin{figure}[!tb]
    \centering
    \includegraphics[width=0.7\linewidth]{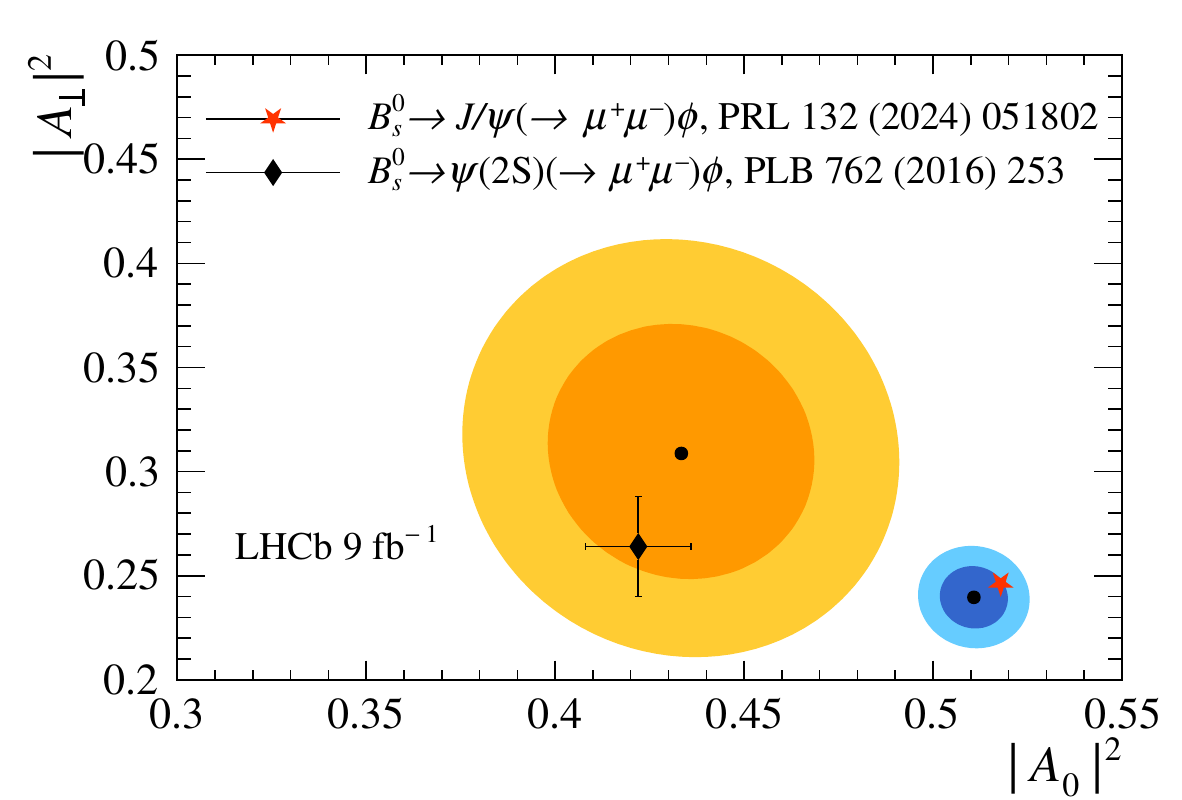}
    \caption{
    Values of the squared transversity amplitudes $|A_{0}|^{2}$ and $|A_{\perp}|^{2}$ for (blue shaded areas) \BsJpsiPhi and (orange shaded areas) \BsPsiTwoSPhi decays determined from \IntFL and \IntSthree. 
    The darker and lighter shaded areas correspond to 68\% and 95\% confidence level intervals, respectively, accounting only for statistical uncertainties. 
    The values of the squared transversity amplitudes for \decay{\jpsi\ \text{and}\ \psitwos}{\mumu} decays from Refs.~\cite{LHCb-PAPER-2023-016} and \cite{LHCb-PAPER-2016-027} are indicated by the markers, where the uncertainties on the \BsJpsiPhi measurement are too small to be visible. 
    }
    \label{fig:validation:control}
\end{figure}

Figures~\ref{fig:validation:jpsi} and~\ref{fig:validation:psitwos} show the results of the fits to the \BsJpsiPhi and \BsPsiTwoSPhi samples, respectively. 
The observables \IntFL and \IntSthree are reinterpreted in terms of the squared transversity amplitudes $|A_0|^2$ and $|A_{\perp}|^2$, correcting for the effect of integration over decay time. 
The values of the squared amplitudes are found to be consistent with, but less precise than, those obtained in the same \Bs decays with \decay{\jpsi\ \text{and}\ \psitwos}{\mumu}, derived from the amplitudes determined in Refs.~\cite{LHCb-PAPER-2023-016,LHCb-PAPER-2016-027}, as shown in Fig.~\ref{fig:validation:control}. 

\section{Systematic uncertainties} 
\label{sec:systematics}

Sources of systematic uncertainty are considered if they could modify the mass or angular distribution of the signal or background components in the analysis. 
The uncertainties assigned for the different sources are determined using pseudoexperiments. 
The pseudoexperiments are generated with an alternative set of assumptions, with sample size much larger than that analysed in this paper, and fitted according to the nominal model. 
A systematic uncertainty is assigned adding the bias on the parameter of interest and its uncertainty in quadrature. 
A summary of the most relevant sources of systematic uncertainty on the observables is presented in Table~\ref{tab:systematics}. 
The total systematic uncertainty on each observable is obtained by summing the contributions from individual sources in quadrature. 
The transformation from \IntAsix to \IntAsixp tends to make uncertainties on \IntAsixp larger than those on other observables, particularly in regions where \IntFL is large. 

A variety of sources of uncertainty related to the efficiency model are considered. 
An uncertainty due to the limited order of the orthogonal functions used in the efficiency model is estimated by trialling different increases in order in the pseudoexperiment generation. 
An uncertainty is assigned to account for the unknown distribution of events in \qsq by evaluating the efficiency model at alternative points in \qsq in the pseudoexperiment generation, shifted from the median by the root-mean-square of the \qsq distribution. 
The small contribution from observables that, due to nonuniform angular efficiency, do not vanish when projecting out the $\Phi$ and two-dimensional \costhetae and \costhetak distributions is evaluated by including these additional terms in the generation. 
To assess the impact of the limited size of the simulated signal sample, new angular efficiency models are derived by bootstrapping the uniformly generated sample. 
Pseudoexperiments are generated according to the new efficiency models and fitted using the nominal one. 
A similar approach is used to assess the impact of the finite size of the simulated samples used to model the semileptonic background and the data sample used to derive the shape of the misidentified background. 
Uncertainties are also considered on the data-simulation corrections applied to the simulated samples. 
The resulting uncertainty is small, and removing the corrections entirely in the pseudoexperiment generation has only a limited impact on the resulting values of the observables. 
The angular efficiency is derived from simulated samples using reconstructed particles whose trajectories are well matched to particles in the signal decay. 
Relaxing the association requirements, in the efficiency model used in the pseudoexperiment generation, also has only a limited impact on the observables.  

For the signal and background modelling, 
uncertainties are assigned based on the assumptions that the signal and different background components factorise in mass and angular distributions. 
To test the assumption on the signal factorisation, signal decays are sampled from the SM simulation and introduced in place of the signal component in the pseudoexperiment generation. 
A similar approach is used for the semileptonic backgrounds. 
For the remaining backgrounds, a moment analysis, similar to that used to derive the angular efficiency coefficients, is used to obtain models that take into account correlations. 
Due to the limited dataset size, the factorisation assumptions on the combinatorial and misidentified backgrounds are the largest sources of systematic uncertainty in the analysis, but are nonetheless well below the statistical uncertainties. 
An uncertainty is also assigned on the signal mass modelling by considering modifications to the tails of the signal mass model. 
Alternate models for the semileptonic background are considered based on kernel-density estimates and, for the angular variables, polynomial models.
Other alternate models are derived by varying the form factors used in generation of the semileptonic samples. 
For the combinatorial background in the high-\qsq region, an alternate mass model is trialled based on a power-law turn-on and fall-off behaviour.   

Uncertainties are also assigned to account for assumptions made when defining the signal angular distribution. 
A systematic uncertainty, due to neglecting angular resolution effects, is estimated by smearing the signal model based on a resolution model in the pseudoexperiment generation. 
The analysis also ignores small contributions from decays to the \KKEE final state where the \Kp\Km pair does not originate from a $\phi$ meson. 
There is no evidence for such contributions in the data. 
To estimate a systematic uncertainty, a $\Kp\Km$ S-wave contribution is introduced at the level of 2\% of the signal based on the analysis of \BsPhiMM decays in Ref.~\cite{LHCb-PAPER-2021-022}.
The uncertainty on \Gammas and \DeltaGammas has a negligible impact on the measured values of the observables but does impact ${\cal D}$ and the interpretation of the measurement. 
The ${\cal O}(1\%)$ production asymmetry between \Bsb and \Bs mesons in proton-proton collisions~\cite{LHCb-PAPER-2014-042} has negligible impact on the measurements.
A systematic uncertainty is estimated by including a small background contribution in the pseudoexperiment generation. 
At high \qsq an uncertainty is also assigned in a similar way to account for backgrounds from partially reconstructed decays of \bquark hadrons to final states including \jpsi and $\phi$ mesons, with too much bremsstrahlung energy recovered.

\begin{table}[!tb]
    \caption{
    Absolute systematic uncertainties on \IntFL, \IntAsixp, \IntSthree and \IntAnine in units of $10^{-2}$ in the \qsq ranges $[0.1,1.1]$/$[1.1,6.0]$/$[15.0,19.0]$\gevgevcccc.
    The contributions to each source of uncertainty are described in the text. 
    }
    \centering
   \scalebox{0.95}{
    \begin{tabular}{lcccc}
    \toprule
    Source  & \IntFL & \IntAsixp & \IntSthree & \IntAnine \\
    \midrule 
    Signal mass model        & 1.2\,/\,2.1\,/\,0.6 & 1.9\,/\,2.8\,/\,1.8 & 1.3\,/\,1.1\,/\,1.7 & 1.7\,/\,1.1\,/\,1.3 \\
    Efficiency model         & 1.3\,/\,1.9\,/\,1.9 & 2.6\,/\,5.0\,/\,2.3 & 2.1\,/\,1.9\,/\,1.9 & 1.8\,/\,1.7\,/\,1.7 \\ 
    S-wave contamination     & 0.5\,/\,1.6\,/\,0.8 & 1.3\,/\,1.8\,/\,0.8 & 1.3\,/\,0.8\,/\,1.4 & 0.9\,/\,0.9\,/\,0.7 \\ 
    Angular resolution       & 0.7\,/\,1.9\,/\,0.4 & 1.0\,/\,2.3\,/\,2.0 & 0.8\,/\,0.7\,/\,0.7 & 1.7\,/\,1.7\,/\,1.1 \\
    Misidentified background & 5.6\,/\,2.8\,/\,3.1 & 2.3\,/\,4.0\,/\,2.3 & 4.3\,/\,5.4\,/\,7.1 & 1.9\,/\,1.5\,/\,2.4 \\
    Leakage background       & 0.0\,/\,1.6\,/\,1.7 & 0.0\,/\,3.2\,/\,2.4 & 0.0\,/\,2.1\,/\,2.9 & 0.0\,/\,1.4\,/\,2.2 \\
    Semileptonic background  & 1.4\,/\,1.1\,/\,0.0 & 2.0\,/\,3.8\,/\,0.0 & 1.4\,/\,3.0\,/\,0.0 & 1.4\,/\,1.4\,/\,0.0 \\
    Combinatorial background & 2.0\,/\,3.2\,/\,1.5 & 1.2\,/\,4.0\,/\,1.9 & 4.3\,/\,0.9\,/\,1.1 & 1.0\,/\,1.3\,/\,2.3 \\
    \bottomrule
    \end{tabular}
    }
    \label{tab:systematics}
\end{table}

\section{Results} 
\label{sec:results} 

\begin{figure}[!tb]
    \centering
    \includegraphics[width=0.32\linewidth]{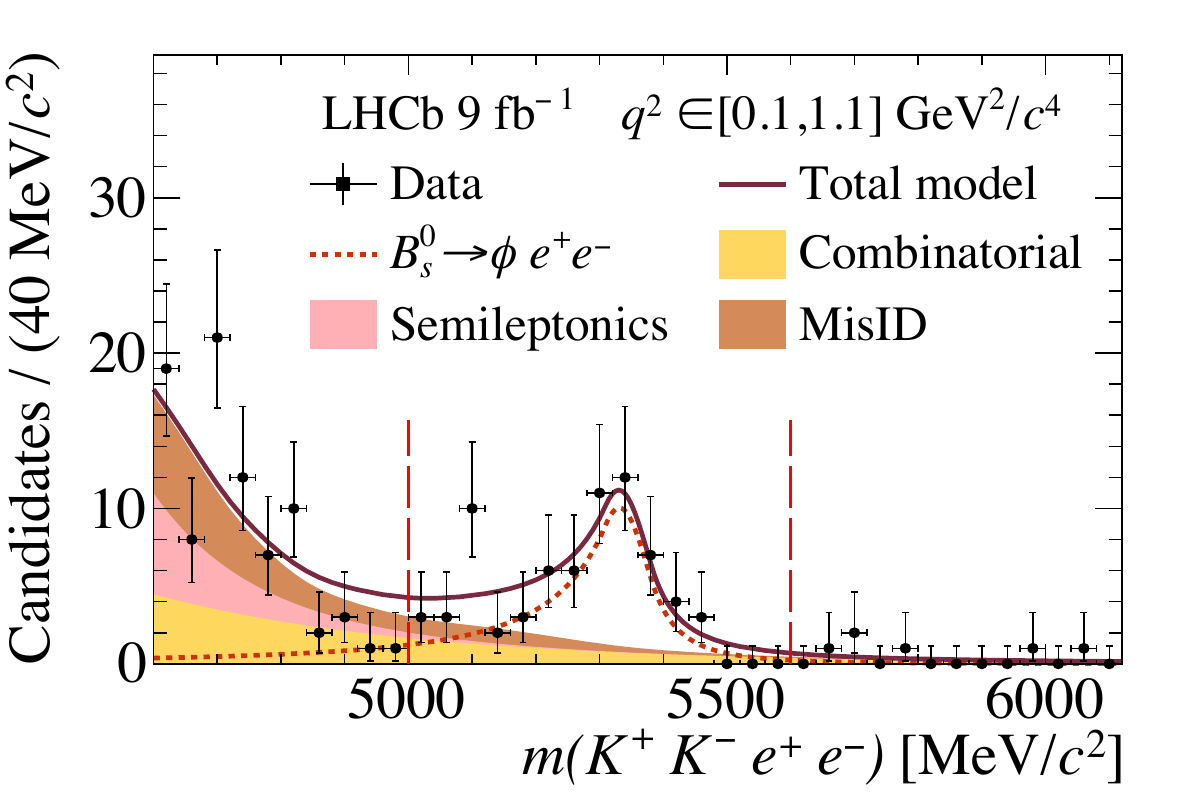} 
    \includegraphics[width=0.32\linewidth]{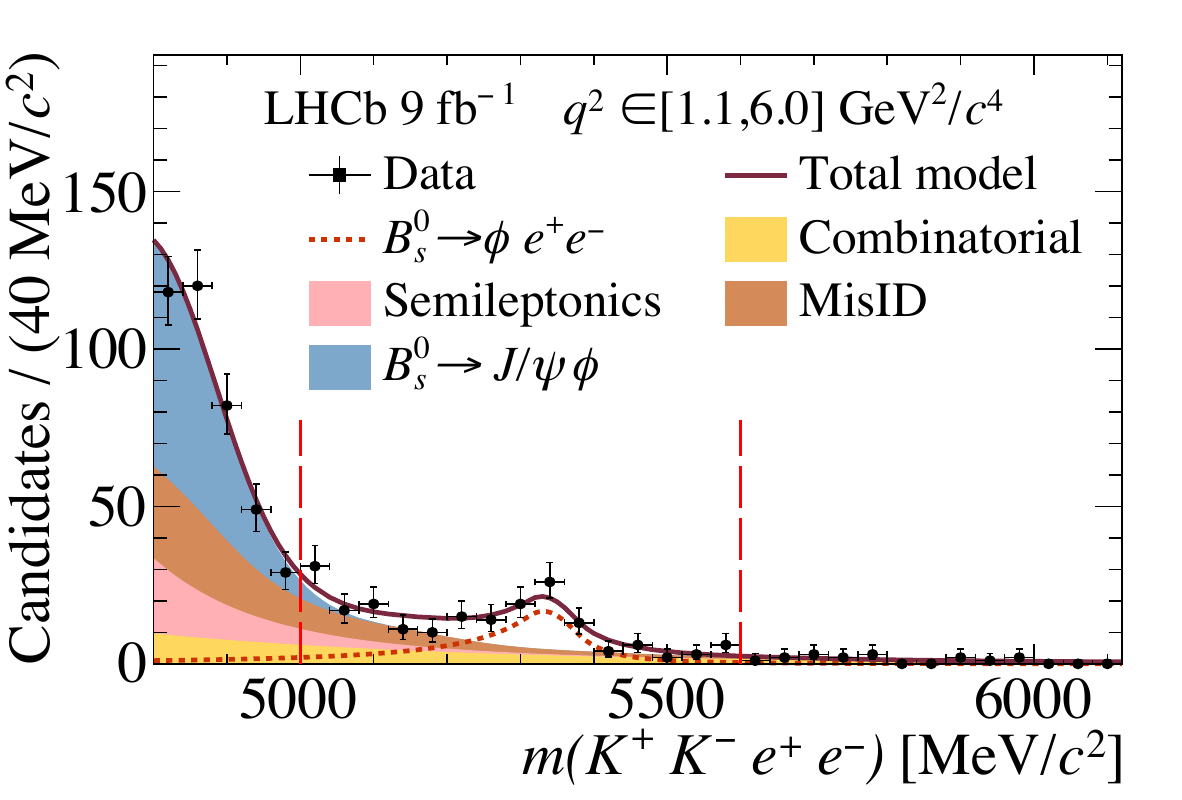} 
    \includegraphics[width=0.32\linewidth]{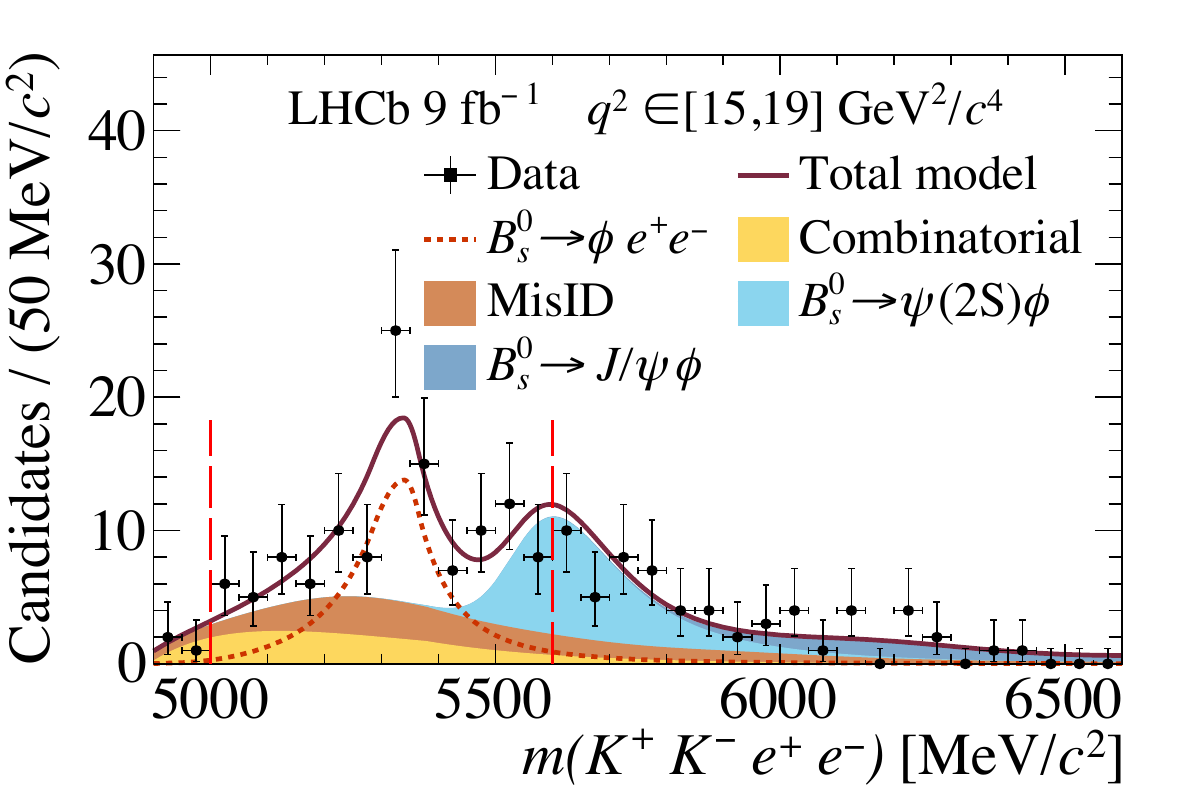} \\ 
    \includegraphics[width=0.32\linewidth]{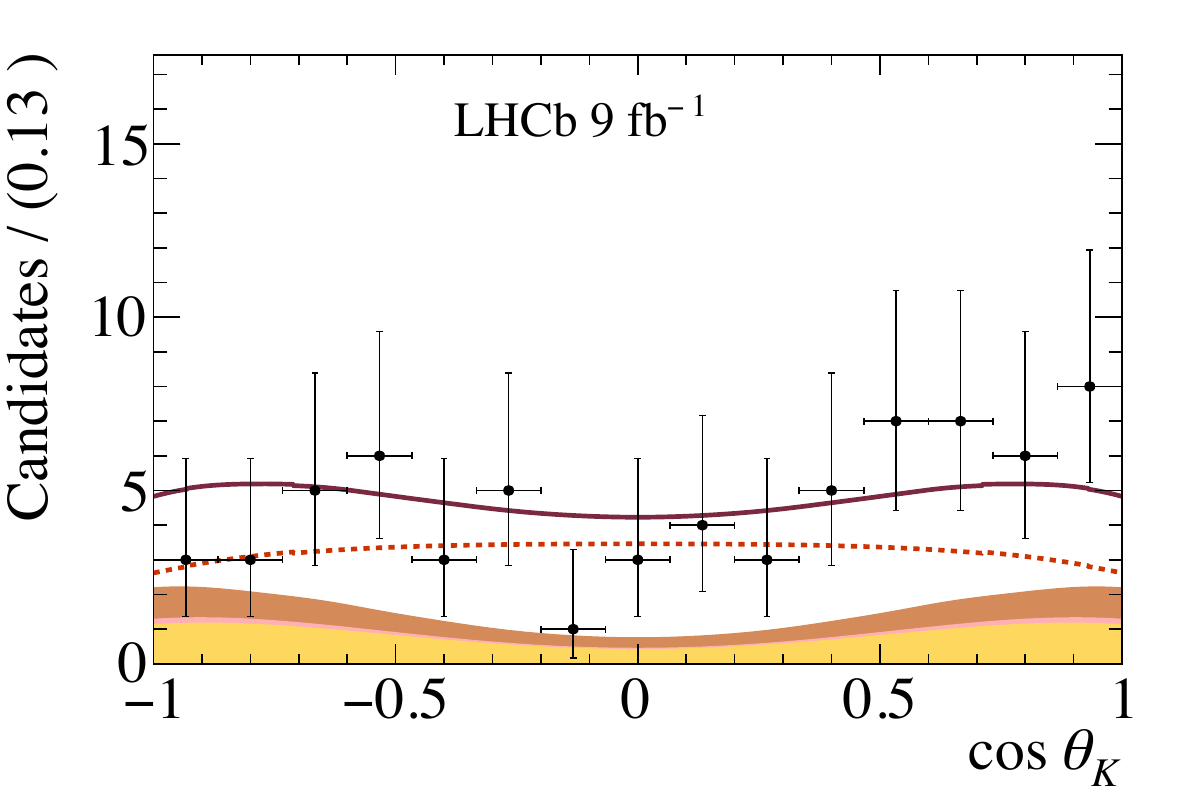} 
    \includegraphics[width=0.32\linewidth]{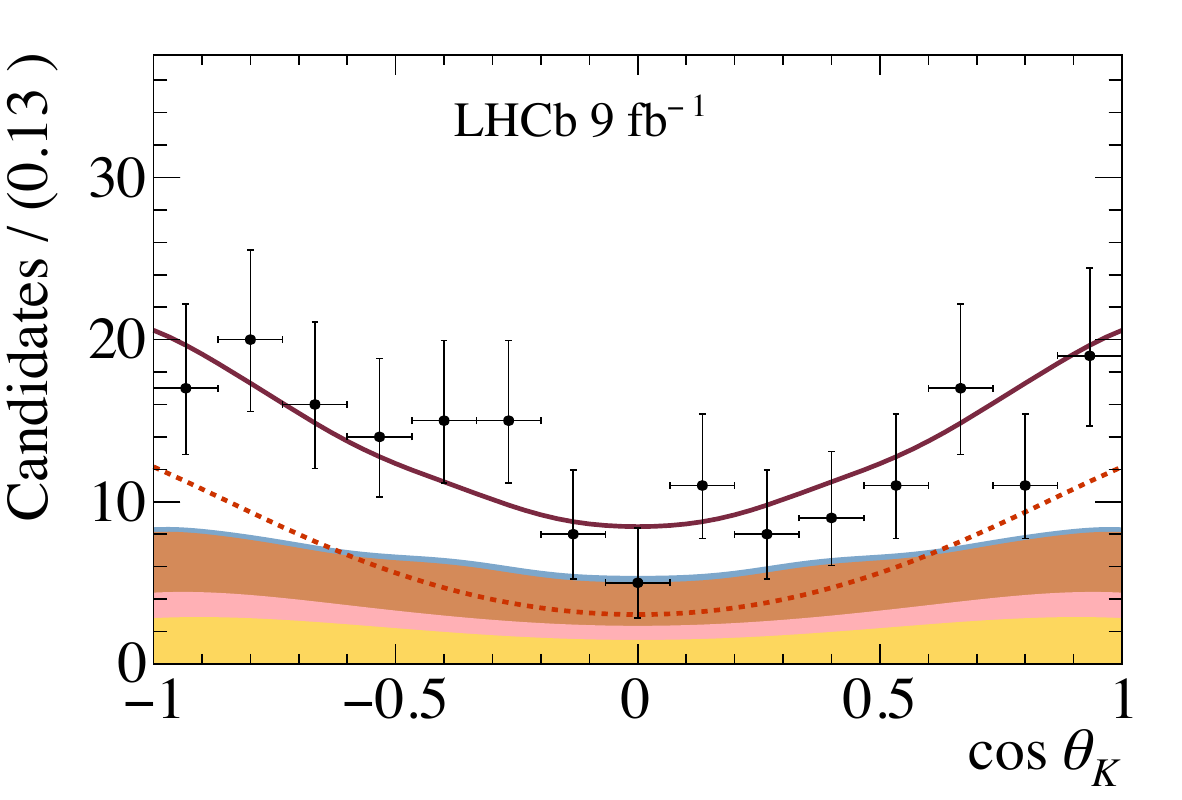} 
    \includegraphics[width=0.32\linewidth]{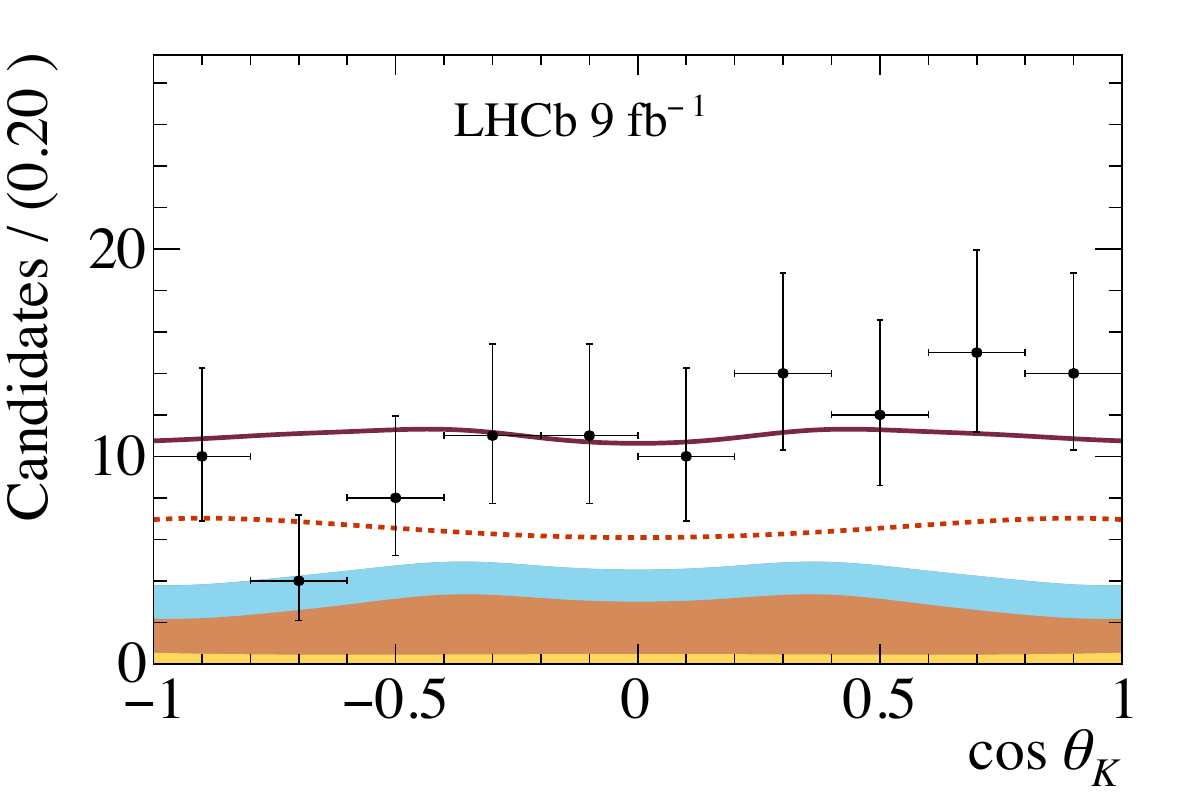} \\ 
    \includegraphics[width=0.32\linewidth]{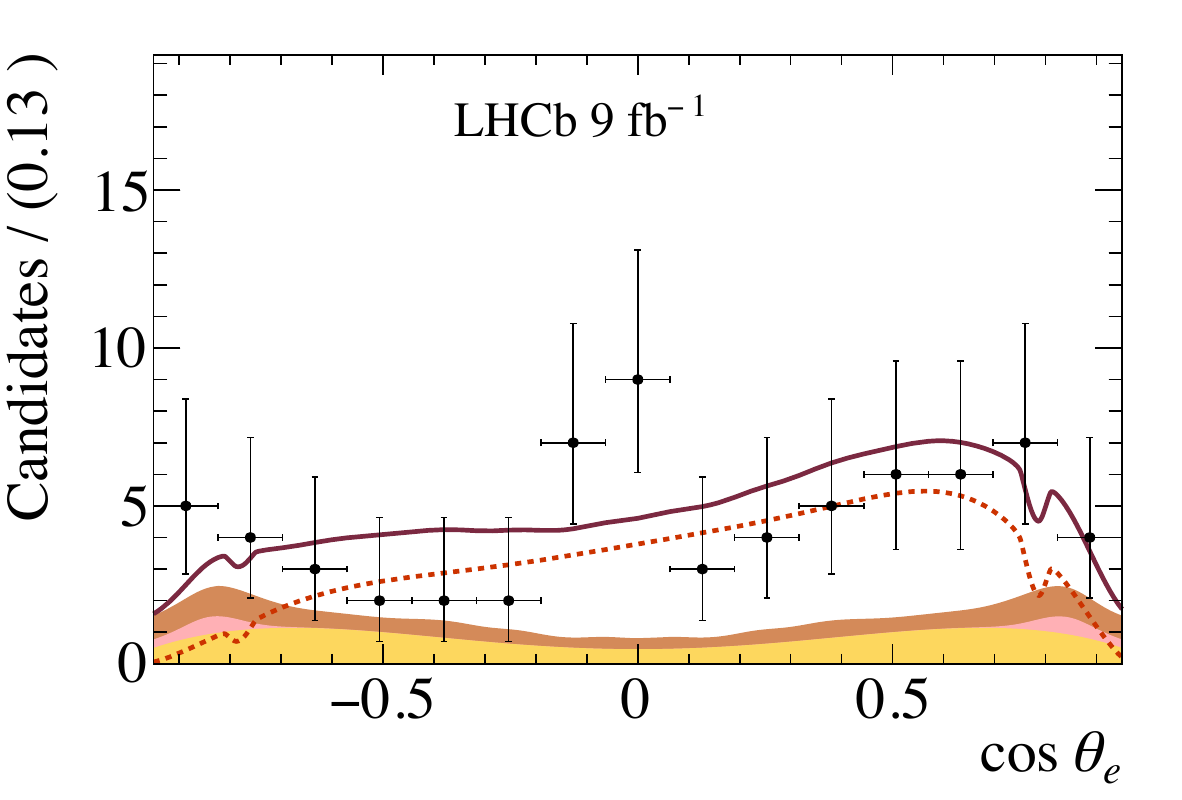} 
    \includegraphics[width=0.32\linewidth]{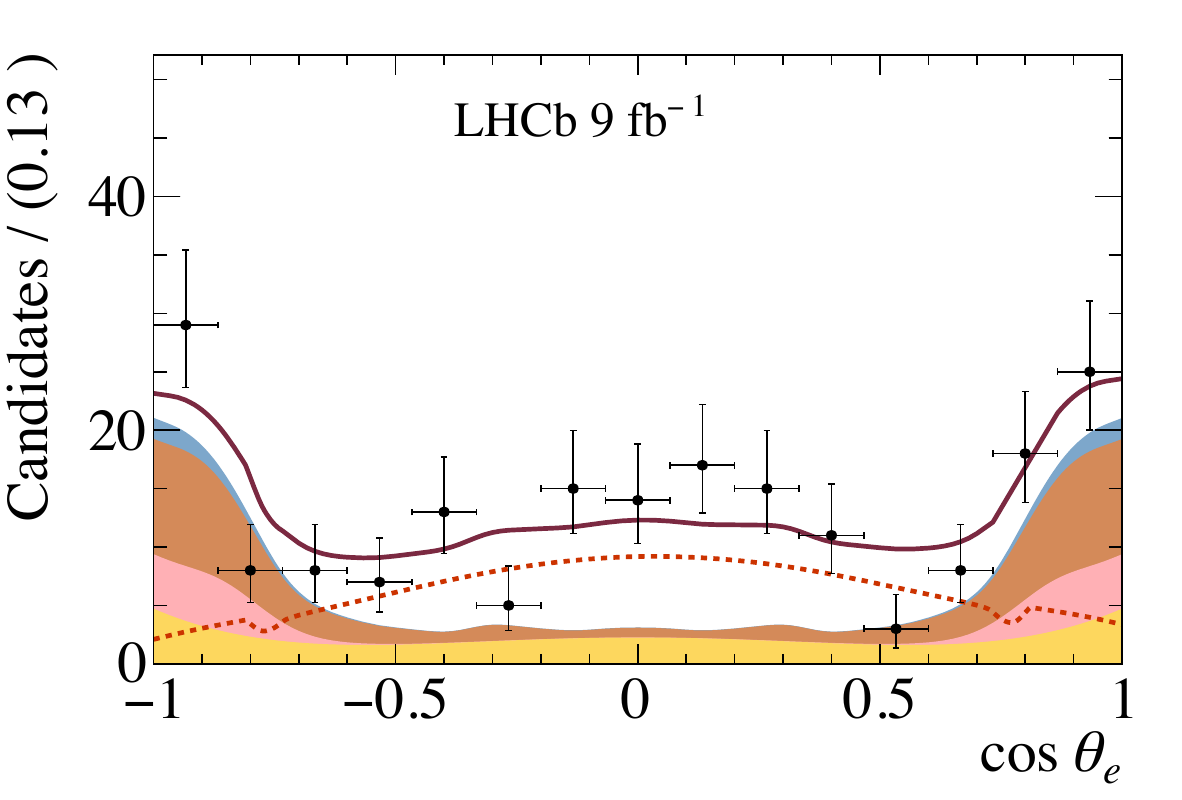} 
    \includegraphics[width=0.32\linewidth]{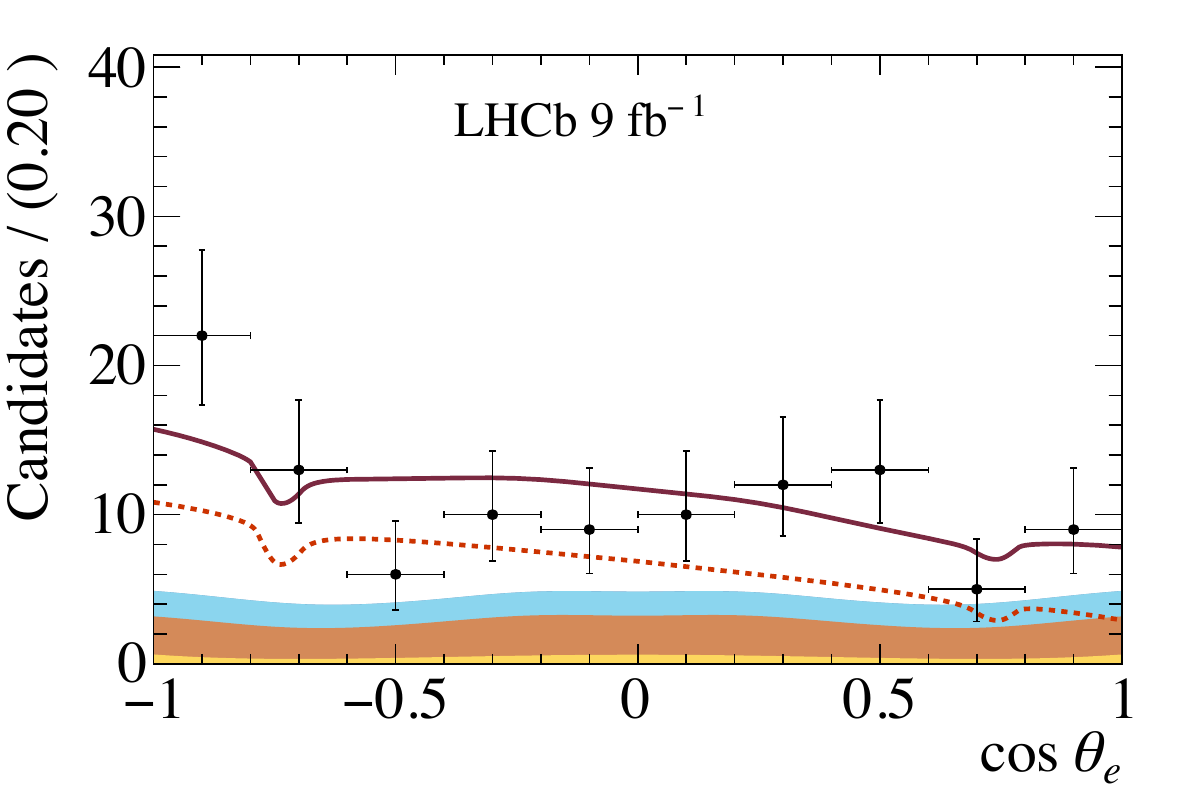} \\ 
    \includegraphics[width=0.32\linewidth]{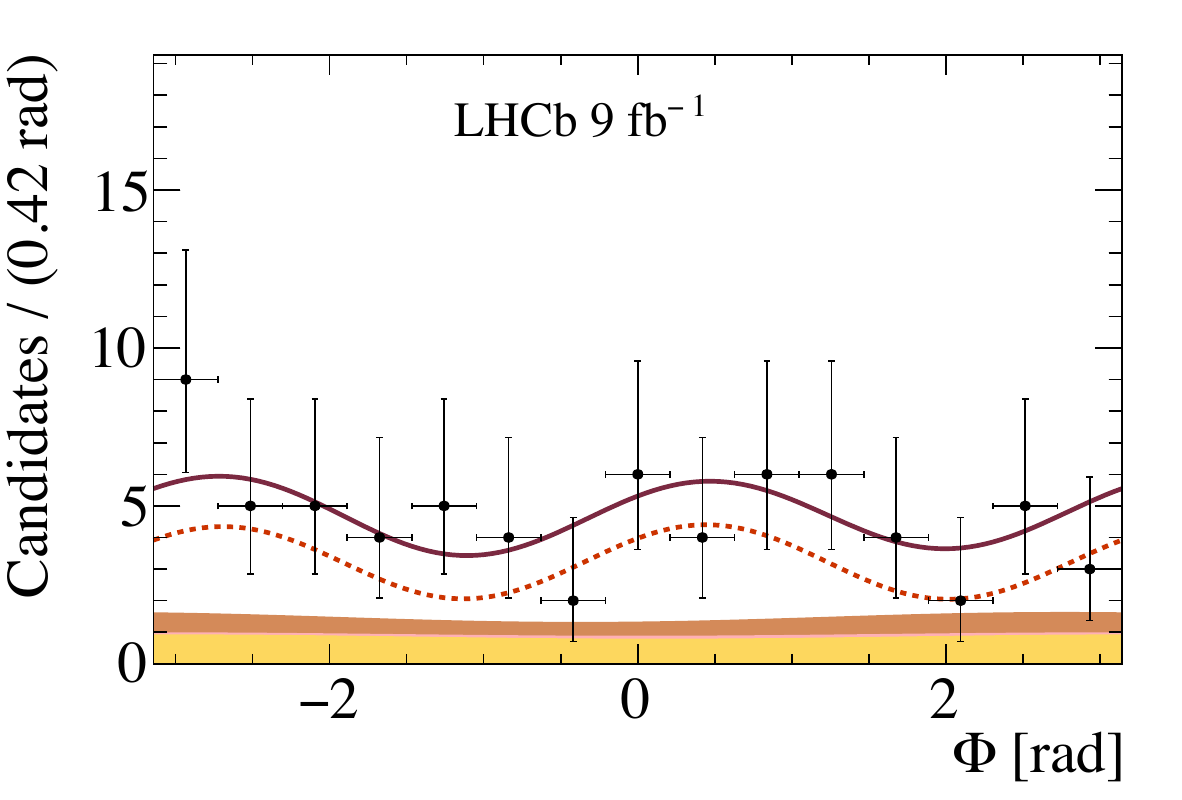} 
    \includegraphics[width=0.32\linewidth]{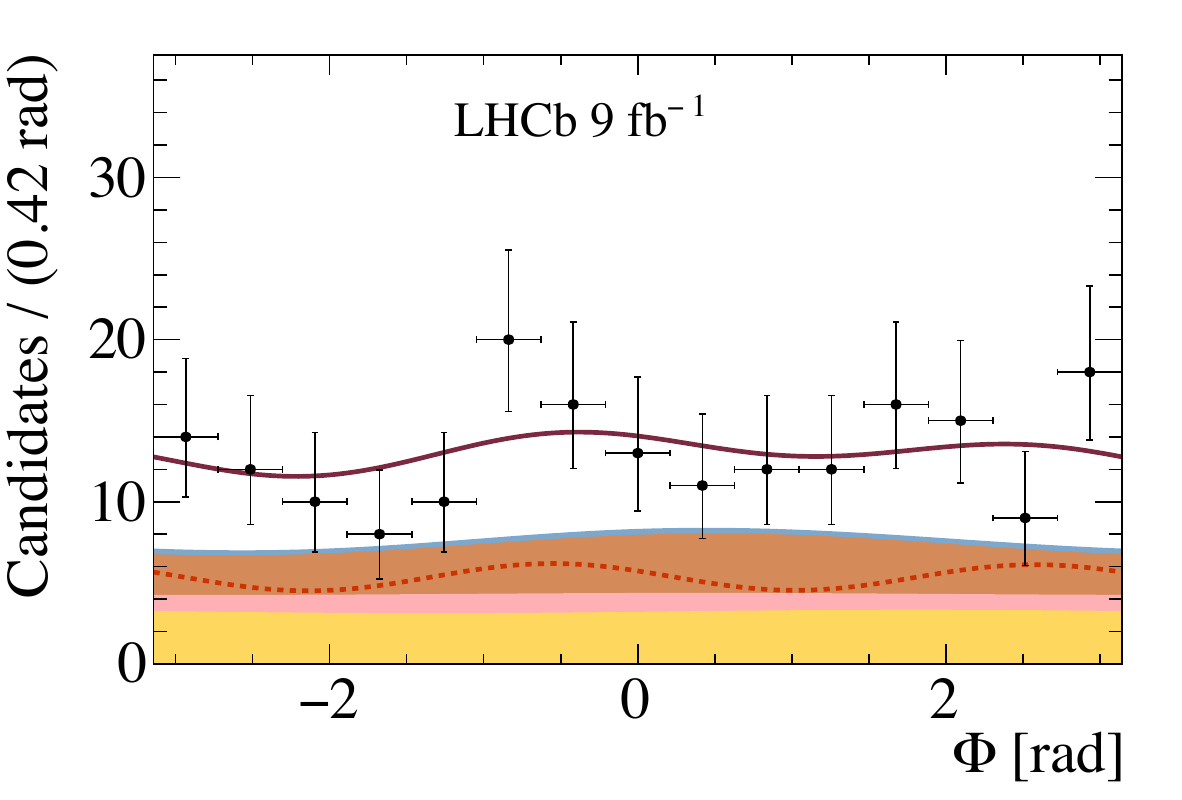} 
    \includegraphics[width=0.32\linewidth]{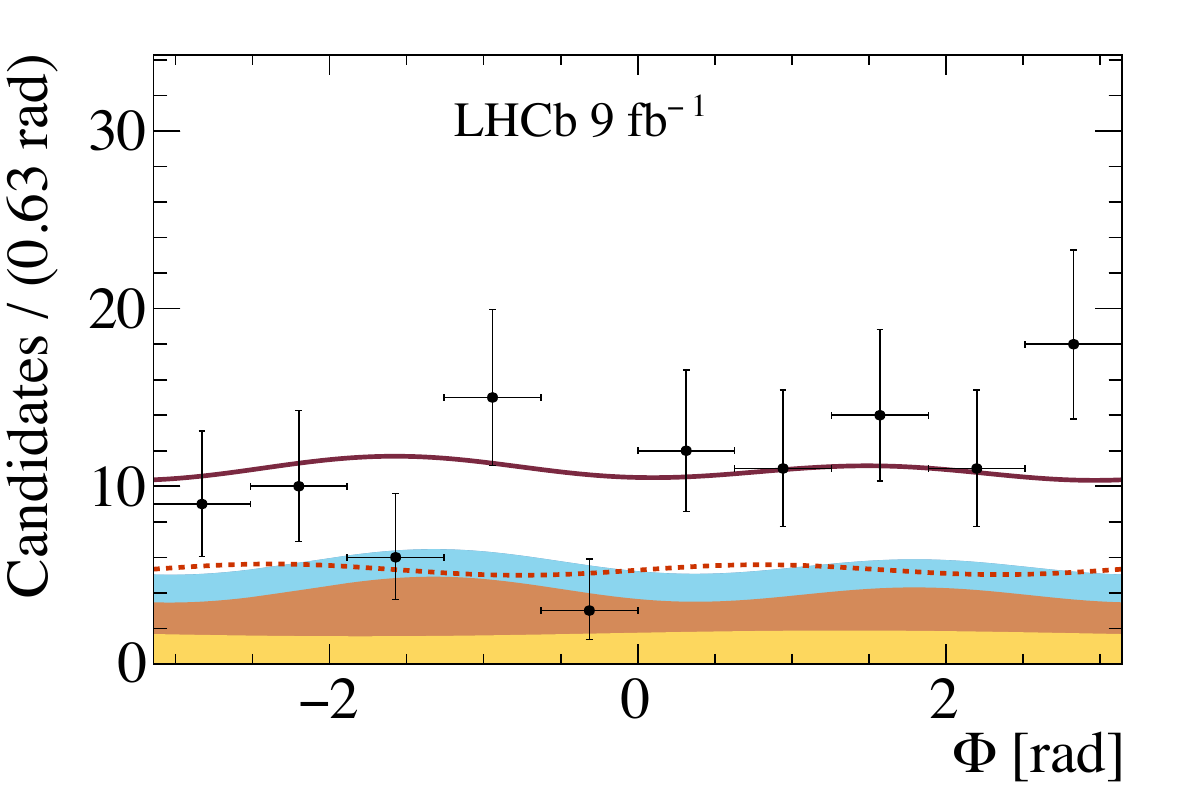} 
    \caption{
    Distributions of selected \BsPhiEE candidates with \qsq in the ranges (left to right) $[0.1,1.1]$, $[1.1,6.0]$ and $[15.0,19.0]\gevgevcccc$ in (top to bottom)~\mKKEE, \costhetak, \costhetae and $\Phi$. 
    The angular distributions are shown in reduced mass windows indicated on the \mKKEE figures by the vertical dashed red lines. 
    The results of the fits with the model described in the text are also shown.
    The dips in the fit results at $|\costhetae| \sim 0.8$ are due to the \Dsm veto.
    }
    \label{fig:results:distributions}
\end{figure}

Figure~\ref{fig:results:distributions} shows the distributions of the selected candidates in \mKKEE, \costhetak, \costhetae and $\Phi$, in each of the \qsq\ ranges, compared with the results of the fits. 
The data are well described by the model. 

The Feldman--Cousins procedure is used to obtain confidence level intervals for each observable and for relevant pairs of observables. 
The likelihood ratio between the tested and best-fit points is used as a test-statistic. 
At each tested point pseudoexperiments are used to determine the expected distribution of the test-statistic and derive the corresponding confidence level of the data. 
In the pseudoexperiment generation, nuisance parameters are fixed to the values returned from the fit to the data at the given values of the parameters of interest, \ie using the plug-in method~\cite{Bodhisattva:2009uba}.

\begin{figure}[p]
    \centering
    \includegraphics[width=0.325\linewidth]{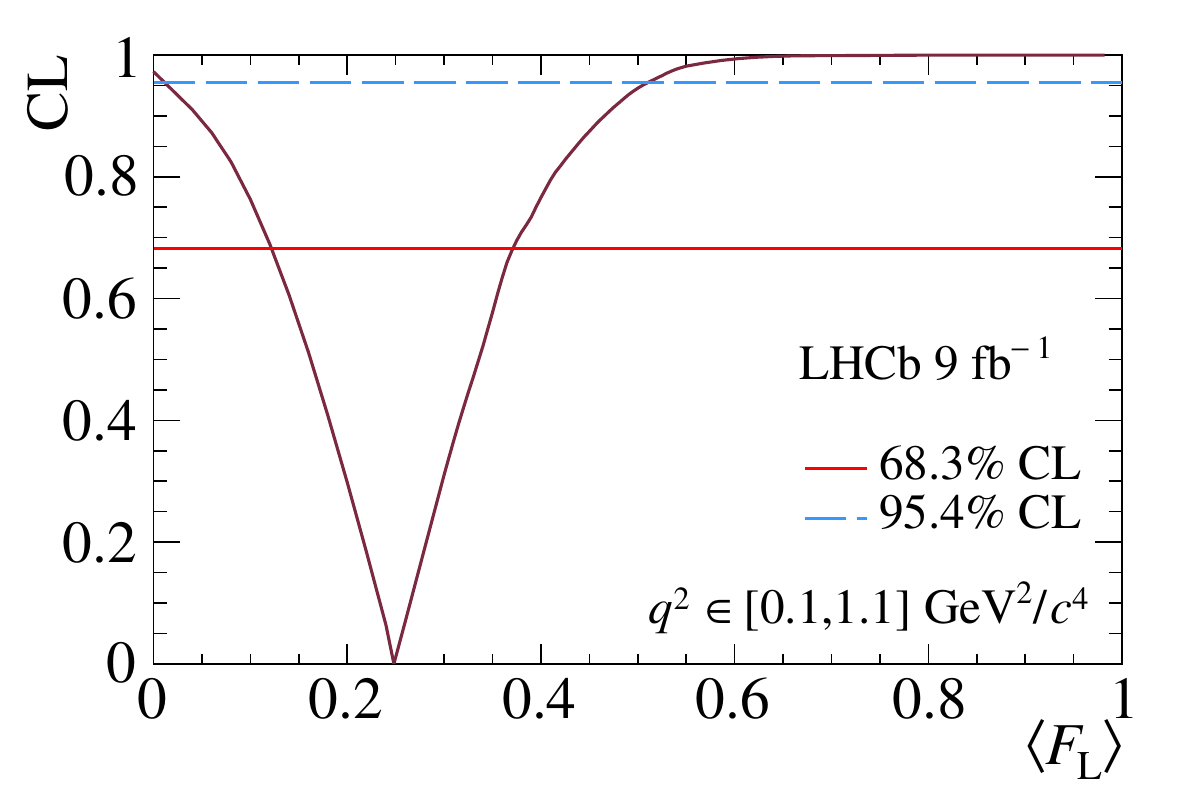} 
    \includegraphics[width=0.325\linewidth]{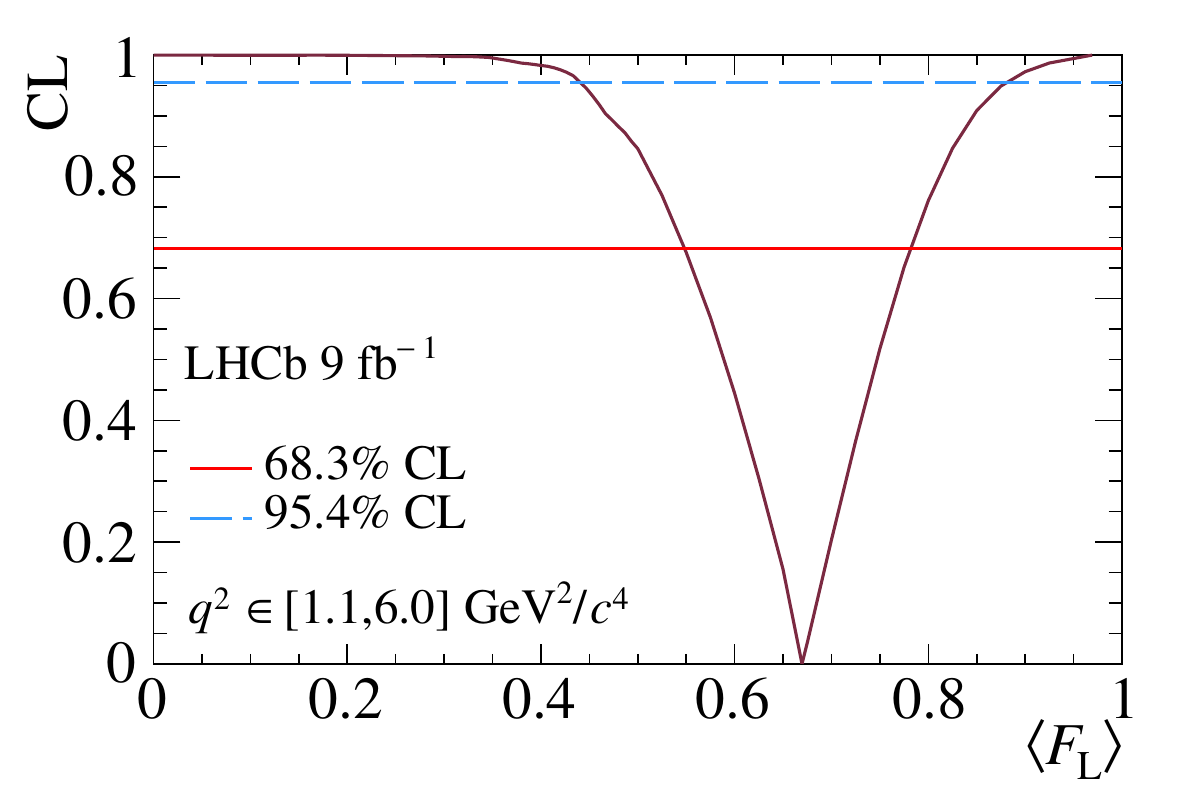} 
    \includegraphics[width=0.325\linewidth]{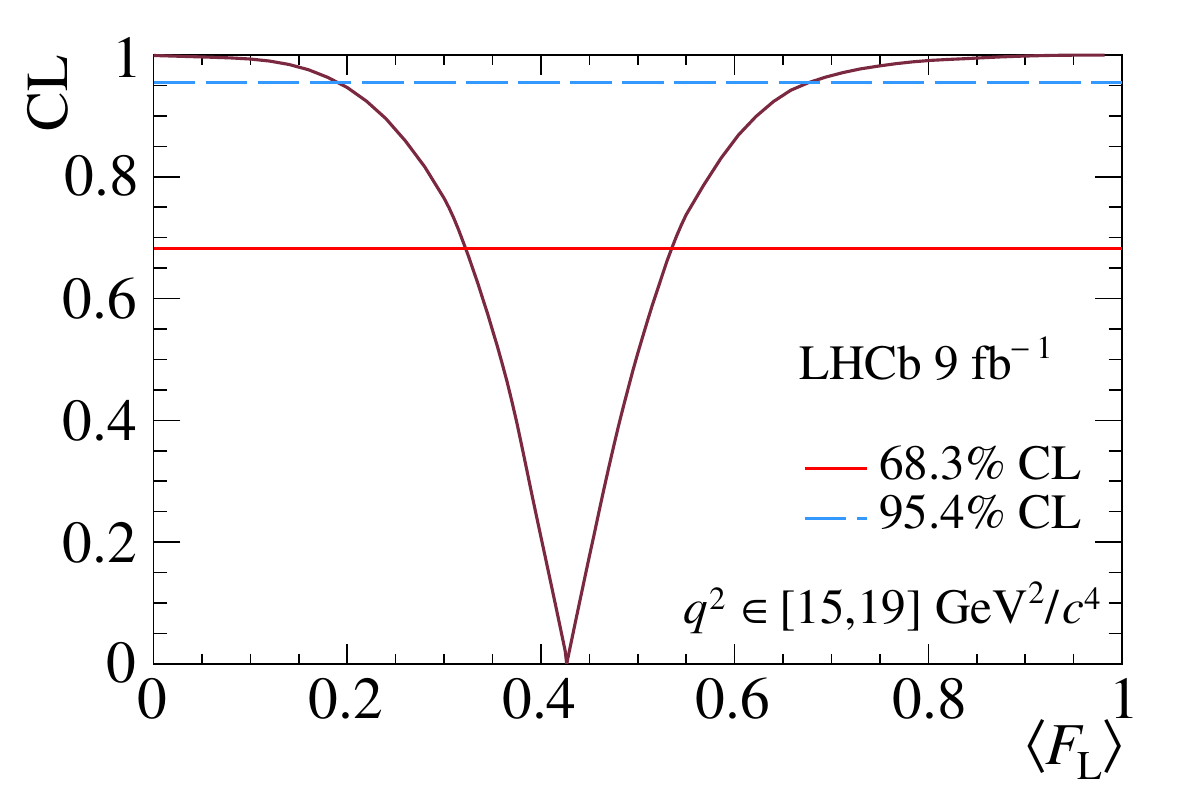} \\
    \includegraphics[width=0.325\linewidth]{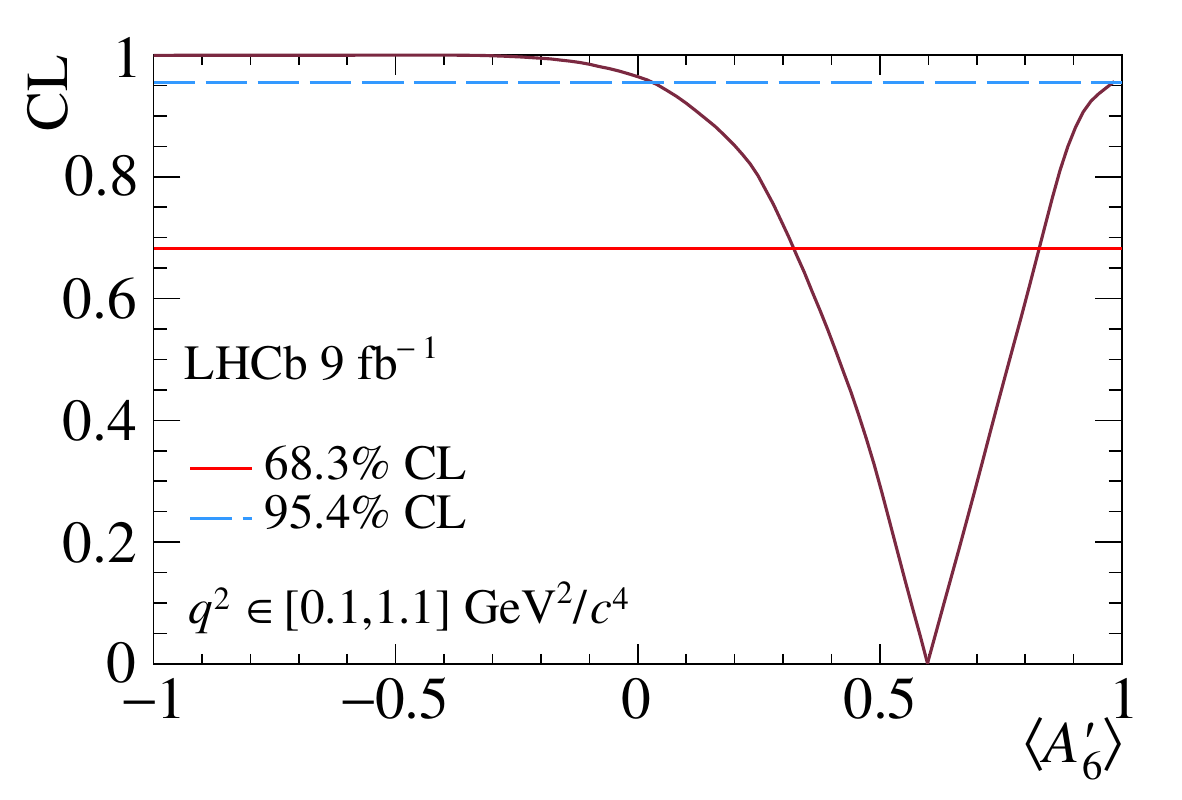} 
    \includegraphics[width=0.325\linewidth]{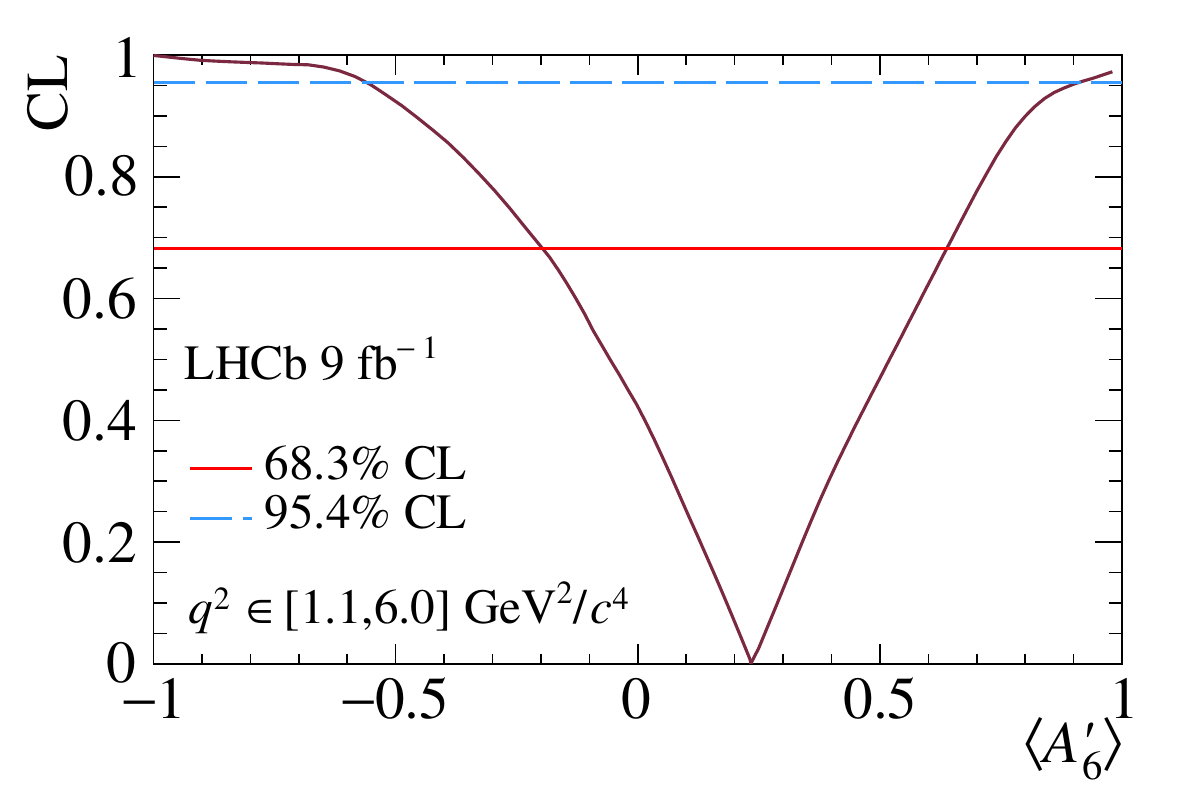} 
    \includegraphics[width=0.325\linewidth]{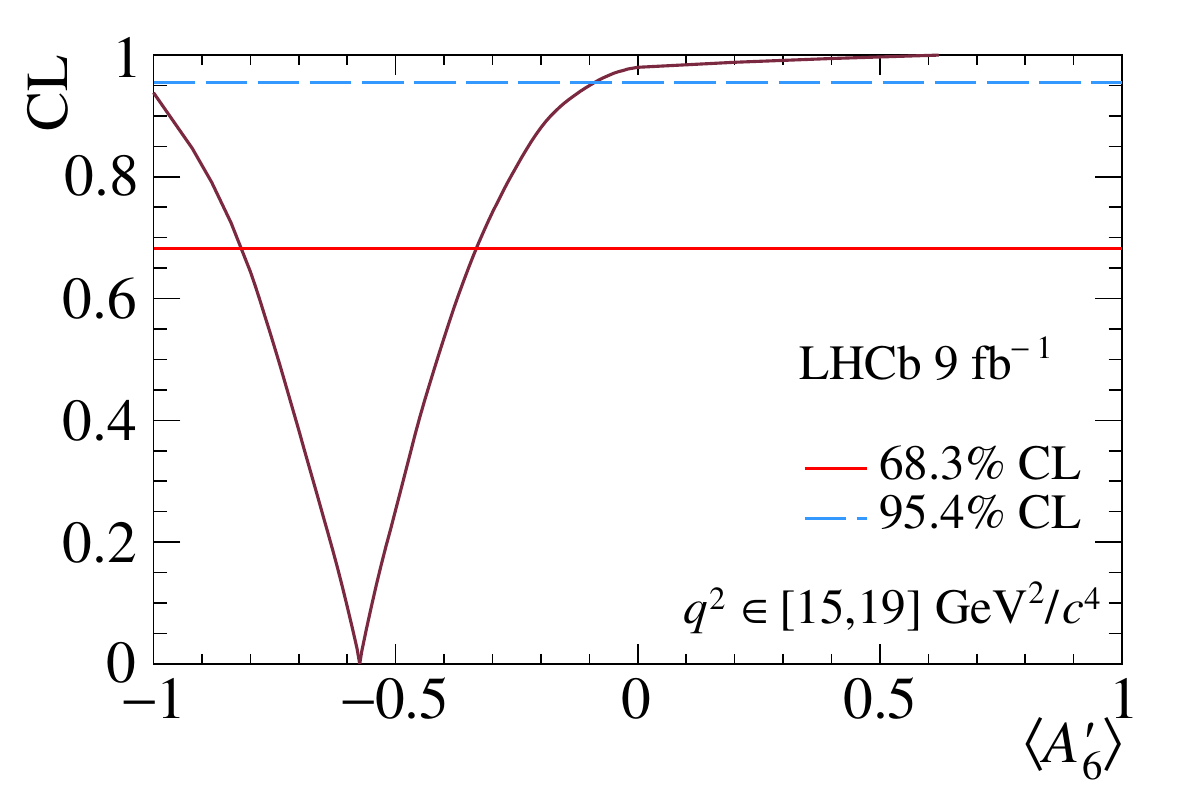} \\
    \includegraphics[width=0.325\linewidth]{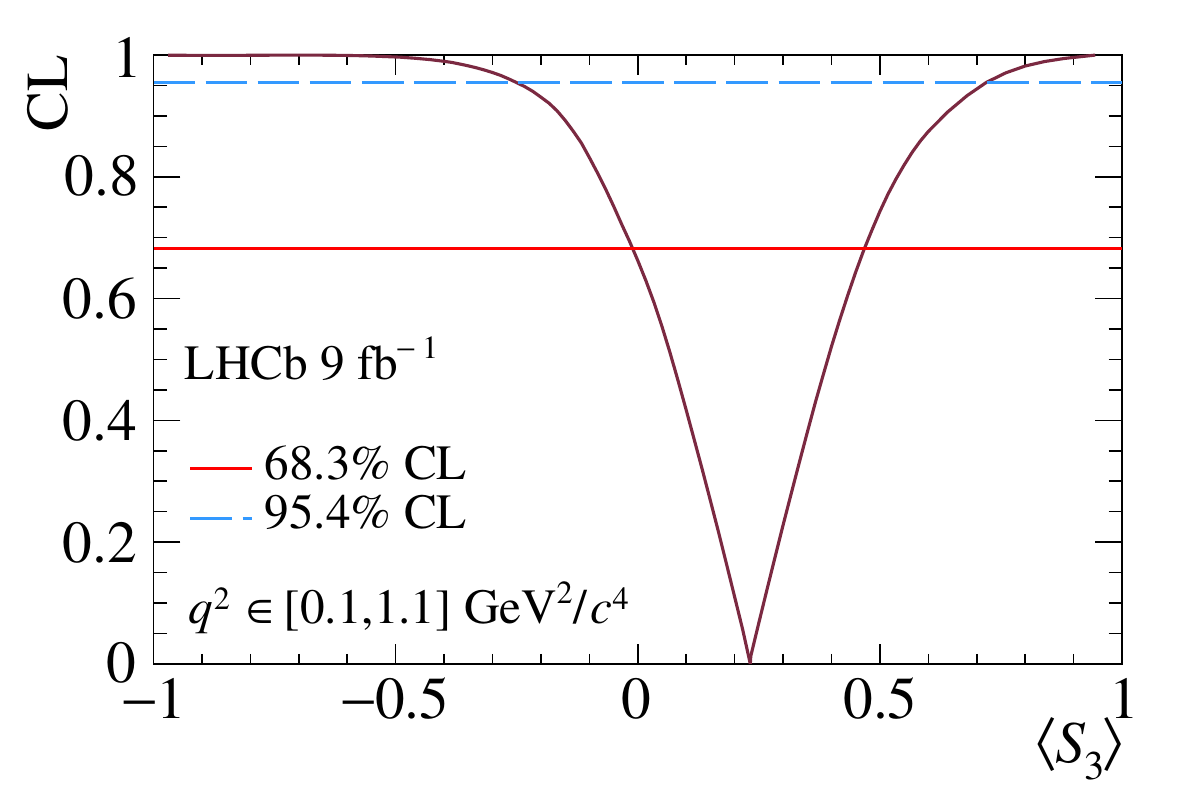} 
    \includegraphics[width=0.325\linewidth]{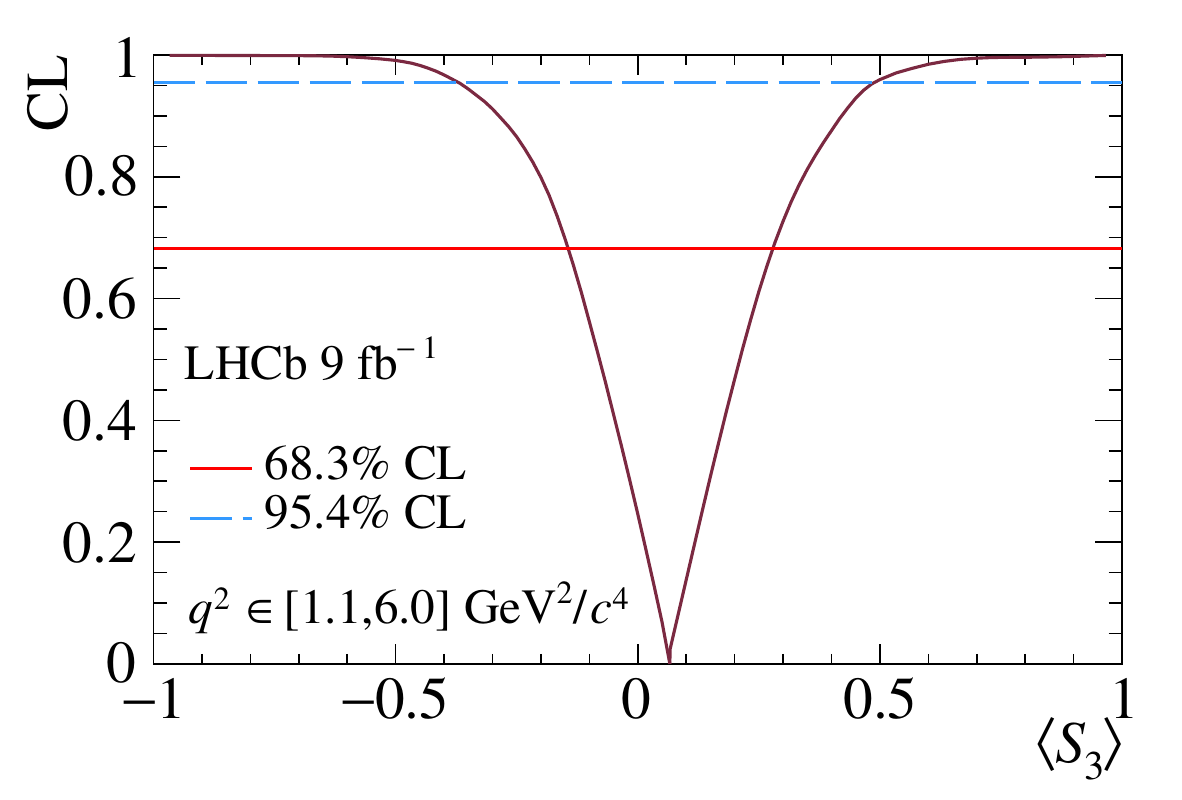} 
    \includegraphics[width=0.325\linewidth]{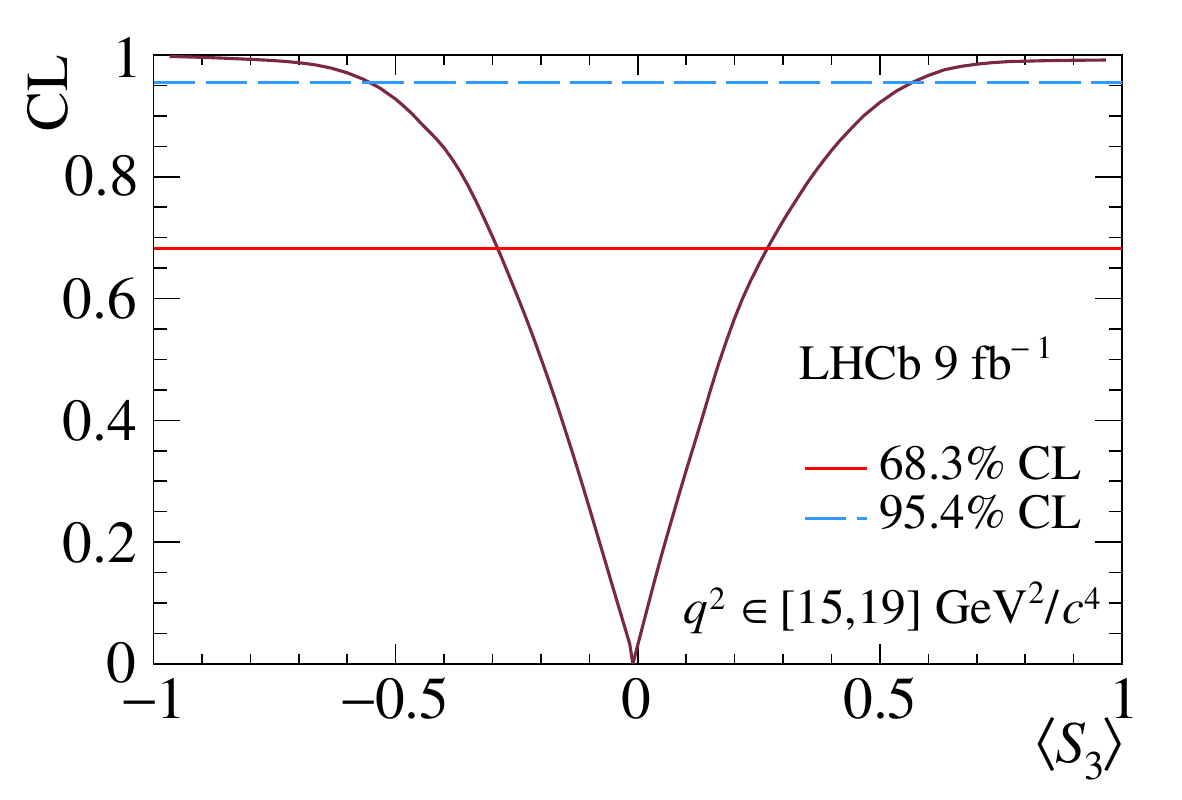} \\
    \includegraphics[width=0.325\linewidth]{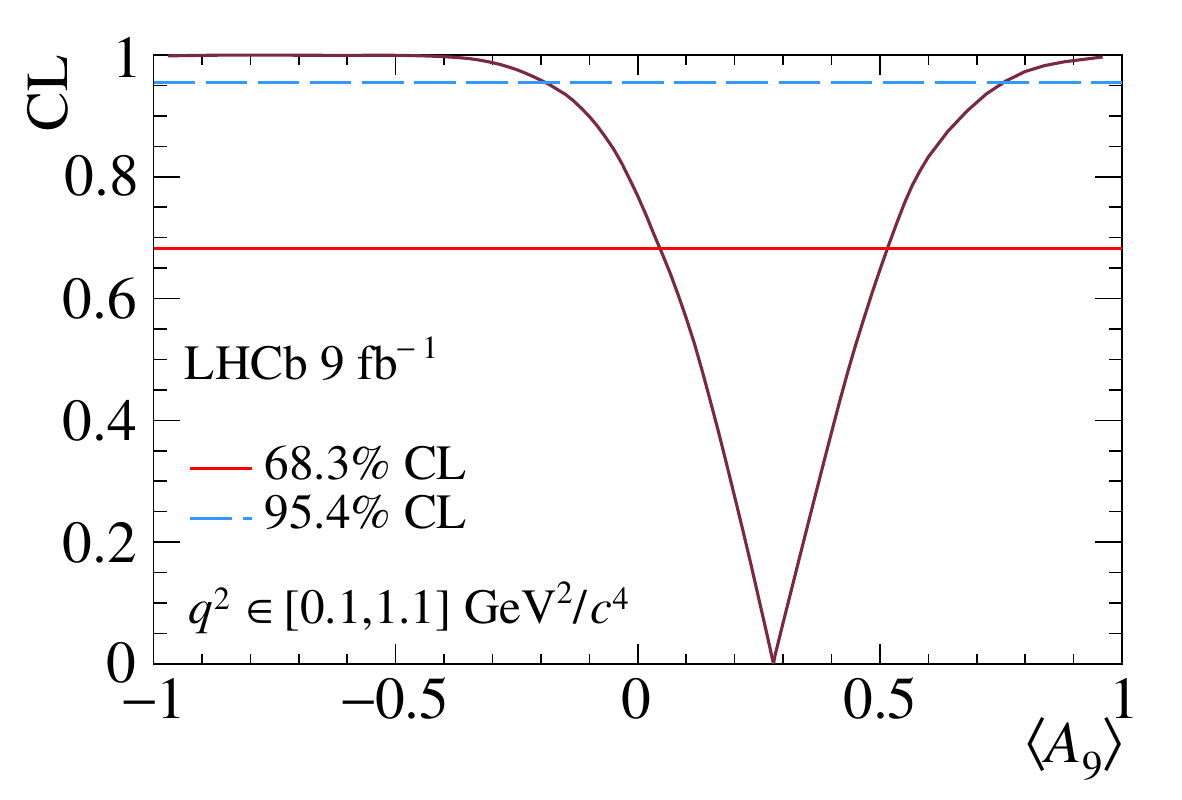} 
    \includegraphics[width=0.325\linewidth]{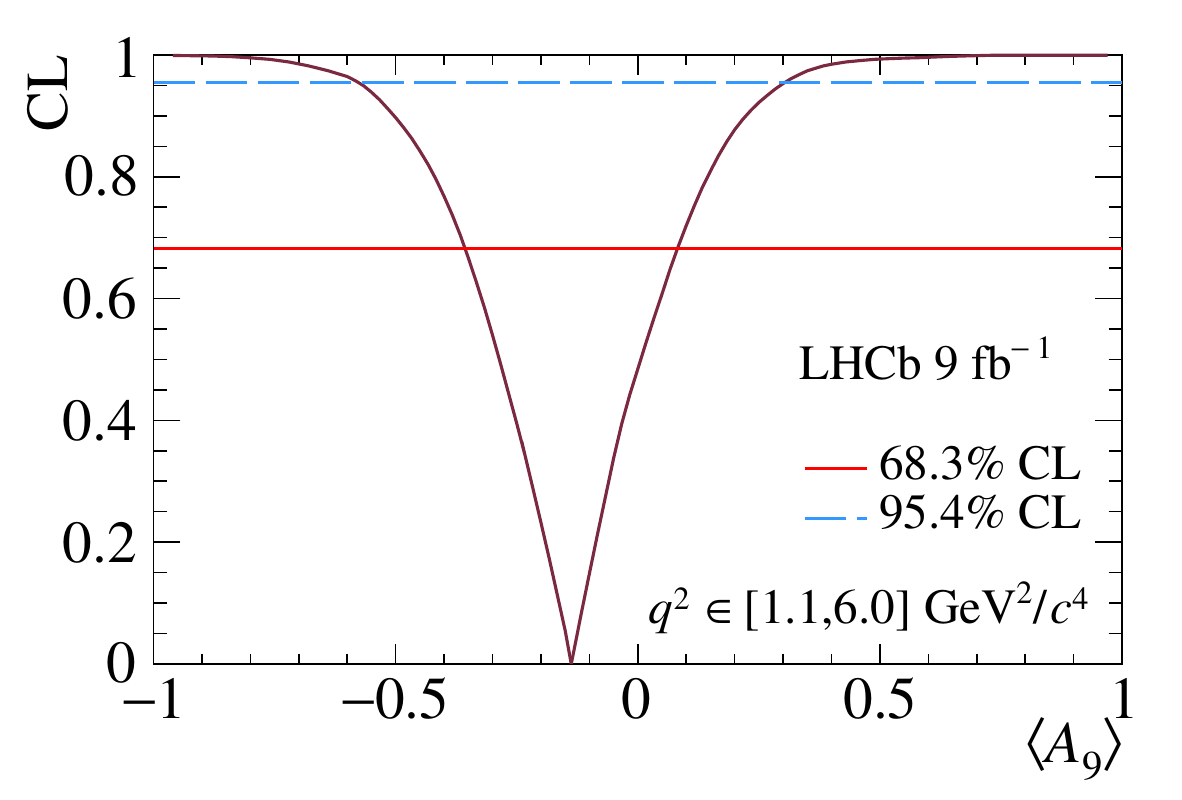} 
    \includegraphics[width=0.325\linewidth]{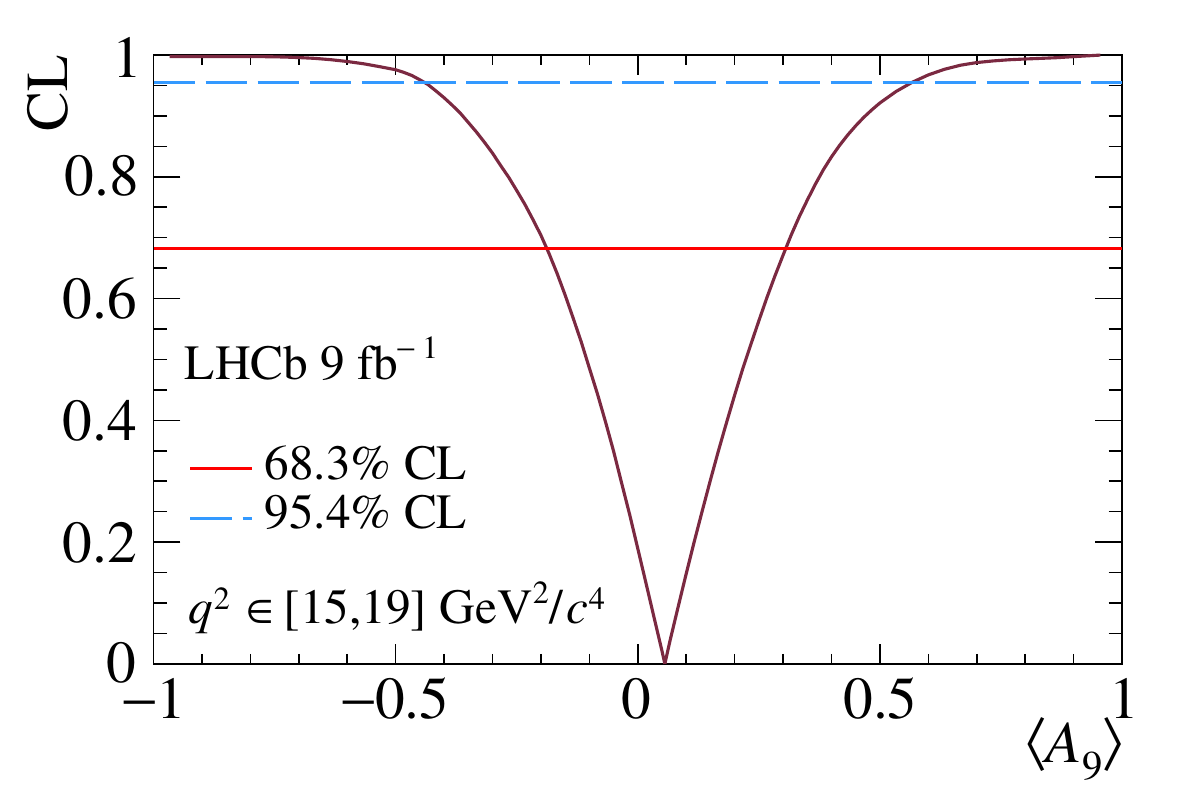} \\    
    \caption{
    Results of the Feldman--Cousins procedure for the observables (top to bottom)~\IntFL, \IntAsixp, \IntSthree and \IntAnine in the \qsq\ ranges (left to right) $[0.1,1.0]$, $[1.1,6.0]$ and $[15.0,19.0]$\gevgevcccc.   
    The $68.3\%$ ($95.4\%$) confidence level is drawn as the red solid (blue dashed) line.
    }
    \label{fig:cl:1d}
\end{figure}

\begin{figure}[p]
    \centering
    \includegraphics[width=0.45\linewidth]{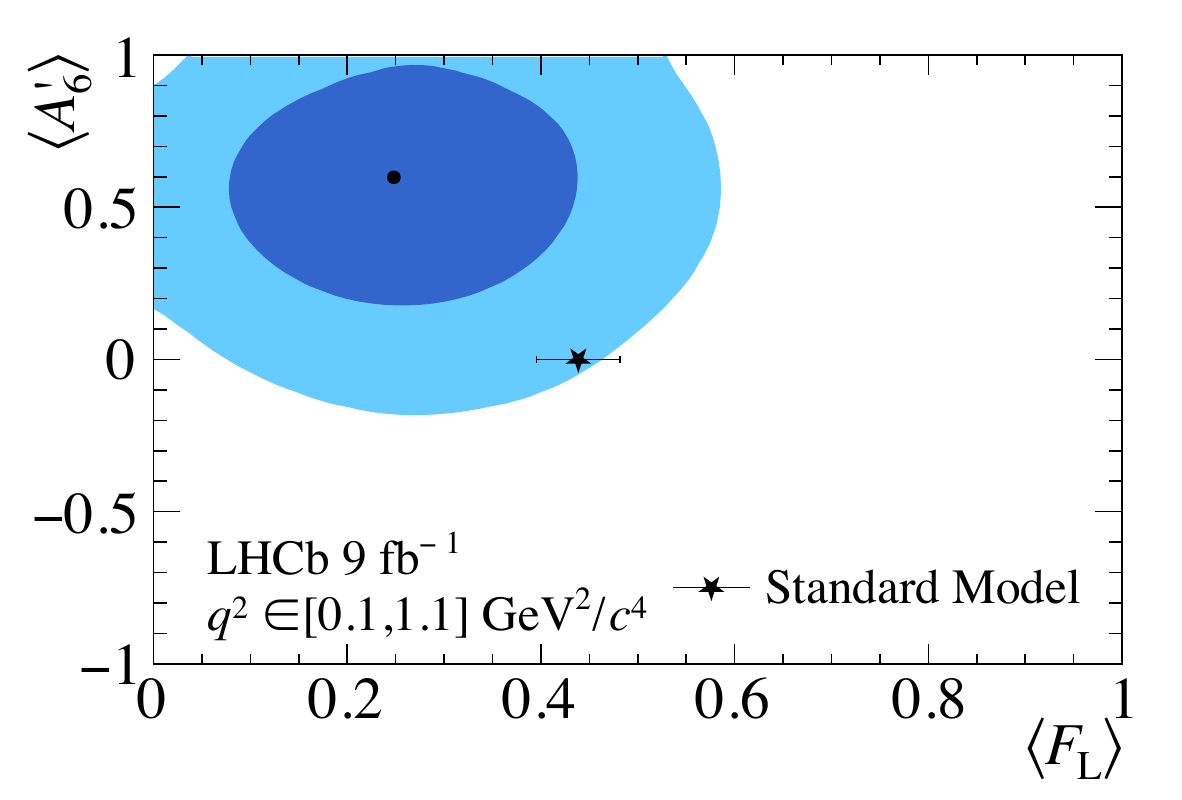}
    \includegraphics[width=0.45\linewidth]{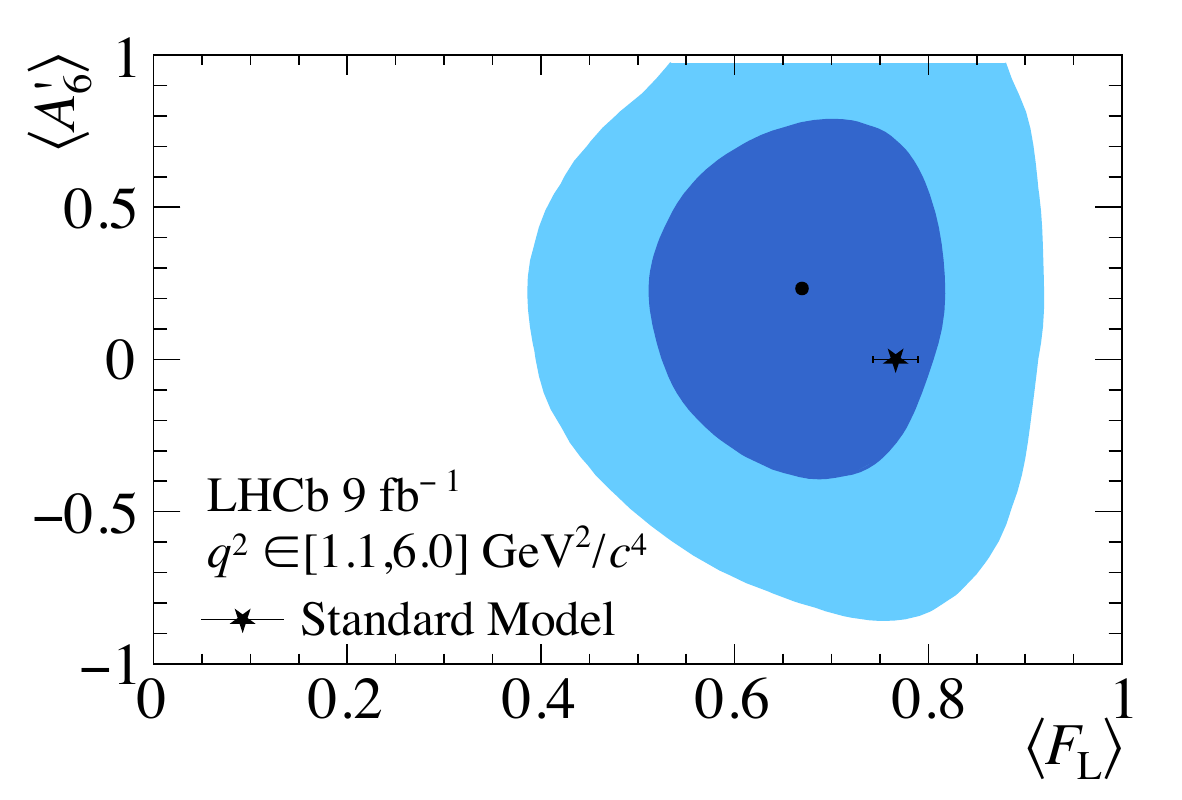} \\
    \includegraphics[width=0.45\linewidth]{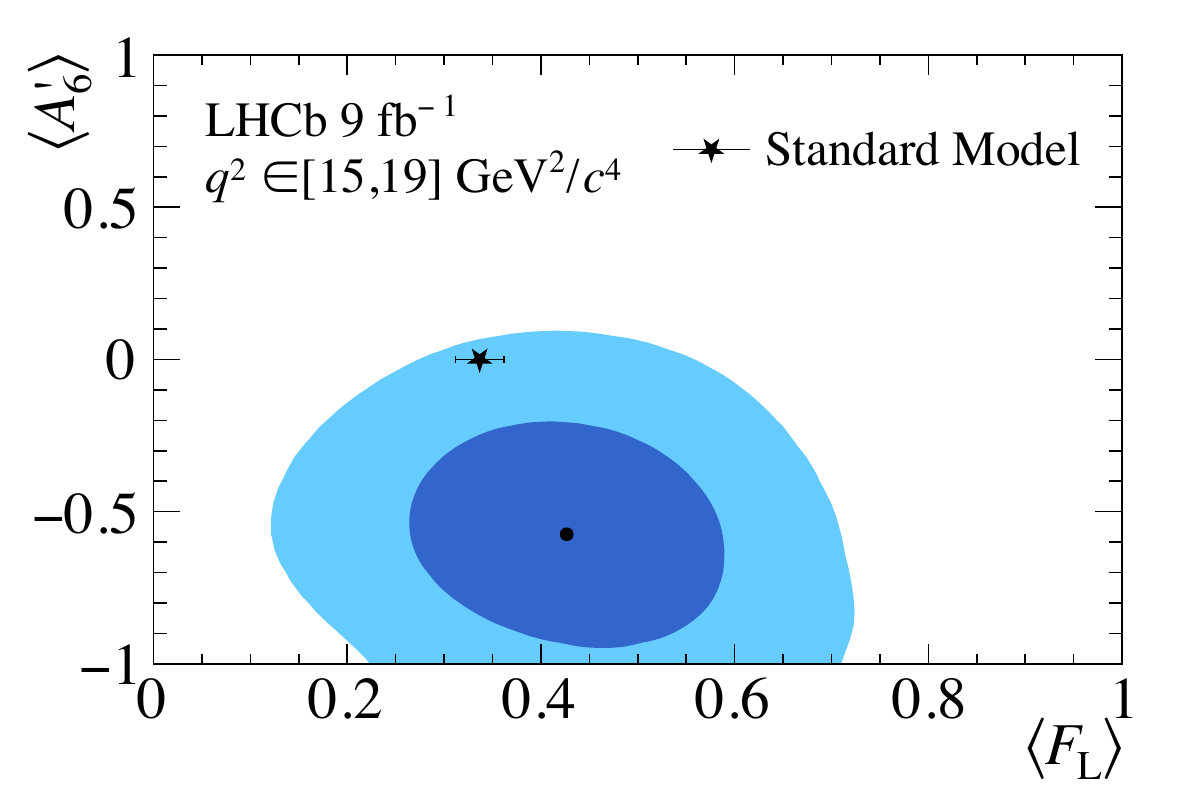}      
    \caption{
    Two-dimensional 68.3\% and 95.5\% confidence level regions for the observables \IntFL and \IntAsixp in the \qsq ranges (top left) $[0.1,1.0]$, (top right) $[1.1,6.0]$ and (bottom) $[15.0,19.0]$\gevgevcccc.   
    Standard Model predictions, obtained from Ref.~\cite{straub:2018flavio}, are also shown.
    }
    \label{fig:cl:2d:fl:a6p}
\end{figure}

\begin{figure}[p]
    \centering
    \includegraphics[width=0.45\linewidth]{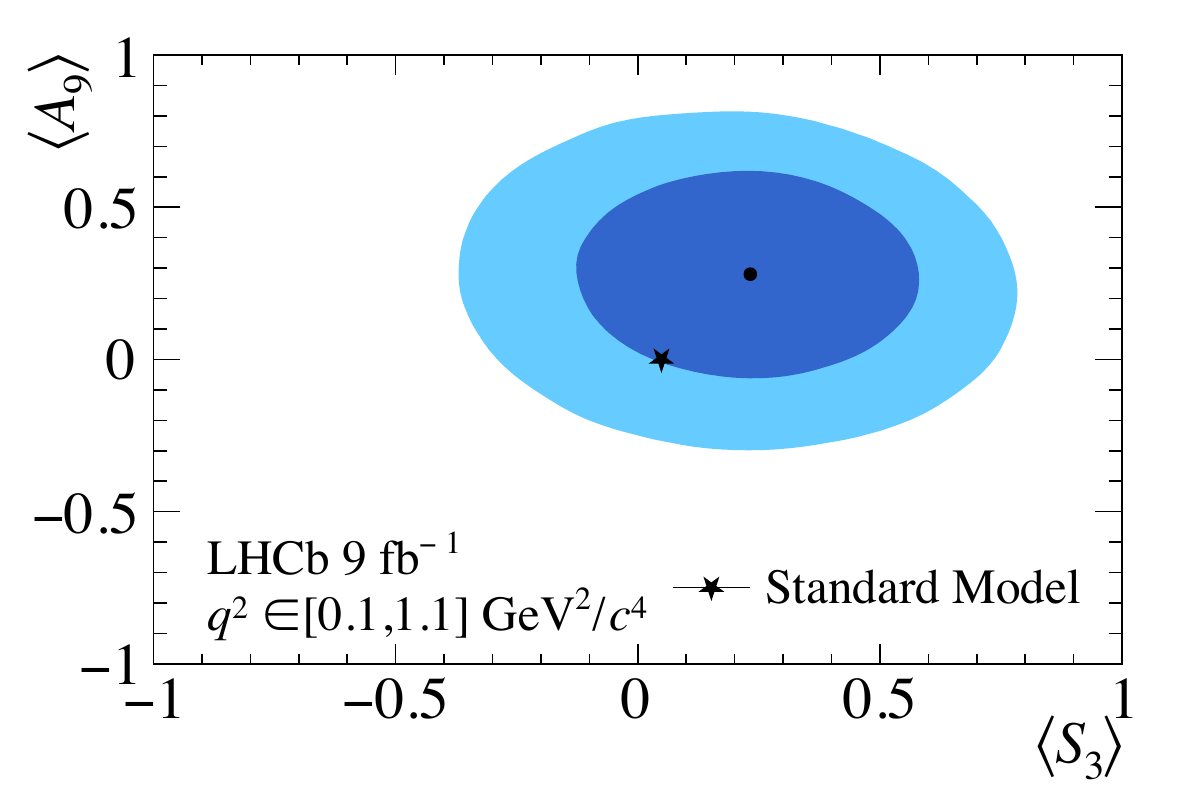}
    \includegraphics[width=0.45\linewidth]{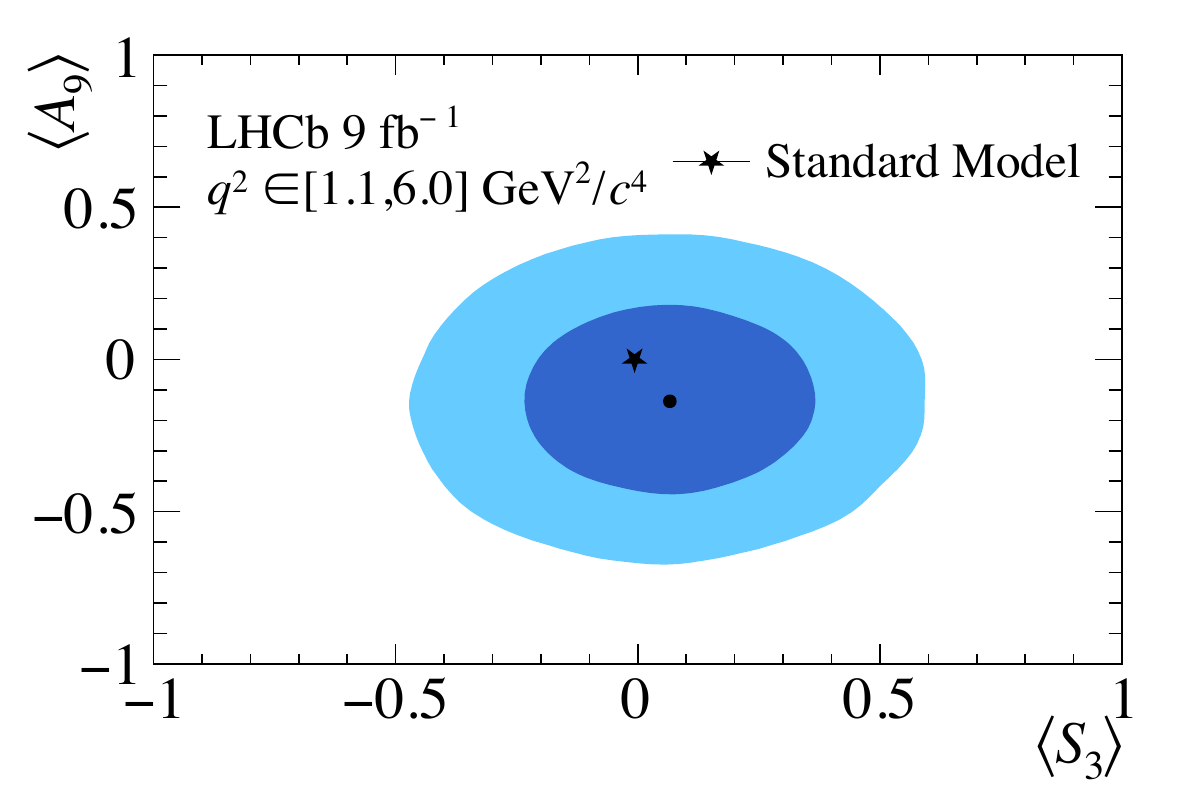} \\
    \includegraphics[width=0.45\linewidth]{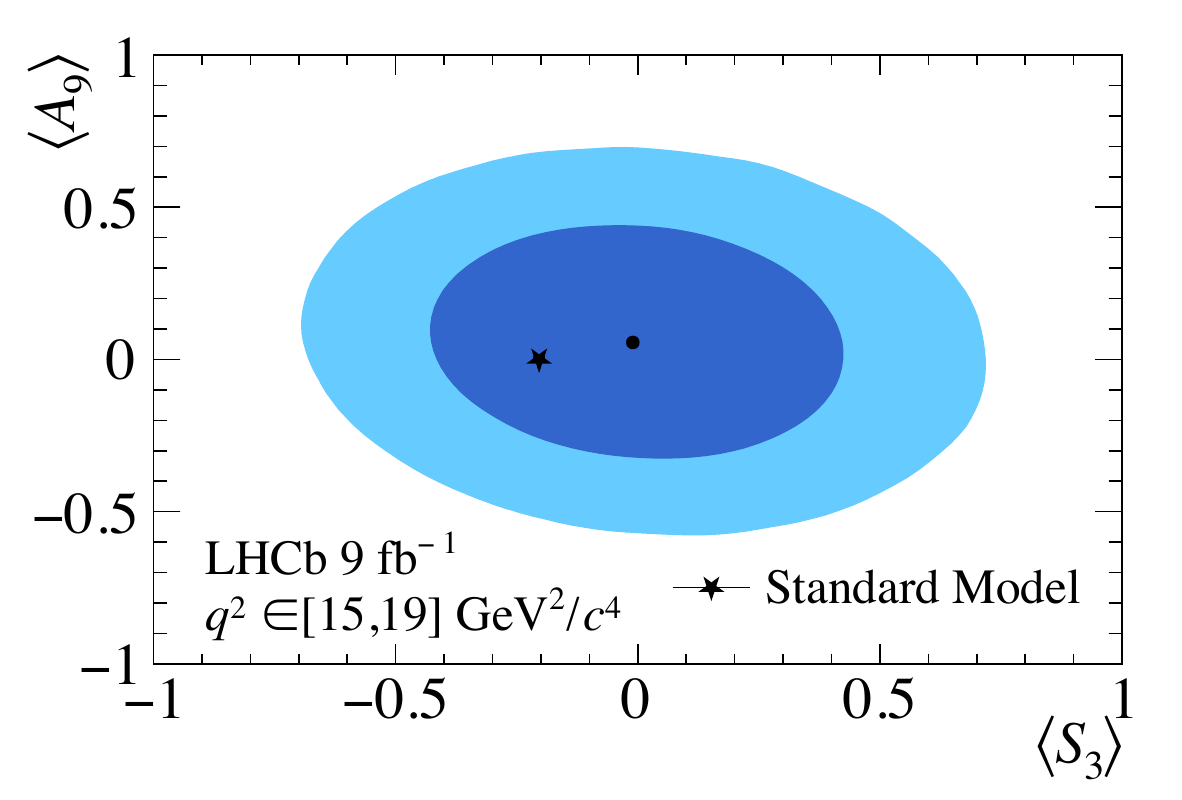}      
    \caption{
    Two-dimensional 68.3\% and 95.5\% confidence level regions for the observables \IntSthree and \IntAnine in the \qsq ranges (top left) $[0.1,1.0]$, (top right) $[1.1,6.0]$ and (bottom) $[15.0,19.0]$\gevgevcccc.   
    Standard Model predictions, obtained from Ref.~\cite{straub:2018flavio}, are also shown.
    }
    \label{fig:cl:2d:s3:a9}
\end{figure}

The resulting intervals on individual parameters are shown in Fig.~\ref{fig:cl:1d}, along with two-dimensional intervals for $\left(\IntFL, \IntAsixp\right)$ and $\left(\IntSthree, \IntAnine\right)$ in Figs.~\ref{fig:cl:2d:fl:a6p} and \ref{fig:cl:2d:s3:a9}, respectively.
The values for the rate-averaged and decay-time-integrated observables and their 68\% confidence level intervals are given in Table~\ref{tab:results}. 

\begin{table}[p]
\caption{
Measured values of the observables in the three \qsq bins. 
The central value indicates the best-fit point, the first set of uncertainties indicate the range of the 68\% confidence level interval about the best-fit point, and the second correspond to the total systematic uncertainty.
}
\centering
\scalebox{0.95}{
\renewcommand{\arraystretch}{1.5}
\begin{tabular}{cccc}
\toprule
Observable & $0.1 < \qsq < 1.1\gevgevcccc$ & $1.1 < \qsq < 6.0\gevgevcccc$ & $15.0 < \qsq < 19.0\gevgevcccc$ \\
\midrule
\IntFL     & $0.25 \,^{+0.12}_{-0.12} \pm 0.06$ &  $\phantom{-}0.67 \,^{+0.12}_{-0.13} \pm 0.06$    & $\phantom{-}0.43 \,^{+0.11}_{-0.10} \pm 0.05$ \\
\IntAsixp  & $0.60 \,^{+0.23}_{-0.28} \pm 0.05$ &  $\phantom{-}0.24 \,^{+0.40}_{-0.42} \pm 0.09$    & $          -0.57 \,^{+0.24}_{-0.25} \pm 0.05$ \\ 
\IntSthree & $0.23 \,^{+0.24}_{-0.24} \pm 0.07$ &  $\phantom{-}0.07 \,^{+0.21}_{-0.21} \pm 0.07$    & $          -0.01 \,^{+0.29}_{-0.28} \pm 0.08$ \\ 
\IntAnine  & $0.28 \,^{+0.23}_{-0.24} \pm 0.04$ &  $          -0.14 \,^{+0.23}_{-0.24} \pm 0.04$    & $\phantom{-}0.06 \,^{+0.25}_{-0.25} \pm 0.05$ \\ 
\bottomrule
\end{tabular}
}
\label{tab:results}
\end{table}

\section{Summary} 
\label{sec:summary} 

An angular analysis of the \BsPhiEE decay has been performed, using a dataset collected with the LHCb detector between 2011 and 2018, corresponding to an integrated luminosity of 9\invfb of high energy proton-proton collisions. 
The analysis is performed in three bins of dielectron mass squared, $\qsq  \in [0.1,1.1]$, $[1.1, 6.0]$ and $[15.0, 19.0]\gevgevcccc$. 
The analysis is performed integrating over decay time, averaging over the range of \qsq\ within each bin and without distinguishing between \Bs\ and \Bsb\ decays.
The observables \IntFL, \IntAsixp, \IntSthree and \IntAnine are measured for the first time in \BsPhiEE decays.

\begin{figure}[!tb]
    \centering
    \includegraphics[width=0.48\linewidth]{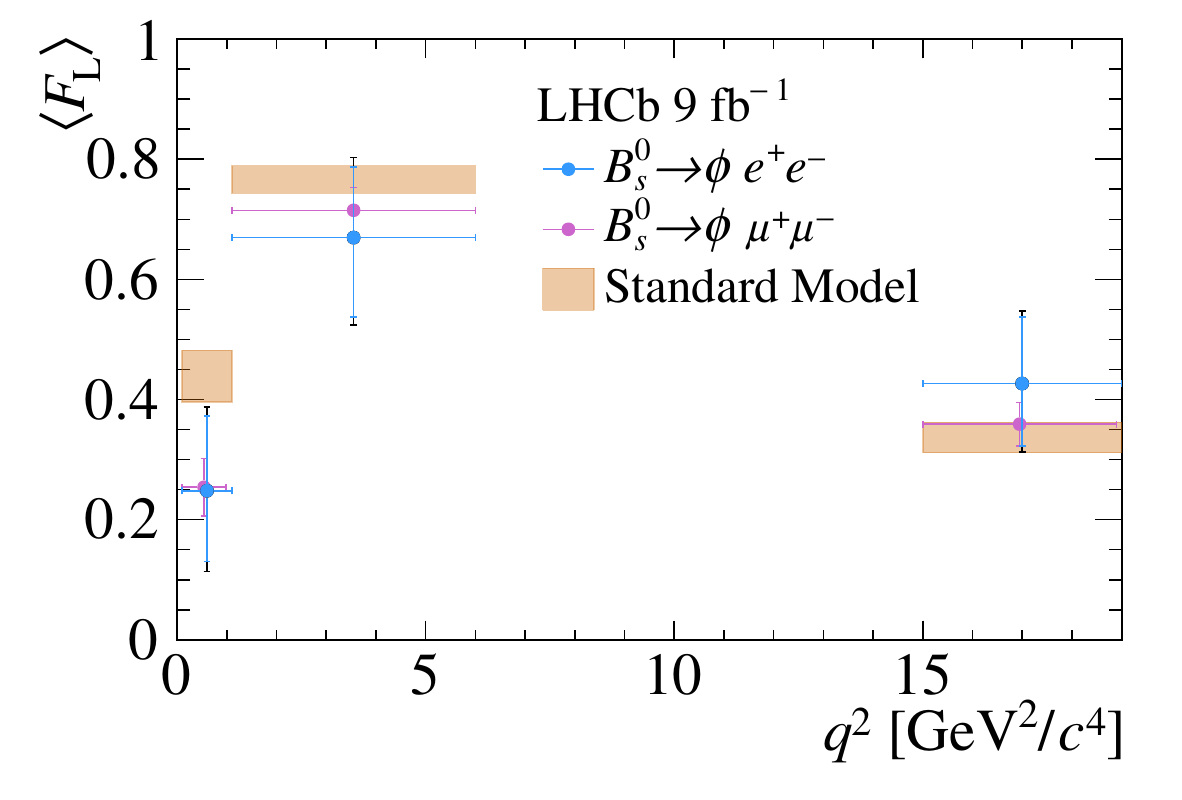} 
    \includegraphics[width=0.48\linewidth]{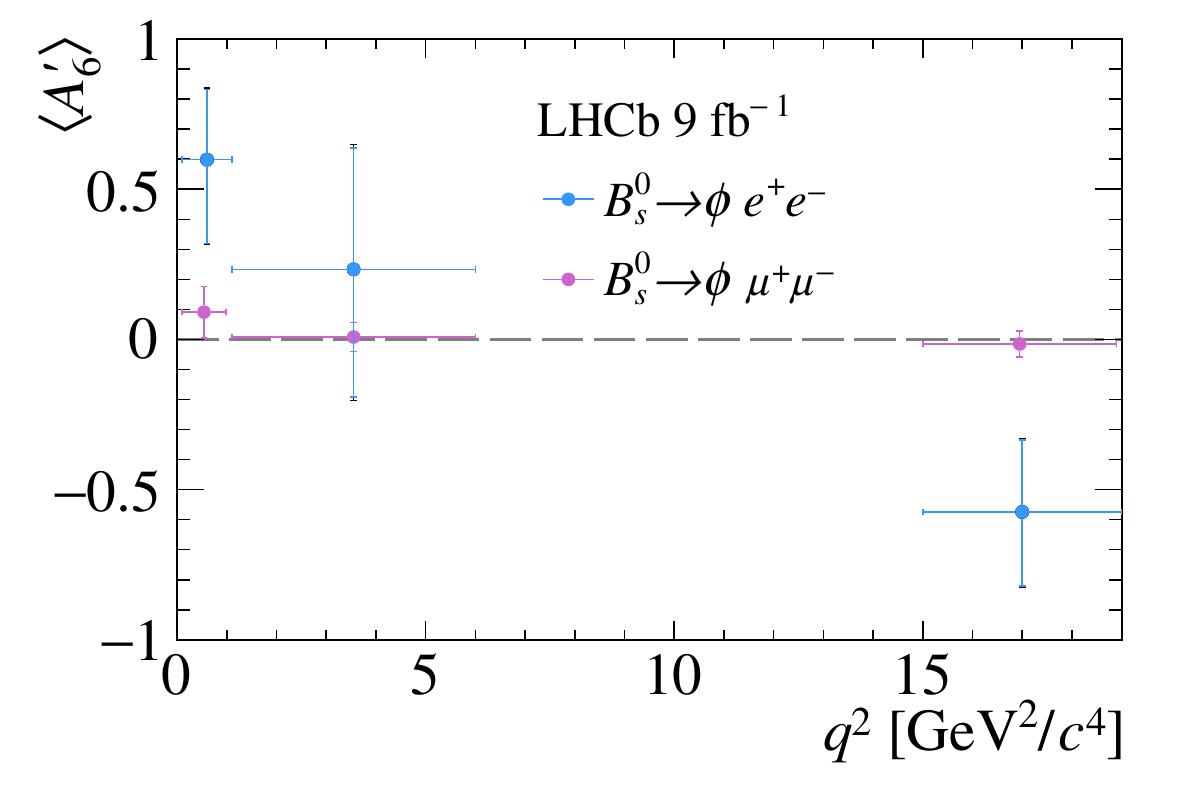} \\  
    \includegraphics[width=0.48\linewidth]{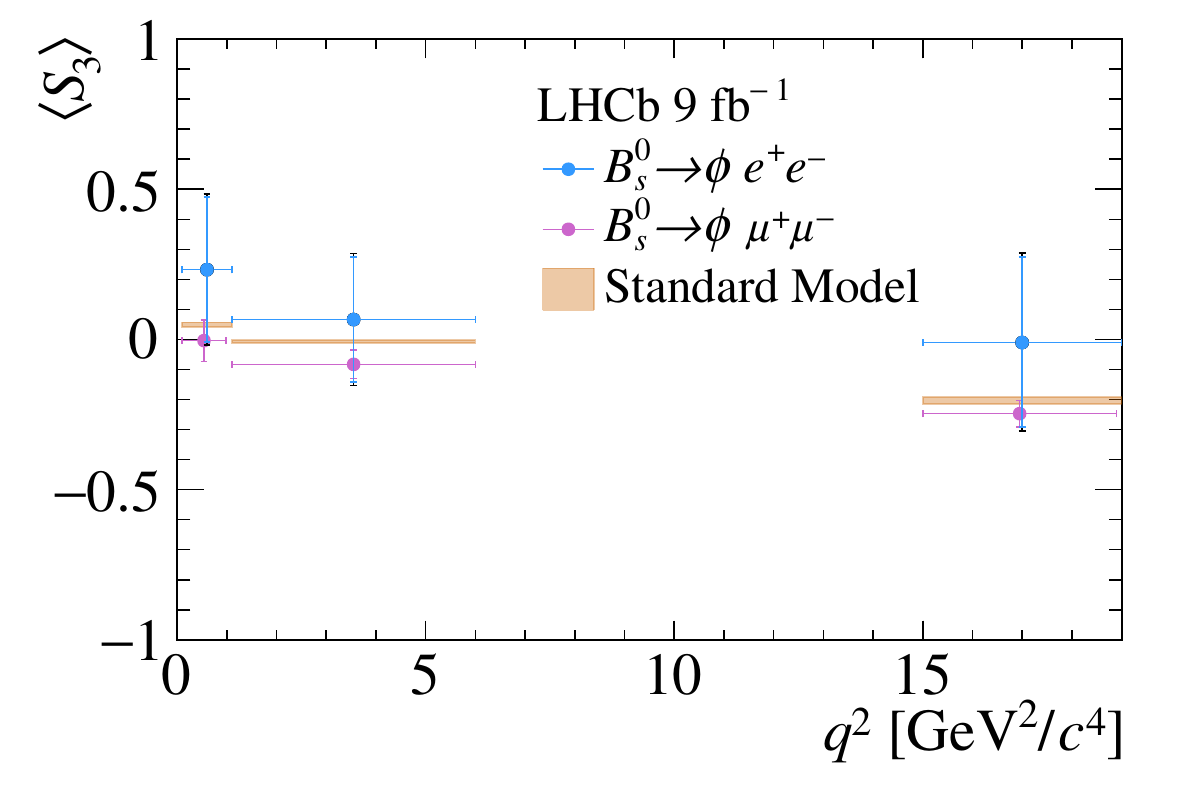} 
    \includegraphics[width=0.48\linewidth]{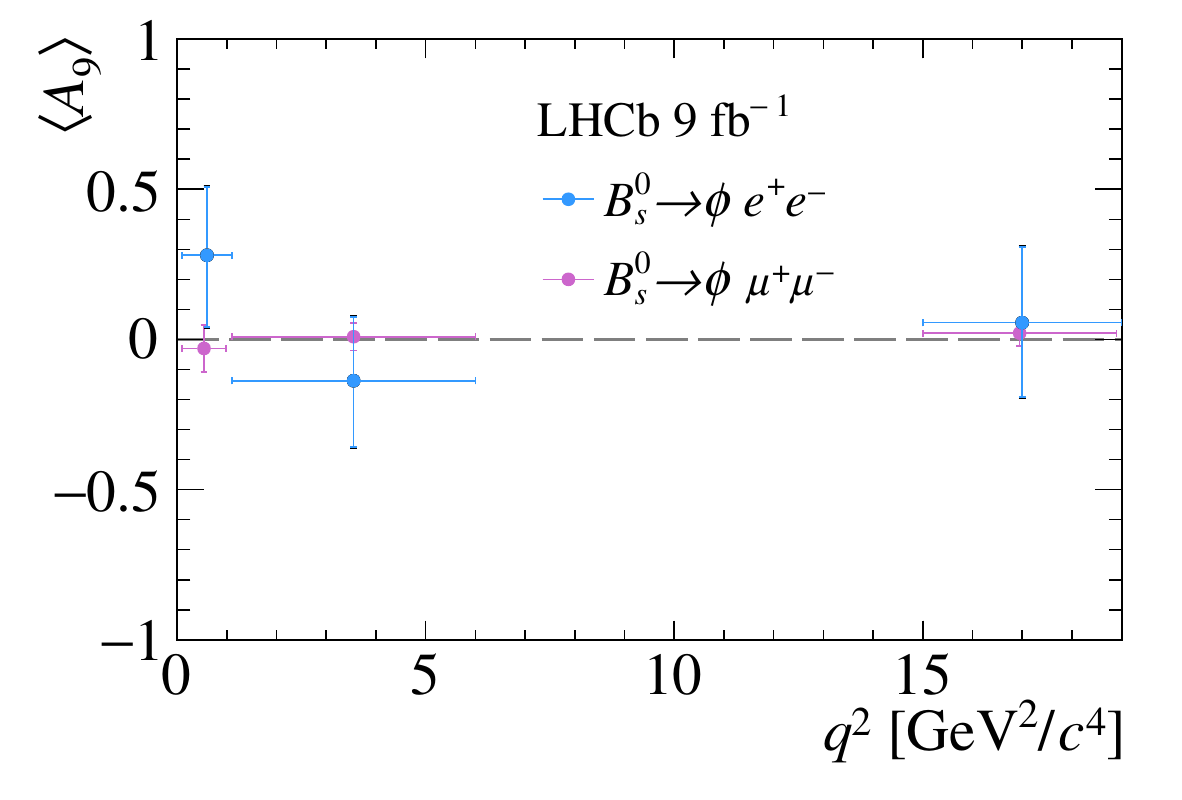} 
    \caption{
    Rate-averaged angular observables as a function of \qsq. 
    The measured values are compared with SM predictions obtained using Ref.~\cite{straub:2018flavio}. 
    The SM predictions for \IntAsixp and \IntAnine are indistinguishable from zero. 
    }
    \label{fig:results} 
\end{figure}

The results for the observables measured in the three \qsq bins are compared to SM predictions in Fig.~\ref{fig:results}. 
The SM predictions are based on Ref.~\cite{straub:2018flavio}, using form factors for the $\Bs \to \phi$ transition from Refs.~\cite{Horgan:2013hoa,Bharucha:2015bzk}, 
modified to account for integration over decay time. 
Values of the angular observables measured in \decay{\Bs}{\phi\mumu} decays~\cite{LHCb-PAPER-2021-022} are also shown for comparison.
The results are seen to be consistent with both the SM predictions and the more precise measurements with \decay{\Bs}{\phi\mumu} decays.
Further improvement in precision, which will allow more stringent tests of the SM, can be anticipated with larger data samples in future.

\section*{Acknowledgements}
%
%
\noindent We express our gratitude to our colleagues in the CERN
accelerator departments for the excellent performance of the LHC. We
thank the technical and administrative staff at the LHCb
institutes.
We acknowledge support from CERN and from the national agencies:
ARC (Australia);
CAPES, CNPq, FAPERJ and FINEP (Brazil); 
MOST and NSFC (China); 
CNRS/IN2P3 (France); 
BMBF, DFG and MPG (Germany); 
INFN (Italy); 
NWO (Netherlands); 
MNiSW and NCN (Poland); 
MCID/IFA (Romania); 
MICIU and AEI (Spain);
SNSF and SER (Switzerland); 
NASU (Ukraine); 
STFC (United Kingdom); 
DOE NP and NSF (USA).
We acknowledge the computing resources that are provided by ARDC (Australia), 
CBPF (Brazil),
CERN, 
IHEP and LZU (China),
IN2P3 (France), 
KIT and DESY (Germany), 
INFN (Italy), 
SURF (Netherlands),
Polish WLCG (Poland),
IFIN-HH (Romania), 
PIC (Spain), CSCS (Switzerland), 
and GridPP (United Kingdom).
We are indebted to the communities behind the multiple open-source
software packages on which we depend.
Individual groups or members have received support from
Key Research Program of Frontier Sciences of CAS, CAS PIFI, CAS CCEPP, 
Fundamental Research Funds for the Central Universities,  and Sci.\ \& Tech.\ Program of Guangzhou (China);
Minciencias (Colombia);
EPLANET, Marie Sk\l{}odowska-Curie Actions, ERC and NextGenerationEU (European Union);
A*MIDEX, ANR, IPhU and Labex P2IO, and R\'{e}gion Auvergne-Rh\^{o}ne-Alpes (France);
Alexander-von-Humboldt Foundation (Germany);
ICSC (Italy); 
Severo Ochoa and Mar\'ia de Maeztu Units of Excellence, GVA, XuntaGal, GENCAT, InTalent-Inditex and Prog.~Atracci\'on Talento CM (Spain);
SRC (Sweden);
the Leverhulme Trust, the Royal Society and UKRI (United Kingdom).


\addcontentsline{toc}{section}{References}
\bibliographystyle{LHCb}
\bibliography{main,standard,LHCb-PAPER,LHCb-CONF,LHCb-DP,LHCb-TDR}

\newpage
\centerline
{\large\bf LHCb collaboration}
\begin
{flushleft}
\small
R.~Aaij$^{38}$\lhcborcid{0000-0003-0533-1952},
A.S.W.~Abdelmotteleb$^{57}$\lhcborcid{0000-0001-7905-0542},
C.~Abellan~Beteta$^{51}$\lhcborcid{0009-0009-0869-6798},
F.~Abudin{\'e}n$^{57}$\lhcborcid{0000-0002-6737-3528},
T.~Ackernley$^{61}$\lhcborcid{0000-0002-5951-3498},
A. A. ~Adefisoye$^{69}$\lhcborcid{0000-0003-2448-1550},
B.~Adeva$^{47}$\lhcborcid{0000-0001-9756-3712},
M.~Adinolfi$^{55}$\lhcborcid{0000-0002-1326-1264},
P.~Adlarson$^{84}$\lhcborcid{0000-0001-6280-3851},
C.~Agapopoulou$^{14}$\lhcborcid{0000-0002-2368-0147}
% ...

{\footnotesize \it

% ...
$^{14}$Universit{\'e} Paris-Saclay, CNRS/IN2P3, IJCLab, Orsay, France\\
% ...
$^{38}$Nikhef National Institute for Subatomic Physics, Amsterdam, Netherlands\\
% ...
$^{47}$Instituto Galego de F{\'\i}sica de Altas Enerx{\'\i}as (IGFAE), Universidade de Santiago de Compostela, Santiago de Compostela, Spain\\
% ...
$^{51}$Physik-Institut, Universit{\"a}t Z{\"u}rich, Z{\"u}rich, Switzerland\\
% ...
$^{55}$H.H. Wills Physics Laboratory, University of Bristol, Bristol, United Kingdom\\
% ...
$^{57}$Department of Physics, University of Warwick, Coventry, United Kingdom\\
% ...
$^{60}$School of Physics and Astronomy, University of Glasgow, Glasgow, United Kingdom\\
$^{61}$Oliver Lodge Laboratory, University of Liverpool, Liverpool, United Kingdom\\
% ...
$^{69}$Syracuse University, Syracuse, NY, United States\\
% ...
$^{84}$Department of Physics and Astronomy, Uppsala University, Uppsala, Sweden, associated to $^{60}$\\
% ...
}
\end{flushleft}

\end{document}